\documentclass[%
 reprint,
 amsmath,amssymb,
 aps,
]{revtex4-2}

\usepackage{graphicx}
\usepackage{dcolumn}
\usepackage{bm}
\usepackage{xcolor}
\usepackage{mathtools}
\usepackage{appendix}

\usepackage[normalem]{ulem} 
\usepackage{afterpage}
\usepackage{ulem}

\newcommand{\ket}[1]{\left\vert #1 \right\rangle}
\newcommand{\bra}[1]{\left\langle #1 \right\vert}

\begin{document}

\preprint{APS/123-QED}

\title{Error-Robust Quantum Signal Processing using Rydberg Atoms}

\author{Sina Zeytino\u{g}lu}
\affiliation{Department of Physics, Harvard University, Cambridge, Massachusetts 02138, USA}
\email{sina\_zeytinoglu@fas.harvard.edu}
\author{Sho Sugiura}%
\affiliation{Physics and Informatics Laboratory, NTT Research, Inc.,
Sunnyvale, California, 94085, USA}
\affiliation{Laboratory for Nuclear Science,
Massachusetts Institute of Technology,
Cambridge, 02139, MA, USA
}%

\date{\today}

\begin{abstract}

Rydberg atom arrays have recently emerged as one of the most promising platforms for quantum simulation and quantum information processing. However, as is the case for other experimental platforms, the longer-term success of the Rydberg atom arrays in implementing quantum algorithms depends crucially on their robustness to gate-induced errors. Here we show that, for an idealized biased error model based on Rydberg atom dynamics,  the implementation of QSP protocols can be made error-robust, in the sense that the asymptotic scaling of the gate-induced error probability is slower than that of gate complexity. Moreover, using experimental parameters reported in the literature, we show that QSP iterates made out of up to a hundred gates can be implemented with constant error probability.
To showcase our approach, we provide a concrete blueprint to implement QSP-based near-optimal Hamiltonian simulation on the Rydberg atom platform. Our protocol substantially improves both the scaling and the overhead of gate-induced errors in comparison to those protocols that implement a fourth-order product-formula.

\end{abstract}

\maketitle


\section{\label{sec:level1}Introduction}

Neutral atoms have become a leading experimental platform for accomplishing useful quantum information processing tasks \cite{brennen1999quantum,briegel2000quantum,jaksch2000fast,brion2007quantum,molmer2011efficient,saffman2016quantum,Weiss2017,adams2019rydberg,henriet2020quantum}, as well as emulating a variety of non-trivial Hamiltonian dynamics \cite{bernien2017probing} and correlated states \cite{browaeys2020many,omran2019generation,Samajdar2021,verresen2020prediction,semeghini2021probing}. In this success, the rich physics of neutral atoms has played an essential role. On the one hand, the tightly-confined hyperfine states of the atoms interact very weakly with the environment \cite{weiss2017quantum}, making these states ideal for storing quantum information \cite{Lukin1998,Lukin2000,bajcsy2003stationary,choi2008mapping}. On the other hand, the extended Rydberg states enable strong interactions between the atoms \cite{ urban2009observation}, allowing fast and high-fidelity multi-qubit gates to be realized \cite{brennen1999quantum,jaksch2000fast,lukin2001dipole}. Moreover, the advances in trapping and manipulating alkali-earth atoms resulted in drastic improvements in the error characteristics of the one- and two-qubit gates on the neutral atom platform \cite{madjarov2020high,Wilson2022,ma2022universal}, making it an important contender to other leading platforms based on trapped ions \cite{haffner2008quantum,home2009complete} and circuit Quantum Electrodynamics \cite{wallraff2004strong,blais2021circuit}. A distinctive advantage of neutral atoms compared to the other platforms is that they can be trapped close to one another, resulting in  a scalable and dynamically reconfigurable \cite{weiss2017quantum,bluvstein2021quantum} architecture. Similarly, the rich internal structure of neutral atoms results in a uniquely versatile setup where both the unitary and dissipative dynamics of the system can be tailored for the specific quantum information task at hand \cite{weimer2010rydberg,wilson2019trapped,barnes2022assembly,ma2022universal,cong2021hardware}. 

Yet, as is the case with all current experimental platforms for realizing quantum computation, Rydberg atoms cannot be controlled without inducing significant unwanted dynamics. Consequently, the protocols implemented for processing quantum information involve errors and the resulting computation is unreliable \cite{Zhou2020,Oh2021,pan2022solving}. While fault-tolerant error-corrected quantum computation is in principle possible \cite{aharonov1997fault,preskill1998reliable,knill2005quantum}, the resources necessary for reaching the error-correction threshold with the error rates achieved in current experiments is daunting \cite{nielsen2002quantum}, despite promising developments \cite{cong2021hardware}.
A direct way to reduce this resource cost is to increase the robustness of the system against errors \cite{nielsen2002quantum}. In particular, it is desirable to realize \textit{error-robust} implementations, where the error probability associated with the implementation scales slower than the gate complexity of the corresponding circuit. Whether the rich physics of the Rydberg atoms can be leveraged to realize error-robust implementations is crucial for the success of the platform. 

Here we design error-robust implementations of a wide range of quantum algorithms on the Rydberg atom platform. 
We achieve such generality by considering implementations of different instantiations of Quantum Signal Processing (QSP) \cite{low2017quantum,gilyen2019quantum}, a framework which unifies Hamiltonian simulation, unstructured search as well as phase-estimation \cite{MartynChuang2021}. In particular, we demonstrate that, assuming an idealized error model based on the physics of Rydberg atoms, the central oracle for the QSP framework, called the block-encoding unitary \cite{Chakraborty2019}, can be implemented with constant error probability with respect to the gate complexity of the corresponding circuit. Moreover, we show that in the parameter regime that is routinely reported in the literature \cite{omran2019generation,Levine2019}, it is possible to realize a hundred-fold reduction of the error probability. 

Our approach consists of two steps. First, we determine the characteristics of an error model which can reduce the error probability for a particular compilation of the block-encoded unitary, given by the Linear Combinations of Unitaries (LCU) \cite{childs2012hamiltonian}. Second, we design Rydberg atom gates that realize the desired biased error model. 
Two main observations help us drastically reduce the error probability associated with the Rydberg atom implementation of LCU, which consists of a state preparation unitary and its inverse, in addition to a sequence of controlled unitaries. First, we observe that the error probability associated with the sequence of controlled unitary operations is reduced drastically if each controlled unitary induces errors only when the control condition is satisfied. Motivated by this observation, we then discuss biased-error controlled unitaries that can be implemented on the Rydberg atom platform.
Consequently, given an ideal implementation of such \textit{biased-error controlled unitaries}, the error probability associated with the LCU protocol scales only with that of the state preparation step.
Motivated by this second observation, we determine a special class of states that can be prepared efficiently using the long-range dipolar interactions between the Rydberg states. In particular, we design a Rydberg blockade gate that prepares any state in the span of computational basis states with one non-zero element in constant time and with constant error probability. We refer to these states as One-Hot amplitude Encoding (OHE) states, and also design schemes for error-robust generation of a more general class of states called $k$-Hot Encoding ($k$HE) states, which are in the span of computational basis states with $k$ non-zero elements. Importantly, the sparse encoding realized by the $k$HE states can be utilized to achieve a scalable architecture. Specifically, when we are interested in general linear combinations of $k$-local Pauli operations, the $k$HE states allow us to use an ancillary register whose size is proportional to that the register used for processing quantum information.
To the best of our knowledge, our results provide the first discussion of error-robust implementations.

The paper is organized as follows. We provide a summary of the main results and insights in Section $\ref{sec:MainResults}$. In Section $\ref{sec:LCU}$, we introduce QSP based on a block-encoding unitary \cite{Chakraborty2019} implemented with LCU \cite{childs2012hamiltonian}. We also show that the structure of the LCU protocol can be leveraged to drastically reduce the effects of errors with low error state preparation and an biased-error controlled unitaries. In Section $\ref{sec:MainStrat}$, we design Rydberg atom gates that have the desired biased error characteristics. We then provide concrete error-robust implementations of QSP protocols on the Rydberg atom platform in Section $\ref{sec:DesignPrinc}$ and show that the error-robustness is scalable in Section $\ref{sec:scalable}$. We showcase our approach in Section $\ref{sec:HamSimImpl}$ by error bounds for the implementation of a QSP-based near-optimal Hamiltonian simulation algorithm and provide a comparison to the numerically optimized fourth-order product formula \cite{childs2018toward}. We conclude with a discussion of our results in Section $\ref{sec:conclusions}$. 

\setbox0\vbox{The sections (you need to make a coherent story out of this)
\begin{enumerate}
    \item \textcolor{red}{The conditions for error-robust implementation of LCU-based QSP} LCU and QSP introduction, we notice the flexibility that LCU gives us to design the ancillary register. We also discuss the tree-like structure of the data that is encoded in the system dynamics through LCU, and argue how this structure provides the conditions under which the implementation of QSP can be made error-robust. 
    \item \textcolor{red}{Rydberg atom gates for error-robust QSP} Next section, we introduce two gates which fulfill the conditions described in the prev. section on the Rydberg atom platform. Because it is not easy to fullfill the conditions naively, the gates are codesigned with the address register. Here we discuss the situation when the error biases are not perfect, which is the case when the Rydberg blockade is not the dominant energy scale. We show that even in this case we get a cubic advantage compared to the conventional gate count. 
    We codify the rules in an EBGC (should I put this after the discussion of the gates? Why did I move it before in the first place?). We detail the restriction of the Rydberg atom gates which limits parallelization, and compare the blockade gates to other mediated gates.
    \item \textcolor{red}{Designing an error-robust ancillary register} Next section, we give a concrete design of the protocols, and discuss the trade-offs between error-robustness, size of the address register, and the depth of the circuit implementing the block-encoding unitary. Our results show that the structure of the ancillary degrees of freedom can be leveraged greatly, especially in qudit systems. 
    \item \textcolor{red}{Scalable implementation} Next section, we consider the scalable implementation of our protocols. The error bounded gate count with the number of subsystems each in the scale of the Rydberg blockade. The circuit depth saturates with the lieb-robinson bound and is therefore optimal. (use the k-Hot encoding here to reduce the error. Do we need the scalable implementation of the V?) The errors are still suppressed in this case and the errors are again capped by $O(n_{\rm site})$.
    \item \textcolor{red}{Hamiltonian simulation comparison} In the last section, we consider the Hamiltonian simulation problem.
    
\end{enumerate} }

\section{Main Results and ideas}
\label{sec:MainResults}

We consider error-robust implementations that arise from the interplay between gate-induced error mechanisms and circuits compiling QSP protocols at multiple layers of abstraction. At the highest level, we determine the characteristics of an idealized error model sufficient for error-robust implementations of QSP protocols. Then, we go down to the hardware level and design Rydberg atom gates which, in a suitable parameter regime, exhibit the characteristics of such an idealized error model. At the system level, we show that the error-robust implementation is scalable, considering the finite range of interactions between the Rydberg atoms. Finally, we highlight the potential of our approach by calculating the error probability for an implementation of QSP-based Hamiltonian simulation. In this section, we provide an informal discussion of the main insights and results pertaining to each level.

\subsection{The conditions for the error-robust implementation of LCU-based QSP}

A great variety of quantum protocols are described as functional transforms $f(A)$ of high-dimensional linear operators $A$. The well-known examples include Hamiltonian simulation, where $f(H)=e^{-iH t}$ \cite{lloyd1996universal} and HHL algorithm for solving linear equations, where $f(A)=A^{+}$ with $+$ denoting the Moore-Penrose pseudo-inverse \cite{harrow2009quantum} \footnote{see Ref. \cite{gilyen2019quantum,MartynChuang2021} for further examples}. The naive \textit{expectation} is that the compilation of such algorithms is simple when $f(\cdot)$ and the input $A$ have simple classical descriptions.

Quantum Signal Processing (QSP) is an \textit{iterative} compilation method that formally fulfills this naive expectation when $f(\cdot)$ is approximated by a low-order polynomial, and $A$ is sparse or approximated by a linear combination of a small number of Pauli strings \footnote{ In contrast, compiling time-dependent Hamiltonian simulation is difficult because then multiple functional transformations $f_t(\cdot)$ and their inputs $A_t$ need to be explicitly specified. }.
Each iteration step of the QSP protocol has two components, called the block-encoding walk operator $W_A$ \cite{Chakraborty2019}, which encodes the linear operator $A$ (i.e., there exists a projector $\Pi$ such that $\Pi W_A \Pi = A$), and the processing unitary \cite{low2017quantum} which encodes a single rotation angle $\phi_i$. For a QSP protocol that terminates after $l$ iterations, the list of angles $\{\phi_i\}$ determines the $l^{\rm th}$-order polynomial approximation of the functional transform $f(\cdot)$. 

The QSP protocols can be simplified drastically when the controlled version of $W_A$ (C$W_A$) is available. Then, the processing unitary is a single-qubit rotation of the control qubit. This is an important simplification from the perspective of error-robust implementation since the single-qubit rotation only contributes a constant to the error probability \textit{per} iteration. 
Consequently, the scaling of the error probability associated with each iteration step of the QSP protocol is the same as the scaling of errors for C$W_A$. \textit{In other words, whether we can achieve an error-robust implementation of the QSP protocol hinges on an error-robust implementation of C$W_A$.}

We find that the Linear Combination of Unitaries (LCU) method is an especially well-suited compilation method for an error-robust implementation of $W_A$. In this method, $A$ is decomposed as a linear combination of $N$ unitary Pauli strings $\{P_i\}$, with the associated coefficients $\{\alpha_i\}$. In the LCU protocol, the data consisting of $\{\alpha_i\}$ and $\{P_i\}$ are encoded by two separate unitaries $V$ and $\bar{U}$, respectively. The state preparation unitary $V$ acts on an ancillary register of size $n_{a}$ ($V\ket{0}^{\otimes n_a}=\sum_{i}\alpha_i \ket{x_i}$), and amplitude-encodes coefficients $\{\alpha_i\}$. On the other hand, $\bar{U}$ takes the different components $\{\ket{x_i}\}$ of the ancillary state as control conditions for applying $\{P_i\}$ to the system register. Formally, $\bar{U}$ can be expanded as $\bar{U}\equiv \prod_{i}^{N}C_{x_i}P_i$. 

We show that the following two conditions are sufficient for an error-robust implementation of $W_A \equiv (2\Pi-\mathbf{I}) V^{\dagger}\bar{U}V$ and its controlled version:
\begin{itemize}
    \item[]Condition 1: For controlled unitaries, error probability is negligible when the control condition is \textit{not} satisfied
    \item[] Condition 2: Controlled version of One-Hot Encoding state-preparation takes constant time/error.
\end{itemize}
Here, we define a $k$-Hot Encoding state as any superposition of bitstrings with $k$ entries in the excited state (e.g., $\ket{1}$).

We show that Condition 1 is sufficient for achieving a dramatically error-robust implementation of $\bar{U}$. On the other hand, through Condition 2, we can design an ancillary register that facilitates the error-robust implementation of $W_A$. We also show that the controlled version of $W_A$ can be implemented without changing the scaling of error probability. Designing a Rydberg atom implementation of C$W_A$ which satisfies these two conditions is the goal of our paper.

\subsection{Designing biased-error Rydberg atom gates}

In order to satisfy Condition 1, we design \textit{single-qubit-controlled} unitary gates which induce errors only when the control condition is satisfied.
Such a single-qubit-controlled unitary was proposed in Ref. \cite{muller2009mesoscopic}. The gate uses the Rydberg-blockade effect in combination with Electromagnetically Induced Transparency (EIT) \cite{boller1991observation,lukin2000nonlinear}, and leverages the rich internal structure of the Rydberg atoms. While the gate was proposed more than a decade ago, to our best knowledge, our work is the first to emphasize its biased error characteristics and use it to achieve error-robust implementations of quantum algorithms.

We demonstrate that the single-qubit controlled gate introduced in Ref. \cite{muller2009mesoscopic} drastically reduces the probability of errors in both the \textit{control} and \textit{target} registers when the control condition is not satisfied (i.e., when the state $\ket{\psi}_c$ of the control atom has vanishing overlap with the control condition, say $\ket{0}_c$,). Similar to other multi-qubit gates that involve the Rydberg-blockade mechanism \cite{jaksch2000fast,lukin2001dipole}, the EIT-based gate protocol starts by exciting the control atom to the Rydberg state if it satisfies the control condition. During this step, the control atom in state $\ket{1}_c$ evolves trivially and does not acquire any gate-induced errors. As a result, the error probability due to the \textit{control} atom is negligible when the control condition is not satisfied.

On the other hand, EIT mechanism ensures that the error probability due to the dynamics of the target atoms can be drastically reduced when the control condition is not satisfied. In particular, when the control atom is not excited to the Rydberg state, the EIT mechanism ensures that the laser field that couples the hyperfine states to shorter-lived excited states is not absorbed (hence the name ``transparency"). Consequently, when the control condition is not satisfied, the evolution of the target atoms is nearly trivial. In contrast, when the control condition \textit{is} satisfied, the Rydberg excitation of the control atom disturbs the EIT mechanism, and the target qubit goes under a non-trivial and error-inducing evolution. As a result, EIT effect enables the Rydberg blockade gates satisfy Condition 1. 

There are two comments in order. 
First, in reality,
the error can never be perfectly biased with respect to the control condition. The ratio of the error probabilities conditioned on the two control conditions is determined by the ratio of two laser intensities in the EIT configuration (Fig. $\ref{fig:Levels}$ b). Specifically, in order to reduce the error probability by a factor of $N$, we need to increase the \textit{intensity} of a laser in by $O(N)$. In other words, the robustness to errors comes at the expense of increased classical resource requirements. Such a trade-off is also present for other controlled unitaries \cite{jaksch2000fast,lukin2001dipole,Levine2019}. However, the EIT-based gate has two characteristics that are advantageous: (i) the EIT-based gate
provides a quadratic advantage in laser intensity compared to conventional gate implementations, where an $N$-fold suppression of errors require an $O(N^2)$ fold increase of the laser intensity, (ii) the EIT-based gate is advantageous even when the laser drive amplitude is much larger than the dipolar interaction strength. 
Second, implementing a unitary that satisfies Condition~1 for general multi-qubit control conditions is not possible by selectively driving atoms as described above. Intuitively, given a multi-atom ancillary register, the local interactions between the laser field and the atoms cannot be configured such that only a single initial state goes through a nontrivial evolution. We address this issue by utilizing a tensor product of $k$ One-Hot Encoding address states. Whether the resulting $k$-Hot encoding state satisfies a $k$-bit control condition can be checked using $k$ single-qubit controlled Pauli operations. This step induces a trivial evolution on all but $k$ control qubits. As a result, a controlled-Pauli operation conditioned on such a $k-$Hot Encoding address state satisfies Condition 1. The non-negligible error probability when the control condition \textit{is satisfied} is only $O(k)$.

Finally, we use the previously reported values of the Rabi frequencies and decay rates to calculate the error probability expected for 100 single-qubit controlled unitaries conditioned on a One-Hot Encoding state to be less than 5 percent. As a result, the combination of our techniques with error-correction promises a significant advance in the realization of fault-tolerant quantum computation \cite{cong2021hardware}. 

\vspace{-0.2cm}
\subsection{Designing error-robust ancillary control register}

We satisfy Conditions 2 for the error-robust implementation of $W_A$ using a novel multi-qubit Rydberg blockage gate, referred to as the One-Hot amplitude-encoding gate $V_{\rm OHE}$. 

We show that a tensor product of $k$ One-Hot Encoding address states can be prepared using $O(k n_{\rm site})$ EIT-based single-qubit controlled $V_{\rm OHE}$ (denoted C$V_{\rm OHE}$) gates, with a total error probability of $O(k)$. Moreover, the reflection unitary required for the walk operator $W_A$ can be implemented in an error-robust way by simply changing the phases of some of the drive lasers implementing $V_{k{\rm OHE}}$. Lastly, the tensor product of $k$ One-Hot Encoding states allows us to encode $N$ amplitudes in a small ancillary register of size $O(k N^{1/k})$. The size of the ancillary register does not satisfy the theoretical lower bound $\Omega(\log{N})$. However, for a system register of $n_{\rm site}$ atoms, as many as $O(n_{\rm site}^{k})$ control conditions can be stored in an ancillary register of size $O(kn_{\rm site})$.

The implementation of C$V_{\rm OHE}$ gates fully utilize the rich physics of the Rydberg atoms, including the long-range dipolar interactions, availability of even and odd parity Rydberg states, as well as EIT. Our results thus highlight the importance of concrete physical processes for realizing error-robust implementations. On the other hand, the scaling results above assume that the range of dipolar interactions is larger than the geometric size of the system and that one laser amplitude in the EIT configuration can be increased as $O(\sqrt{N})$. To codify the rules for calculating the error probability under these assumptions, we define the Error Bounded Gate Count (EBGC). Our main result is that when EBGC is valid and $N = O(n_{\rm site}^k)$, the LCU-based walk operator can be implemented with constant error and $O(k n_{\rm site})$ ancillae.
 
\subsection{Scalable implementation and Hamiltonian simulation}

The designs discussed so far assumed that the interaction range of the dipolar interactions between the Rydberg atoms is infinite. However, in reality, the dipolar interactions are effective only up to a fixed length scale, the so-called Rydberg blockade radius. When the finite range of the Rydberg blockade effect is taken into account, the scaling of the error probability with increasing system size depends on the number of subsystems $n_{\rm sub}$ whose geometric size is smaller than  the Rydberg blockade volume. We show that as long as the EBGC is valid, it is possible to implement each iteration of the QSP protocol with error probability that scales with $O(n_{\rm sub})$. Because the EBGC scaling is independent of the number of gates acting on each subsystem, the resulting implementation is error-robust. 

Finally, we showcase our approach and compare the error-robustness of the Rydberg implementation of the QSP-based Hamiltonian simulation algorithm to that of a simulation algorithm based on the fourth-order product formula. For a fair comparison, we implement the product formula algorithm using the biased-error Rydberg atom gate-set designed for QSP protocols. Hence, implementations of the two algorithms enjoy increased robustness to errors. Still, when EBGC is valid, the scaling of error probability is the same as the optimal gate complexity, and the associated overhead is reduced with respect to the fourth-order product formula by more than an order of magnitude. 

\section{Block encoding by LCU}
\label{sec:LCU}

Here we discuss the method of LCU \cite{childs2012hamiltonian}, which offers a generic and constructive strategy to implement block-encoding unitaries for linear combinations of multi-qubit Pauli operators. In order to assess the time and space complexities of the LCU method, we introduce the scaling variable $N$ which denotes the number of Pauli operators that constitute the target operator $A$. In particular, we decompose $A$ as 
\begin{align}
    A = \sum_{i=1}^{N}|\alpha_i|^2 P_i,
    \label{eq:BlockEncOp},
\end{align}
where we set $\sum_{i=1}^{N}|\alpha_i|^2=1$.
In the context of Hamiltonian simulation, the number of coefficients required to implement a $k$-local Hamiltonian on a system consisting of $n_{\rm site}$ qubits is $N = O(n_{\rm site}^k)$, while for geometrically local Hamiltonians where the number of atoms within an interaction range is $N_{I}$, we have $N = O(N_{I}^k n_{\rm site})$. It is important to note that in this decomposition we assume that the coefficients $\{\alpha_i\}$ are given and cannot be further compressed into a smaller set. 

In the following, we first review the LCU method formally, and then discuss how its structure can be interpreted as a in a circuit that loads the classical data describing $A$ into a quantum processor.

\subsection{Algorithm:}

The LCU decomposition of the block-encoding unitary in Eq. ($\ref{eq:GoalUnitary1}$) consists of three unitaries \cite{childs2012hamiltonian}.
\begin{align}
U = V^{\dagger}\bar{U}V.
\label{eq:BlockEnc}
\end{align}
The block-encoding unitary acts on $n_a$ ancilla qubits and $n_{\rm site}$ system qubits.
The unitary $V$ rotates the $n_a$-qubit initial ancilla state $\ket{0}^{\otimes n_a}$ to a linear combination of the computational basis states $\{\ket{x_i}\}$ which encode the pre-computed classical coefficients $\alpha_i$ 
\begin{align}
    \ket{\Psi}_{a}\equiv V \ket{0}^{\otimes n_a} =
    \sum_{i=1}^{N} \alpha_i \ket{x_i}.
    \label{eq:anc_prep}
\end{align}
The operator $V$ can be understood as an amplitude-encoding state-preparation unitary \cite{Rebentrost2014}. We note that the number of ancilla qubits $n_a>\lceil\log{N}\rceil$ depends on the choice of the basis $\{\ket{x_i}\}$. 

Then, we apply the following conditional unitary operation
\begin{align}
    \bar{U} \equiv \sum^{N}_{i} \ket{x_i}\langle x_i| \otimes P_i.
    \label{eq:Ubar}
\end{align}
The action of $\bar{U}$ entangles each Pauli operator with an orthogonal \textit{address} state of the ancilla register
\begin{align}
  \sum_{i=1}^{N} \alpha_i \ket{x_i} \otimes\ket{\Psi_{\rm sys}} \xrightarrow{\bar{U}} \sum_{i=1}^{N}  \ket{x_i} \otimes \left(\alpha_i P_i\right)\ket{\Psi_{\rm sys}}.
   \label{eq:qROM}
\end{align}


Finally, a block-encoding of a superposition of multi-qubit Paulis $\{P_i\}$ is obtained by rotating the address space by an application of $V^{\dagger}$ 
\begin{align}
    \nonumber &V^{-1}\bar{U} \ket{\Psi}_{a}\otimes \ket{\Psi_{sys}} = \sum_{i}^{N} |\alpha_i|^2 \ket{0}^{\otimes n_a} \otimes P_i \ket{\Psi_{sys}} + \ket{\Psi^{\perp}}\\
    &= \ket{0}^{\otimes n_a} \otimes \left[ A \ket{\Psi_{sys}} \right] + \ket{\Psi^{\perp}}
    \label{eq:GoalUnitary1}
\end{align}
where the unnormalized wavevector $\ket{\Psi^{\perp}}$ satisfies 
$\left(\left(\ket{0}\bra{0}\right)^{\otimes n_a}\otimes\mathbf{1}\right) \ket{\Psi^{\perp}} \equiv \Pi_0 \ket{\Psi^{\perp}} = 0 $. Consequently, $\Pi_0 U \Pi_0 = A$, and the block-encoding unitary has the form
\begin{align}
    U \dot{=}\left( \begin{array}{cc}
    A  & *\\
    *    & *
    \end{array}\right).
\end{align}
We remind the reader that the unitarity of $U$ implies that the Hermitian operator block-encoded in this way satisfies $||A|| \leq 1$. Moreover, the block-encoding unitary implemented through LCU is Hermitian (i.e., $U^{\dagger}=U$). 

\subsubsection{Freedom to design the address register} As we noted before, the ancillary Hilbert space is not constrained in the above discussion. While the original discussion of block-encoding unitary sets $n_a = \lceil\log{(N)}\rceil$ \cite{Chakraborty2019}, we refrain from this choice. 
Indeed, we show that by designing the address register (i.e., the bitstrings $\{x_i\}$) is useful for constructing error-robust implementations of the block-encoding unitary.

Indeed, there are infinitely many ancillary states which result in a block-encoding of the same signal operator. To see this, divide the ancilla register into two parts $a_1$ and $a_2$ consisting of $n_{a_1}$ and $n_{a_2}$ ancillary qubits, respectively. Then we can construct two state-preparation unitaries $V_a$ and $V_p$ that are equivalent from the perspective of LCU-based block-encoding, if $\bar{U}$ acts on only the system and second ancilla register $a_2$:
\begin{align*}
    &\ket{0}^{\otimes n_{a_1}}\ket{0}^{\otimes n_{a_2}}\xrightarrow{V_{\rm p}} \ket{0}^{\otimes n_{a_1}}\sum_{i}^{N}\alpha_i\ket{x_i}\\
    &\ket{0}^{\otimes n_{a_1}}\ket{0}^{\otimes n_{a_2}}\xrightarrow{V_{\rm a}} \sum_{i}^{N}\alpha_i\ket{\Psi_i}\ket{x_i},
\end{align*}
where the states $\{\ket{\Psi_i}\}$ can be any state of the Hilbert space of $n_{a_1}\geq n_{a_2}$ qubits. This property will be important in our discussion of the scalable and error-robust implementation of the state-preparation unitary in Section $\ref{sec:scalable}$.

\subsection{Processing of block encoded matrices by QSP}
\label{sec:QSP}


Next, we review QSP framework introduced in Refs. \cite{low2017quantum,low2019hamiltonian}. From the perspective of compilation of quantum subroutines, QSP can be understood as a efficient way of manipulating a block-encoded operator $A$ to realize the block-encoding of a polynomial function $P(A)$. The polynomial $P(A)$ is defined through an ordered list of $n$ angles $\{\phi_i\}$, whose size determines the order of the polynomial as well as the query complexity of QSP. Here, we only give a brief discussion of the QSP protocol such that the requirements for its error-robust implementation are evident. For an introduction to QSP see Appendix \ref{app:QSP}.

\begin{figure*}
    \centering
    \includegraphics[width=\textwidth]{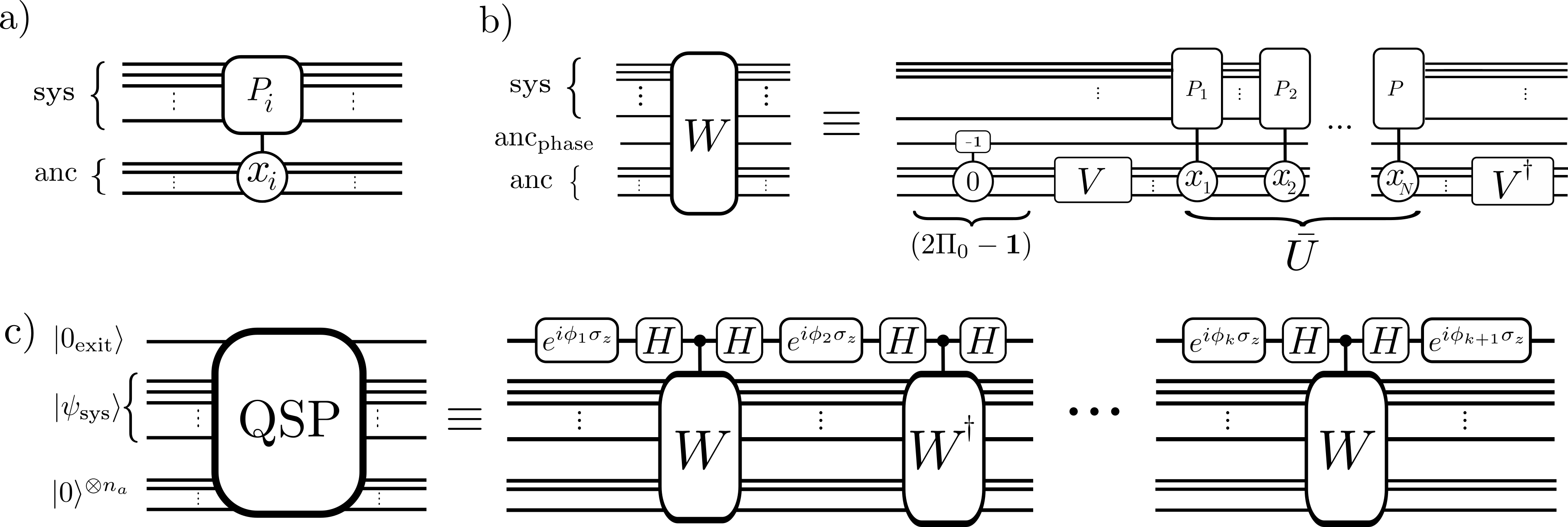}
    \caption{The circuit diagrams for the protocols discussed in this work. a) The multi-qubit controlled Pauli operation $C_{x_i}P_i$ where the conrol register is ancillary and the target register is the system which is controlled. b) The decomposition of the QSP iterate $W$ (notice the factor of $i$ difference with the definition in Ref. \cite{low2017quantum}), consisting of a multi-qubit controlled phase gate where the target register is a single ``phase" ancilla and the LCU-based block-encoding protocol.
    c) The circuit decomposition of the QSP method for producing polynomials of block-encoded matrices, which consists of controlled versions of the walk operator $W$ conditioned on the state of an additional ``exit ancilla". In this work we show how to realize error-robust implementations of these circuits on the Rydberg atom platform.}
    \label{fig:QSP}
\end{figure*}

The QSP protocols proceed by iterating between a controlled oracular unitary C$W$ derived from the block-encoding unitary $U$ in Eq. ($\ref{eq:BlockEnc}$), and a signal processing step, which consists of single qubit rotations on the ``exit" ancilla that controls $W$ (see Fig. \ref{fig:QSP}). Formally, the QSP protocol has the form 
\begin{align}
\mathcal{U} = \left[\prod_{i=1}^{n} e^{i\phi_i\sigma_x^{(e)}} \mathrm{C}_{e}W \right]e^{i\phi_0\sigma_z^{(e)}},
\end{align}
where $\sigma_x^{(e)}$ acts on the exit ancilla.
The phases associated with the single qubit rotations in the processing step define the polynomial function $P(A)$ that is block-encoded by the resulting unitary transformation. In the case of a qubitized block-encoding unitary $U=U^{\dagger}$ oracular unitary $W$ is simply expressed as 
\begin{align}
    W=(2\Pi_0 - \mathbf{I})U,
    \label{eq:QubitWalk}
\end{align}
where $\Pi_0$ is the projector to the all-zeros address state. 

\subsection{Requirements for an error-robust implementation of the QSP protocols}

The QSP protocols can be thought of as a compilation strategy for quantum algorithms. The structure of the LCU-based QSP protocols allows one to reduce the adverse effects of errors when (i) the required controlled unitaries are implemented in a way that the errors are induced only when the control condition is satisfied, and (ii) the state preparation unitary $V$ can be implemented with constant error scaling. 

To see how these requirements result in a drastic reduction of errors in the implementation of the QSP oracle $W$, first consider the two components of the QSP protocols that use controlled gates extensively:  the unitary $\bar{U}$ in Eq. ($\ref{eq:BlockEnc}$) and the reflection operator $2\Pi_0-\mathbf{I}$. It is crucial to notice that when each of the controlled Pauli gates in Eq. ($\ref{eq:Ubar}$) are implemented in a way that errors are induced only when the address register is in the desired state, then the total error probability associated with $\bar{U}$ is constant with respect to $N$ and scales linearly with the highest weight $k=O(n_{\rm site})$ of the Pauli strings $\{P_i\}$. A similar implementation of the controlled unitary implementing the reflection operator $2\Pi_0-\mathbf{I}$ result in a constant error per reflection gate. As a result, the oracular unitary $W$ which involves only $k$-local Paulis can be implemented with constant error if the state preparation unitary $V$ can be implemented with constant error. 

An implementation of each QSP iterate which has only a constant error probability entails that the error probability of implementations of QSP protocols has the same scaling as that of the query complexity, which is optimal query with respect to the approximation error when $P(A)$ approximates a smooth function \cite{low2017quantum}. In summary, the (i) biased-error controlled unitaries and (ii) constant error state preparation unitary are sufficient for implementing QSP protocols with near optimal scaling of the error probability with respect to the approximation error. In the next section, we design the ancillary address register for the QSP protocol in a way that allows the above requirements to be satisfied for the implementations of $\bar{U}$, $2\Pi_0-\mathbf{I}$, and $V$ on the Rydberg atom platform. 

\section{Rydberg atom Gates}
\label{sec:MainStrat}

In this section, we introduce the building blocks for error-robust implementions of QSP protocols on the Rydberg atom platform. We start the section with an introductory discussion of what constitutes an error-bounded gates, and how to calculate the error-bounded gate count (EBGC) of a particular protocol implemented using idealized versions of the proposed Rydberg gates. Crucially, EBGC does not correspond to the gate complexity of the circuit decomposition of the protocols in terms of the Rydberg gates, as it takes into account the information of the input states. Indeed, that the error probability does not have to scale as the gate complexity is what makes error-robust implementations possible.

We briefly review the relevant level diagrams and single-qubit gates in Section $\ref{subsec:LevelDiag}$. In Sections $\ref{sec:AmpEncGate}$ and $\ref{sec:CmUk}$, we introduce two multi-qubit gates utilizing the Rydberg blockade mechanism. Each multi-qubit gate serves a different function in the error-robust implementation of the LCU-based block-encoding unitary. The first multi-qubit gate, which we name ``One-Hot Encoding" (OHE) gate, (see Section $\ref{sec:AmpEncGate}$) allows us to load the classically-stored coefficient data $\{\alpha_i\}$ efficiently to orthogonal ancillary address states. The OHE gate is the building block of the state preparation unitary $V$ of the LCU protocol [see Eq. ($\ref{eq:BlockEnc}$) ]. Surprisingly, when the Rydberg blockade radius is infinite, the gate takes constant time and EBGC. In Section $\ref{sec:CmUk}$, we introduce a multi-qubit controlled Pauli operation, which can be expressed formally as, 
\begin{align}
    C_{x_i}P_i \equiv \ket{x_i}\bra{x_i}\otimes P_i + \left[\mathbf{1} - \ket{x_i}\bra{x_i}\right] \otimes \mathbf{1},
\end{align} 
where the bitstring $x_i$ will be referred to as the address or the control condition. Intuitively, the unitaries $\{C_{x_i}P_i\}$ are the building blocks of $\bar{U}$ in Eq. ($\ref{eq:Ubar}$) and they ``load" the classical data describing the Pauli strings $\{P_i\}$ in the decomposition of the block-encoded operator $A$ (see Section $\ref{sec:LCU}$) into quantum mechanical address states $|x_i\rangle$. 

The results of this section sets the stage for a concrete blueprint of an efficient and scalable implementation of the QSP-based optimal Hamiltonian simulation of Refs. \cite{low2019hamiltonian,haah2021quantum}, including the geometric arrangement Rydberg atoms and pulse sequences.


\subsection{Error-bounded gate counts (EGBCs) and the subadditivity of errors}
\label{sec:subadditivity}


In the following, we define an error-bounded gate count (EBGC) to quantify the way that the error probability grows as a function of scaling variables $n_{\rm site}$ and $N$. Conventionally, the gate counts are equated to the size of a quantum circuit. Here, the relationship between the circuit size and the error probability is established by via the subadditivity property of errors \cite{childs2017lecture}, which gives an upper bound for the spread of the errors introduced with each additional gate. However, the subadditivity bound may be extremely loose for a given protocol as it completely disregards both the structure of input states as well as the structure of the errors specific to an experimental implementation, which may be biased to introduce increase the error probability differently for different input states. Here, on the other hand, we count gates in a way that is dependent on their input states, with the aim of capturing when biased error model can be leveraged to achieve an error probability that scales slower than the gate complexity as a function of $N$.


The gate counting method, which we call the Error-Bounded Gate Count (EBGC) is based on an idealization of the Rydberg atom gates proposed in this work. It considers only the fundamental sources of error, given by non-adiabatic contributions and radiative decay processes, and assume that the error rates of each source is the same. In principle, the unwanted transitions due to blackbody radiation can also be included, given that we use optical pumping methods to convert such errors to dephasing errors \cite{cong2021hardware}. Our method assumes that the errors due to laser phase and amplitude fluctuations, as well as those due to the finite temperature atomic motion and the associated Doppler shift can all be eliminated \cite{Petrosyan2017,Low2012}. The finite lifetime of the hyperfine states is neglected given the orders of magnitude separation between this lifetime and the time it takes to implement the proposed gates \cite{yavuz2006fast}. We emphasize that although our error-bounded gate count is specific to the Rydberg atom platform, the strategy to design control protocols that take advantage of the biases in the relevant error model can be applied to any experimental platform. 

\subsubsection{Subadditivity of errors}
To put the discussion on firm footing, we sketch the proof of subadditivity of errors, and underline its shortcomings. Consider a circuit $C$ that can be described by an ordered product of $T$ unitaries $\{W_i\}$ [not to be confused with the walk operator $W$ in (\ref{eq:QubitWalk})] $C=\prod_{i}^{T}W_i$, and an imperfect implementation $\tilde{C}$ of $C$, where each $W_i$ is replaced by $\tilde{W}_i$. We assume $\tilde{W}_i$ to be unitary for simplicity. Now, given the same input state $\ket{\phi_0}$, we are interested in the difference between the outputs $\ket{\phi_T}$ and $\ket{\tilde{\phi}_T}$ of $C$ and $\tilde{C}$, respectively. Define
\begin{align}
    \ket{\phi_1}&\equiv W_1\ket{\phi_0}\\
    \ket{\tilde{\phi}_1}&=\tilde{W}_1\ket{\phi_0}=\frac{1}{\mathcal{N}}\left(\ket{\phi_1} + \ket{E_1}\right),
\end{align}
where we define the error vector $\ket{E_1}$ and the normalization $\mathcal{N}$. The size of the error vector satisfies the following inequality
\begin{align}
    \left|\left|\frac{\ket{E_1}}{\mathcal{N}}\right|\right|=\left|\left| \left(\tilde{W}_1-W_1 \right)\ket{\phi_0}\right|\right| \leq \left|\left| \left(\tilde{W}_1 - W_1 \right)\right|\right|_{\rm sup} \equiv \epsilon_1,
\end{align}
where the error $\epsilon_1$ associated with $\tilde{W}_1$ is defined via the spectral norm, which, crucially, is completely oblivious to the input vector $\ket{\phi_0}$. The worst case scenario is that all errors from each $W_{i}$ constructively interfere. Since $\{W_i\}$ are all unitary, the errors introduced by the $i^{\rm th}$ step is not amplified for any later step, and we obtain the inequality
\begin{align}
    \left|\left|\left(C-\tilde{C}\right)\ket{\Phi_0}\right|\right|\leq \sum_{i}\left|\left|  \tilde{W}_i - W_{i} \right|\right|_{\rm sup}=\sum_i \epsilon_i.
    \label{eq:subadditivity}
\end{align}
As a result, decomposing each $W_i$ using a universal gate-set with known error rates, we can relate the size of the circuit to the total error of the circuit. 
However, we emphasize again that in the above discussion the definition of errors $\epsilon_i$ in Eq. ($\ref{eq:subadditivity}$) is independent of the structure of the input state. To understand the shortcomings of this definition, notice that in the context of the LCU protocol, the omission of the particularities of $\ket{\phi_0}$ corresponds to forgetting about the fact that the ancillary registers are initiated in the $\ket{0}^{\otimes n_{a}}$ state and that we know how this initial state transforms at each step of our circuit. Our goal, on the other hand, is to use our knowledge of the trajectory of ancilla qubits to design error robust protocols. Hence, if we want to verify if any of our proposed implementations are error-robust, we need to make sure that we know how to calculate a bound for error probability given the knowledge of the states of the ancillary address register. 

EBGC is the tool that we develop to this end. In particular, we use the error-bounded gate count to take into account our knowledge of the biases of the error model and the knowledge of the input state at each step. Not surprisingly, we show that for most of our protocols, we obtain a better scaling of the number of gates than as indicated by Eq. ($\ref{eq:subadditivity}$). In the following, we introduce the rules for calculating the gate count for single-qubit rotations and controlled unitaries in the form $C_{x_i}U_1 \cdots U_k$ in an \textit{ad-hoc} manner. We support the models and assumptions that go into the EBGC with the physical error mechanisms relevant to the Rydberg atom system in Sections $\ref{subsec:LevelDiag}$, $\ref{sec:AmpEncGate}$, $\ref{sec:CmUk}$, and $\ref{sec:CVOHE}$.

\subsubsection{Error-bounded gate count (EBGC)}
\label{sec:EBGC}
We distinguish three factors which determine EBGC. These factors constitute the additional knowledge which makes error-robust implementations possible: (i) the rotation angle of single-qubit rotations (ii) the dimensionality of the local Hilbert space of each Rydberg atom, and (iii) the dependence of the errors introduced during controlled unitary operations on the state of the control register.
In the following, EBGC is normalized such that the Rydberg atom implementation of a CNOT gate requires at most 1 error-bounded gate.

As for the first factor, we observe that our protocols often use a continuous family of gates, such as single-qubit rotations by an arbitrary angle. In our error model, we assume that the error rate increases monotonically with the rotation angle. For example, given the single qubit rotation
\begin{align}
  R_{\theta}\ket{0} \equiv \cos{(\theta)}\ket{0} + \sin{(\theta)}\ket{1},  
\end{align}
the error associated with implementation of $R_{\theta}$ on the Rydberg atom platform is proportional to $|\theta/\pi|$.
More precisely, we assign an EBGC of $\left|\frac{\theta}{3\pi}\right|$ to $R_{\theta}$. Notice that this rule associates $1/3$  error-bounded gates (in units of the error probability of a CNOT gate) for each single qubit Pauli operator.

Second, the protocols discussed in the rest of the paper take advantage of the fact that each Rydberg atom has more than two-states. A local Hilbert space of more than two-dimensions entails that the experimentalist can choose laser pulses which only acts on a two-dimensional subspace of the local Hilbert space. As a result, the errors are introduced only when the Rydberg atom is in a state with a non-zero overlap with the subspace influenced by the laser pulse. Consider as an example a laser pulse sequence implementing the unitary that transfers an atom from the logical hyperfine state $\ket{1}$ to the Rydberg state $\ket{R}$ [the level diagram associated with each atom is discussed in more detail in Section $\ref{sec:Levels}$]. Given the initial state $\sqrt{1-|\alpha|^2}\ket{0} + \alpha \ket{1}$, the transfer has an EBGC of $1/3|\alpha|^2$ error-bounded gates.

We also use a generalization of this rule to count the number of gates associated with our multi-qubit One-Hot amplitude-encoding gate $V_{\rm OHE}$ in Section $\ref{sec:AmpEncGate}$ and its controlled counterpart in Section $\ref{sec:CVOHE}$. The most important property of these gates is that they utilize the strong Rydberg blockade effect in order to constraint the dynamics of, say, $N$ atoms onto a two-dimensional qubit-like subspace, and the EBGC calculates the gate count similarly to that of a single-qubit gate. As a result, the EBGC of $V_{\rm OHE}$ is independent of the number of qubits involved, and it is equivalent to that of a single CNOT gate.

Lastly, our gate count makes sure that the cost of controlled unitaries $C_{x_i}P_i$ are assessed in accordance with a physical error model in the limit that dipolar interactions set the highest energy scale. The EBGC sums up the error probability due to errors in the target and control registers separately.
While the errors in the control register occur while checking whether a control condition $x_i$ is satisfied, the errors in the target register are assumed to be introduced only when the state of the control (address) register satisfies the control condition. Hence, the contributions to the total error probability should be weighed by the probability that the control condition is satisfied. In Section $\ref{sec:CmUk}$, we discuss the concrete experimental protocol which can realize such a biased error model, assuming that the system is in a certain parameter regime.

As a concrete example, consider a single CNOT gate, where the control register is initially in $\ket{\psi_{c}}=\sqrt{1-|\alpha|^{2}} \ket{0_c} + \alpha \ket{1_c}$. We assume that the contribution to the error probability from the \textit{control} register during the CNOT gate operation scales with
\begin{align}
    \bra{\psi_c} \hat{n}_{1}^{(c)} \ket{\psi_{c}} \equiv\bra{\psi_c}1_c\rangle \langle 1_c \ket{\psi_{c}}=|\bra{1_c}\psi_c\rangle|^2.
    \label{eq:n1}
\end{align} 
Moreover, if the input state of the \textit{target} register is not known, the errors introduced to the target register is proportional to the probability that the control condition is satisfied (i.e., $\bra{\psi_c} \hat{n}_{1}^{(c)} \ket{\psi_{c}}$).
Hence, in this case, the EBGC count assigns an error probability of $|\alpha|^2$ to the CNOT gate implemented on the Rydberg platform, given that the control atom is in state $\ket{\psi_c}$. 

The knowledge of the target register's state can be also be used to reduce the EBGC (see Section $\ref{sec:MainStrat}$). In particular, implementing the controlled unitary which exits the target atom from $\ket{1}$ to $\ket{R}$ conditioned on the state of a control atom. Given the target input state $\ket{\psi_t} = \sqrt{1-|\beta|^2}\ket{0} + \beta \ket{1}$, results in an EBGC of 
$\frac{1}{3}|\alpha|^2(2+|\beta|^2)$. In the following, we denote this gate as $CX^{(R)}$. 
Notice that this gate count is identical to that of the CNOT gate when $|\beta|^2 =|\alpha|^2=1$, when the error probability of the Rydberg atom implementation of the CNOT gate is maximized. 

Extending EBGC for single-control multi-target unitaries of the form $CU_{1}\cdots U_k\equiv CP_i$ is straightforward. In this case, assuming no knowledge of the target register, the error introduced into the target register is proportional to the $k$ times the probability that the control condition is satisfied. 
As before, the EBGCs are subject to modification when the state of the \textit{target} register is known. 


The gate counts are summarized in Table $\ref{tab:gatecount}$, for a given input state $\ket{\psi_c}$ of the control register and the control condition $\ket{x_0}$. The unit of the gate count is determined by the maximum error cost of a CNOT gate, which is 3 single-qubit gates in our gate count \cite{jaksch2000fast}. We evaluate the depth to implement each gate using the time unit $t_{\rm{step}}$, given by the time it takes to achieve a complete transfer of the $\ket{0}$ state to $\ket{1}$ state. In Section $\ref{sec:EITGate}$, we discuss the parameter regime that the EBGC is valid.

\begin{table*}[]
    \centering
    \begin{tabular}{c|c|c|c|c}
         & $R_{\theta}$ & $V_{\rm OHE}$& $C V_{\rm OHE}$ & $C U_1 \cdots U_k$  \\  \hline
        {\rm gates}& $\left |\theta/(3\pi)\right|$  & 1 & 4/3 $|\bra{1}\psi_c\rangle|^2$ & $(2+k)/3 |\bra{1}\psi_c\rangle|^2$ \\  \hline
         ${\rm depth}$ & $\left |\theta/\pi\right|$ & 2& 4 & 3 \\  \hline
    \end{tabular}
    \caption{The EBGCs for the native gates of the Rydberg system. All gate countd are normalized by the maximum error probability of a single CNOT gate. The input state of the control register is $\ket{\psi_c}$, and $\theta$ is the single-qubit rotation angle. The error model takes into account only errors caused by the radiative decay rate of the Rydberg states and the non-adiabatic errors due to imperfect blockade. Most importantly, the cost of single-qubit-controlled unitary depends on the probability that the control condition is satisfied.}
    \label{tab:gatecount}
\end{table*}

Our gate count not only assesses an experimental scenario, but also guides us to design algorithms with lower EBGC by taking full advantage of the structure of the errors relevant for that experimental scenario. More specifically, EBGC allows us to demonstrate that the structure of the errors relevant for the proposed Rydberg atom gates can leveraged to design error-robust implementations of quantum algorithms.



\subsection{Rydberg Interactions, Level Diagrams and Single Qubit Rotations}
\label{subsec:LevelDiag}

\subsubsection{Dipolar interactions:} Although all the gates that we will be discussing rely on the same Rydberg blockade mechanism as discussed in Ref. \cite{jaksch2000fast,lukin2001dipole,weimer2010rydberg}, we require both short- and long- range dipolar interactions in order to implement the full variety of multi-qubit gates that we utilize in this work. The two main factors which effect the range of dipolar interaction between Rydberg atoms are (i) whether the dipolar interactions are of long-ranged resonant dipole-dipole type or of short-ranged Van der Waals type and (ii) the dipole moments associated with different Rydberg states \cite{adams2019rydberg,Low2012}.
While the long-ranged dipolar interactions between the Rydberg states are useful for the One-Hot amplitude encoding gate we discuss in Section $\ref{sec:AmpEncGate}$, the possibility of controlling the range of short-ranged interactions will play an important role in implementing a parallelized version of our scheme in Section $\ref{sec:HamSimImpl}$. Fortunately, the required characteristics can be in principle realized with the current experimental setups \cite{Morgado2021,adams2019rydberg,Low2012}. 

\subsubsection{Level Diagrams and Single Qubit Rotations:} 
\label{sec:Levels}

\begin{figure*}[htp]
    \centering
    \includegraphics[width=0.7\textwidth]{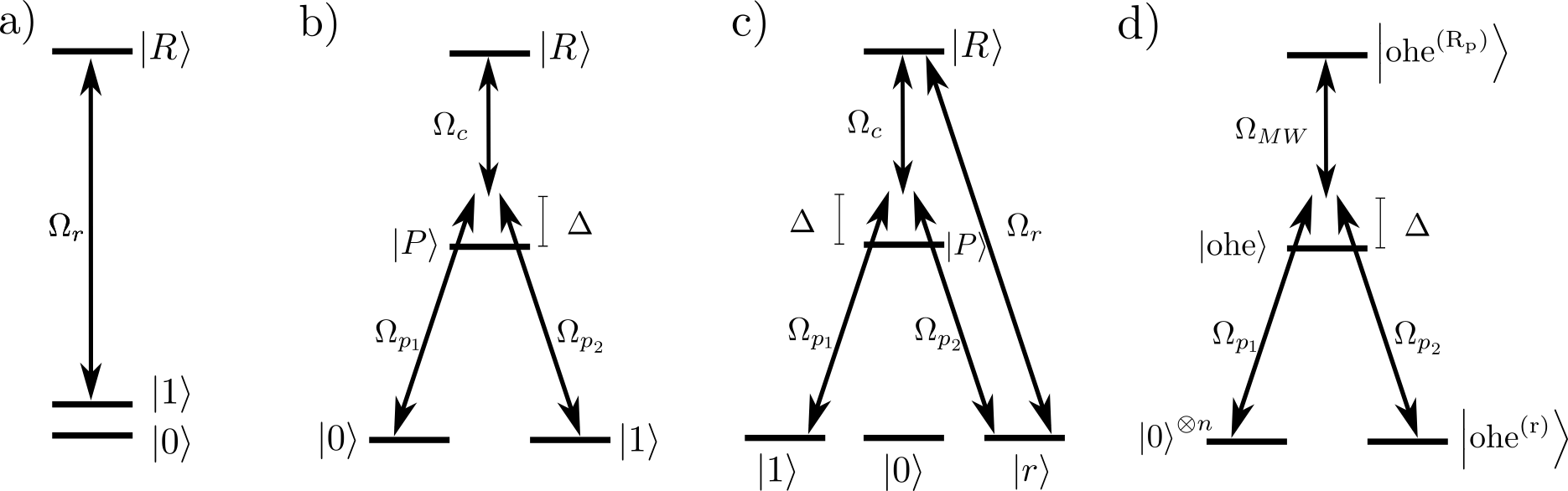}
    \caption{ The level diagrams used in the proposed protocols. a) The level diagram
    for the Rydberg atoms in the control registers, where a hyperfine state is excited to the Rydberg manifold. b-d) Three different level diagrams for the Rydberg atoms in the target registers, each use the EIT mechanism [see $b)$] to introduce errors that are biased on the state of the control register. c) The level diagram for the target atom of the $CX^{(R)}$ gate. The transfer between $\ket{1}$ and $\ket{r}$ is controlled by the energy shift of state $\ket{R}$, followed by an excitation to the Rydberg state. d) the level scheme of the multi-qubit control register for the $CV_{\rm OHE}$ gate. The coupling between the hyperfine level $\ket{r}$ and the Rydberg state $\ket{R}$ is omitted for simplicity. The state $\ket{\rm{ohe}}$ the One-Hot encoding gate of the Rydberg states. Similarly, $\ket{\rm{ohe}^{(\psi)}}$ replaces the Rydberg states in $\ket{{\rm ohe}}$ with the single-qubit state $\ket{\psi}$. The Rabi frequency $\Omega_{MW}$ of the microwave drive couples $\ket{R}$ and $\ket{R_p}$. }
    \label{fig:Levels}
\end{figure*}

The four level diagrams that are relevant to our implementation are shown in Fig. $\ref{fig:Levels}$ c). The diagrams consist of three types of states. Although these diagrams greatly simplify the experimental reality, the three types of states provide sufficient correspondence between our work and the experimental setup. First, we have  long-lived hyperfine states $\ket{0}$, $\ket{1}$, and $\ket{r}$, which make up the two logical states and an auxilliary state for each Rydberg atom. Second, we have an intermediate state $\ket{P}$ which is useful to implement rotations within the hyperfine manifold, but which have a much shorter lifetime than the hyperfine states due to a larger radiative decay rate. The intermediate state is also crucial for the realization of the EIT scheme that we will discuss in the Section $
\ref{sec:EITGate}$. Lastly, the high-energy Rydberg states $\ket{R}$ which not only have a shorter lifetime than the hyperfine states, due to radiative decay, but also evolve under an interacting Hamiltonian, which can be written as 
\begin{align}
    H_{R} = \sum_{i} J_{ij} \ket{R_iR_j}\bra{R_iR_j},
\end{align}
where $\ket{R_iR_j}\equiv \ket{R_i}\otimes \ket{R_j}$ is the two-particle state where the $i^{\rm th}$ and $j^{\rm th}$ atoms located at positions $r_i$ and $r_j$ are in the Rydberg state. Although in reality the interaction strength has the form $J_{ij}\propto \frac{1}{|r_i-r_j|^{\nu}}$ with $\nu \in \mathbf{N}$, it is reasonable to model such a spatial dependence as a step function which takes the value $J$ when $|r_i-r_j|<R_b$  and vanishes otherwise. We refer to the distance $R_b$ as the ``blockade radius". The interaction strength $J$ is finite. As a consequence, even when the radiative decay rate is not taken into account, the two-qubit blockade gate cannot be implemented perfectly. The errors due to the imperfect blockade will be referred to as non-adiabatic errors, whose error-probability is $\propto J^2/\Omega^2$, where $\Omega$ is the characteristic Rabi frequency of the laser drive connecting the low energy states to the Rydberg state. In the following, we assume that these non-adiabatic errors are as large as the errors introduced by the radiative decay rate, unless otherwise specified. 

For the implementation of single-qubit rotations, we choose to use $\ket{P}$ as the intermediate state (see Fig.~$\ref{fig:Levels}$ a). Specifically, we can drive transitions between the logical states $\ket{0}$ and $\ket{1}$ using a Raman scheme which virtually excites the short-lived intermediate state $\ket{P}$. The errors associated with the virtual occupation of $\ket{P}$ motivate our rule for counting single-qubit gates in Section \ref{sec:EBGC}. Specifically, given $\ket{0}$ as our initial state, the errors scale with the time that the short-lived state is virtually occupied during the rotation to the superposition state $\sqrt{1-|\alpha|^2}\ket{0} + \alpha \ket{1}$, resulting in the EBGC of $\arcsin(\alpha)/(3\pi)$ as in Table $\ref{tab:gatecount}$. The EBGC does not change as long as we choose one of 3 hyperfine states of the Rydberg atom (i.e., $\ket{0}$, $\ket{1}$, and $\ket{r}$).


\subsection{One-Hot amplitude encoding gate}
\label{sec:AmpEncGate}
In the following, we introduce a new gate which can be thought of as a multi-qubit generalization of the single-qubit gate. The reason that $V_{\rm OHE}$ is a generalization of the single-qubit gate is that the long-range Rydberg interactions constrain the many-body Hilbert space relevant for the evolution to a \textit{two-dimensional} subspace. Consequently, both the single-qubit gate and the One-Hot encoding gate are used to store classical information encoded in the duration $t_0$ and the amplitude $
\Omega$ of the laser drive in quantum mechanical degrees of freedom. 
More specifically, the single-qubit rotation loads a single amplitude $\alpha \equiv \arcsin{(\Omega t)}$ on a single qubit. Similarly, the One-Hot amplitude encoding gate $V_{\rm OHE}$ is a way of loading $M$ amplitudes 
$\{\alpha_i\}$ where $\alpha_i \propto \Omega_i$ into $M$ qubits in constant time. Because our scheme implements $M$ amplitudes in the computational basis states with only one excitation (i.e., one qubit in the $\ket{1}$ state), we refer to it as the ``One-Hot" amplitude-encoding gate.
From a physical point of view $V_{\rm OHE}$ gate achieves to load all of the information encoded in the relative local intensity of the laser field into orthogonal computational basis states of a quantum register. 

\begin{figure}
    \centering
    \includegraphics[width=0.45\textwidth]{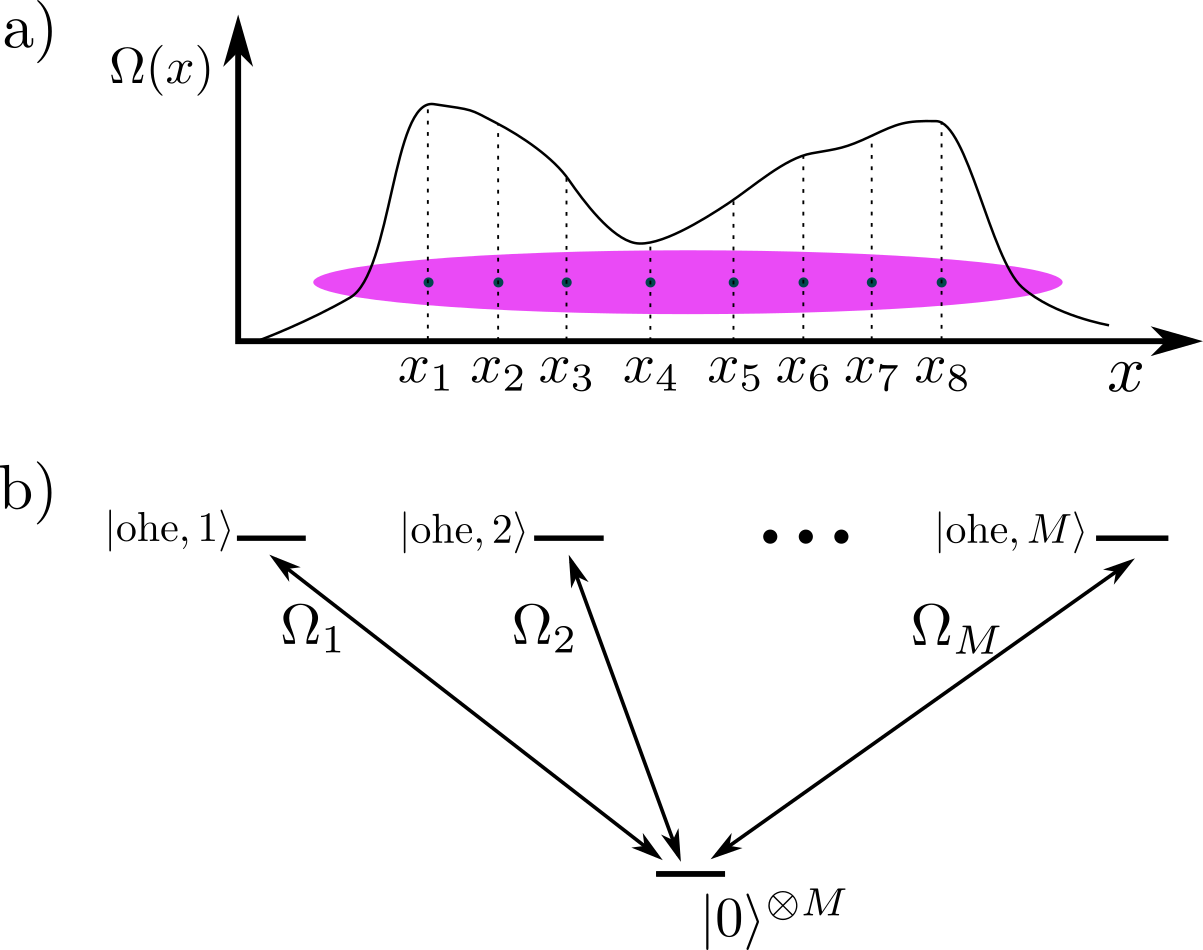}
    \caption{ a) The experimental scheme for implementing $V_{\rm{OHE}}$. All atoms are within each others blockade radii. The spatial dependence of the drive amplitude is utilized to encode complex-valued classical data stored in the laser amplitudes $\Omega_i\equiv \Omega(x_i)$ into a quantum register. b) The level diagram of $M$ Rydberg atoms for the configuration in a). The One-Hot encoding computational basis states $\ket{\rm{ohe},i}$ have a single Rydberg excitation on the $i^{\mathrm{th}}$ atom.}
    \label{fig:VImpl}
\end{figure}

The sequence of unitaries that implement $V_{\rm OHE}$ builds on a similar gate discussed in the context of preparing the W state on the Rydberg platform \cite{Unanyan2002}. Starting from the state $\ket{0}^{\otimes M}$, we coherently drive the ancillae with $M$ amplitudes $\{\Omega_0 \alpha_i\}$ where $\left(\Omega_0/J\right)^2\ll 1$. Starting from the $\ket{0}^{\otimes M}$ state, and assuming that each Rydberg level causes an energy shift of $J$ on the Rydberg states of all other qubits, the dynamics is constrained to a two-dimensional Hilbert space spanned by
$$\ket{0}^{\otimes M} \quad \mathrm{and} \quad \sum_{i} \alpha_i \ket{{\rm ohe}^{(R)},i},$$ 
where we define the One-Hot encoding basis states $\ket{{\rm ohe}^{(R)},i}\equiv\ket{0\cdots0 R_i0\cdots 0}$, each of which has only one Rydberg excitation.  
Projecting the drive Hamiltonian $H_d=\sum_{i=1}^{M} \Omega_0\alpha_i \sigma^{+}+h.c.$ onto this subspace yields the effective Hamiltonian
\begin{align}
    \nonumber \bar{H}&\equiv P_{1}H_dP_1 \\
    &=\Omega_0\left( \sum_{i=1}^{M} \alpha_i \ket{{\rm ohe}^{(R)},i} \right)\bra{0^{\otimes M}} + h.c.,
\end{align}
which is analogous to a Pauli operator in the constrained Hilbert space (notice $\bar{H}^2 =\mathbf{1}$). A schematic for the implementating $V_{\rm{OHE}}$ is given in Fig. $\ref{fig:VImpl}$.

Hence, given the initial state $\ket{0}^{\otimes M}$, evolving the system under $\bar{H}$ for time $t^* = \frac{\pi}{|\Omega_0|}$, prepares the following OHE state 
\begin{align}
    \nonumber  U_{0r}\ket{0}^{\otimes M} &=  e^{-it^* \bar{H} }  \ket{0}^{\otimes M}\\ 
    &=\sum_{i=1}^{M}\alpha_i \ket{{\rm ohe}^{(R)},i}\equiv \ket{\rm{ohe}^{(R)}}.
    \label{eq:ProjEvo}
\end{align}
While the time to implement $U_{0r}$ scales as $O(1/\sqrt{M})$, when each atom is driven by an independent laser of fixed amplitude the run-time of One-Hot encoding gate is increased (i.e., $\sqrt{M}$-fold) due to the requirement that our final state needs to be within the long-lived logical subspace of each atom. In other words, we are required to transfer each ancilla atom excited to their Rydberg state to the long-lived hyperfine $\ket{1}$ $\ket{r}$ states using the following evolution operator
\begin{align}
    U_{\sigma r} = \exp{\left( i t_1^* \sum_{i=1}^{M} (\Omega_1\ket{r_i}\langle \sigma_i| + h.c.) \right)},
\end{align}
where $\sigma=\{1,r\}$ and $t_1^* = \left|\frac{\pi}{\Omega_1}\right|$. Assuming that the Rabi frequencies of local drives are the same, $\Omega_0/\Omega_1 = O(\sqrt{M})$, as
the second part of the evolution does not take advantage of the collective enhancement of the effective Rabi frequency in the presence of blockade interactions. Given this bottleneck, we chose the single-qubit drive strengths in the implementation of $U_{0r}$ as $\Omega_0=O(1)$ such that the runtime of the $V_{\rm OHE}$ gate is $2 t^{*}_1 = 2t_{\rm step}$. Thus, $V_{\rm OHE}$ has an implementation depth of 2.  
We emphasize that this result holds only in the limit of infinite blockade radius.
We discuss the case of finite maximum blockade radius in Section $\ref{sec:scalable}$.

To arrive at the relevant EBGC, we consider two sources of errors: (i) those that result from the radiative decay rate of the atoms in their Rydberg states and (ii) the non-adiabatic errors that result from the imperfect blockade interactions. Because we have at most one atom in the Rydberg state during the implementation of $V_{\rm OHE}$, the errors due to the radiative decay mechanism is the same as those associated with a single-qubit gate where the initial $\ket{0}$ state is completely transferred to the $\ket{1}$ state. On the other hand, the non-adiabatic errors resulting from the finite value of the strength $J$ of dipolar interactions grow as $O\left((\Omega_0/J)^2\right) = O(1)$, since the bottleneck induced by $U_{1r}$ entails that we set $O(\Omega_0)=O(1)$, as explained in the previous paragraph. Including the errors introduced by the radiative decay of the Rydberg states during $U_{1r}$, the number of gates involved in implementing $V_{\rm OHE}$ is 3/3=1. It is important to emphasize that the above error cost of $V_{\rm OHE}$ is calculated assuming that the coupling between the Rydberg and hyperfine manifolds is induced by a single photon transition, as the introduction of intermediate states which do not experience an energy shift due to dipolar interactions result in radiative errors that scale as $O(M)$. Such an excitation can be realized in the alkali-earth metal atoms as in Ref.~\cite{madjarov2020high}. 


Before we continue, we emphasize another important property of the $V_{\rm OHE}$ gate which allows for a hardware efficient implementation of the reflection operator $2(\ket{0}\bra{0})^{\otimes n_a} - \mathbf{I}$ required for the QSP walk operator. Let us define $\tilde{V}_{\rm OHE}\equiv U_{0r} U^{\dagger}_{1r}$. Then, we have
\begin{align}
    \nonumber & \tilde{V}_{\rm OHE} \ket{{\rm ohe},l} = U_{0r} \ket{{\rm ohe}^{(R)},l}\\
    \nonumber & = U_{0r}\left [ \alpha_l \ket{{\rm ohe}^{R}} + \sum_{j=1}^{n_a-1}\big\langle {\rm ohe}^{(R)}\ket{{\rm ohe}_{\perp}^{(R)},l}\ket{{\rm ohe}_{\perp}^{(R)},j}\right]\\
    &= i \alpha_l \ket{0}^{\otimes n_a} + \sum_{j=1}^{n_a-1}\big\langle {\rm ohe}^{(R)}\ket{{\rm ohe}_{\perp}^{(R)},l}\ket{{\rm ohe}_{\perp}^{(R)},j}
    \label{eq:VOHETilde}
\end{align}
where $\left\{\ket{\rm{ohe}^{(R)}},\left\{\ket{{\rm ohe}_{\perp}^{(R)},j}\right\}\right\}$ is a set of $n_a$ orthogonal OHE states. To reach the final equality, we used the fact that the action of $U_{0r}$ on $\left\{\ket{{\rm ohe}_{\perp}^{(R)},j}\right\}$ is trivial. Given Eq. ($\ref{eq:VOHETilde}$), we find that 
\begin{align}
  \tilde{V}_{\rm OHE}\bar{U}V_{\rm OHE} = (\mathbf{1}-2\Pi_0) V_{\rm OHE}^{\dagger}\bar{U} V_{\rm OHE}.
  \label{eq:ImplementReflection}
\end{align}
The equation above allows us to realize a hardware-efficient implementation the reflection operation using Rydberg atoms. 


\subsection{Biased-error controlled unitary gates $C U_1\cdots U_k$ }
\label{sec:CmUk}
While the unitary $V_{\rm OHE}$ offers a way of loading the classical data stored in $\vec{\alpha}$ into quantum degrees of freedom, the controlled unitary gates load the classical data describing the Pauli strings $\{P_{i}\}$ ( see Section $\ref{sec:LCU}$) into the quantum processor. In particular, implementation of each single-qubit rotation $U_{j}\equiv e^{i\theta_j \hat{n}_j\cdot \vec{\sigma}_{j}}$ on the $j^{\rm th}$ target atom load the information regarding the position of that single qubit as well as the axis $\hat{n}_j\in \{\hat{x},\hat{y},\hat{z}\}$, and the angle $\theta_j$ associated with its rotation. By conditioning products of single-qubit rotations on the ancillary address states $|x_i\rangle$, we can make sure the relevant Pauli string can be retrieved conditionally on orthogonal address states. 

In Section $\ref{sec:subadditivity}$, we considered an error model which assumes that the error probability is completely conditional on the state of the control qubit. Here, we describe the concrete protocol for which such an error model is valid. In particular, we discuss the multi-target controlled unitary proposed in Ref. \cite{muller2009mesoscopic}, which utilizes an interference phenomenon called Electrodynamically Induced Transparency (EIT) to ensure that the evolution of the target atoms can be made near trivial and error-free when the control condition is not satisfied. This protocol thus motivates the way we count the gates for each single-control conditional unitary using EBGC (see Section $\ref{sec:EBGC}$). In the physical implementation the errors are not perfectly biased, and the contribution to the error probability when the control condition is not satisfied is not completely negligible. The ratio between the error contributions when the control condition is satisfied and not satisfied can be increased by increasing the strength of a laser drive. However, unlike the conventional dependence of the error probability to the amplitude of the drive laser, where the error probability is inversely proportional to the \textit{amplitude} of the laser drive, our scheme realized a two-qubit gate where the error probability is inversely proportional to the \textit{intensity} of the laser drive, providing a quadratic advantage. Moreover, the drive amplitude can be increased up to an order of magnitude above the strength $J$ of dipolar interactions. While these caveats are crucial for experiments on the Rydberg atom platform, we emphasize that the result ``LCU-based block-encoding unitary can be implemented with constant error scaling with respect to the system size" is independent of how the biased-error controlled-Pauli operations can be implemented on an experimental platform.

As it will become apparent from the following discussion, the single-qubit controlled Pauli operations conditioned on One-Hot encoding states can be used to realize an error-robust implementation of the LCU-based block-encoding unitary. However, if we only consider One-Hot Encoding address states, the error robust implementation comes at the expense of an address registers of size $N$, which is not scalable. In order to reduce the size of the ancillary address register, we propose to use $k$-Hot Encoding ($k$HE) address states. An $n$-qubit $k$-Hot Encoding state $\ket{k{\rm he}^{(1)}}$ is defined as a linear superposition of computation basis states which have $k$ atoms in the $\ket{1}$ state. Crucially, using $\ket{k{\rm he}^{(1)}}$ as address states increases the number of address states exponentially with $k$, while increasing the size of the address register only linearly in $k$. We discuss the Rydberg atom gates necessary for preparing a $k$HE state as well as implementing $k$-qubit controlled Pauli operations conditioned on the $k$HE address state in Section $\ref{sec:CVOHE}$.


\begin{figure*}
    \centering
    \includegraphics[width=0.9\textwidth]{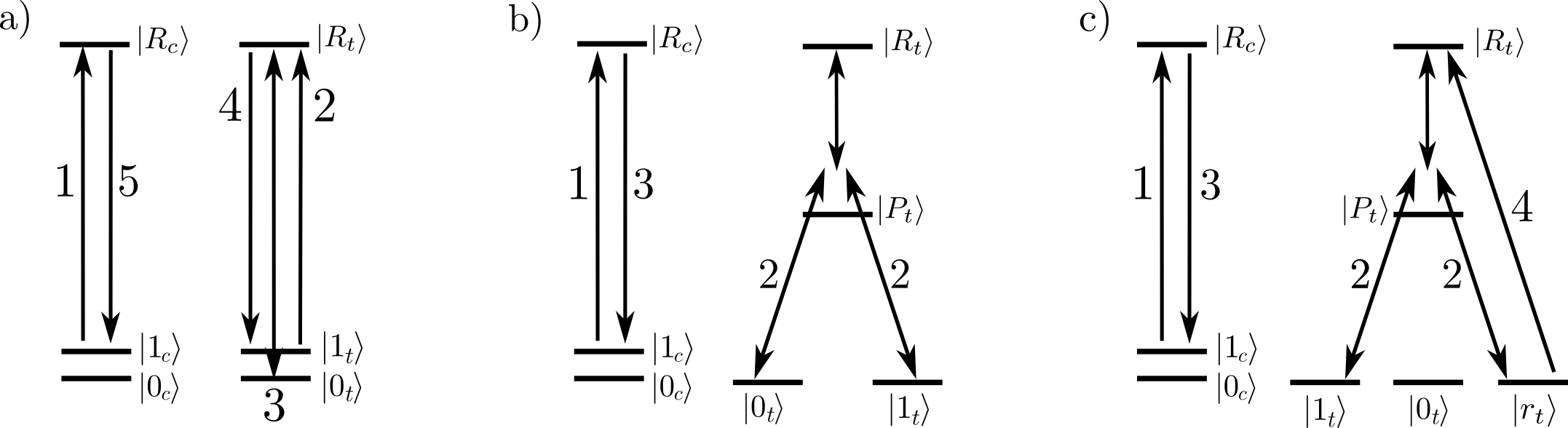}
    \caption{The pulse sequences for implementing the three conditional unitaries that are used to implement the algorithms in the  QSP framework.
    The strengths of each laser pulse are given in Fig. $\ref{fig:Levels}$. figure: a) The pulse sequence for the conventional two-qubit conditional gate (CNOT) on the Rydberg atom platform implemented via the blockade mechanism. Notice that there is substantial error probability both when the control condition is satisfied and violated. b) Implementation of the EIT-based blockade gate which results in an error model that is strongly biased on the state of the control atom. Notice that the control laser on the target register is always on, and hence does not have a number associated to it. c) The laser pulse sequence for implementing the conditional excitation to the Rydberg state, denoted $CX^{(R)}$ in the main text. The same pulse sequence is used to implement the $CV_{\rm OHE}$ gate.  }
    \label{fig:CZ}
\end{figure*}

\subsubsection{EIT-based single-control multi-target unitary on the Rydberg platform}
\label{sec:EITGate}
The EIT-based controlled unitary operations utilize interference to ensure that if the control condition is not satisfied, the evolution of the target atoms initiated in the logical subspace stays in a non-radiative ``dark" subspace, thereby drastically reducing the errors due to radiative decay from both the intermediate state $\ket{P}$ and the Rydberg state $\ket{R}$. Moreover, the Rydberg interactions become relevant to the evolution only when the control condition is satisfied and therefore, the non-adiabatic corrections are only relevant to this case. As a result, the EIT-based blockade gates introduce errors in a biased way, and the error probability depends on whether the control condition is satisfied or not. Thus, the central quantity of our analysis is the ratio $\eta_{\epsilon}$ of the error probabilities conditioned on the satisfaction of the control condition. 

We show that the ratio $\eta_{\epsilon}$ can be reduced by a factor of $N$ by increasing the \textit{amplitude} of a laser drive by $O(\sqrt{N})$. This is in contrast to the more well-known Rydberg-blockade based gates where the \textit{total} error probability decreases linearly with increasing drive amplitude \cite{jaksch2000fast,Levine2019}. Moreover, the error probability of these conventional Rydberg-blockade gates are dominated by errors due to the unwanted population of the Rydberg state when the drive amplitude is of the order of the dipolar interactions $J$. In contrast, the EIT-based scheme 
allows for the strength of the drive amplitude to be increased above the blockade interactions strength $J$ while still resulting in an error-robust implementation. 
The results of this section are the core justification of our error model for single-qubit-controlled unitaries discussed in Section $\ref{sec:subadditivity}$.

Next, we demonstrate that when $\eta^{-1}_{\epsilon}= N$, we can achieve a constant error implementation of $\bar{U}$ for $N$ Pauli strings (see Section $\ref{sec:LCU}$), using $N$ EIT-based single-qubit controlled Pauli gates acting on an $N$-qubit One-Hot Encoding address state $V_{\rm OHE}\ket{0}^{\otimes N}$. The implementation uses $O(N)$ ancillary address qubits, and is therefore not scalable. We address this problem by designing protocols to utilize $k$HE states in Sections $\ref{sec:CVOHE}$ and $\ref{sec:DesignPrinc}$. 


\textit{Protocol:} 
We start the discussion of the EIT-based blockade-gates with the implementation of a CNOT gate~\cite{muller2009mesoscopic}. The scheme uses the level scheme in Fig. $\ref{fig:Levels}$ $(a)$ and $(b)$ for the control and target qubits, respectively. The target qubit is continuously driven by a control field $\Omega_c$ during the three-step protocol. In the first step, the control atom is excited to a Rydberg state if it satisfies the control condition. Secondly, lasers inducing the two probe Rabi frequencies $|\Omega_{p1}|e^{i\theta_1}$ and $|\Omega_{p2}|e^{i\theta_2}$ are shone on for the target atom. The frequencies of the control and probe frequencies are such that the excitation to the Rydberg state is two-photon resonant. Denoting the detuning between the hyperfine and intermediate state as $\Delta$, and defining the radiative decay rates $\gamma_R$ and $\gamma_P$ of the states $\ket{R}$ and $\ket{P}$, respectively, we consider an experiment satisfying a set of inequalities that define the perturbative regime, $\{\Delta,J\}\gg\{ \Omega_{1},\Omega_{2},\Omega_c\}\gg\{\gamma_{R},\gamma_P\}$. The second step of the protocol takes time $\tau_g \equiv \pi \frac{\Delta}{\Omega_p^2}$.

Two scenarios are relevant for the second step of the EIT-based blockade gate 
\begin{enumerate}
    \item If the control atom is not in the Rydberg state, then both logical states of the target atom evolve adiabatically in a dark (non-radiative) subspace spanned by 
\begin{align*}
    \ket{\psi_d}& = \frac{1}{\sqrt{1-x^2}}\left(\ket{\psi_l} + x\ket{R}\right)\\
    \ket{\tilde{\psi}_d} &=\frac{1}{\sqrt{2}}\left(\ket{0}-e^{i(\theta_1-\theta_2)}\ket{1}\right),
    \label{eq:DarkSpace}
\end{align*}
and eventually return back to the initial state \cite{muller2009mesoscopic}. In the above expression we defined $x\equiv|\sqrt{2}\Omega_p/\Omega_c|$ as a time-dependent dimensionless quantity. The logical state $\ket{\psi_l} = \frac{1}{\sqrt{2}}(\ket{0}+e^{i(\theta_1-\theta_2)}\ket{1})$ is orthogonal to $\ket{\tilde{\psi}_d}$. Most importantly, the two dark-states have no contribution from the short-lived intermediate state $\ket{P}$. Hence, the errors during the adiabatic evolution come solely due to decay of the small occupation in the Rydberg state $\ket{R_t}$, which scales as $x^2$. 

\item If, on the other hand, the control condition is satisfied and the control atom is excited to its Rydberg state, then the EIT condition that ensure an evolution only within the dark-state manifold is no longer satisfied, and the transitions between the two logical states $\ket{0_t}$ and $\ket{1_t}$ are mediated by the virtual excitation of the short-lived state $\ket{P}$, introducing errors due to the finite decay rate $\gamma_{P}$.
\end{enumerate}

The last step of the gate simply brings the control atom back to the hyperfine manifold. Whether the Pauli gate applied on the target qubit is $\sigma_x^{(t)}$ or $\sigma_{y}^{(t)}$ is determined by the phase difference between the control pulses $\Omega_{p1}$ and $\Omega_{p2}$. The multi-target generalization of the EIT-based controlled unitary is obtained by simply increasing the number of target qubits within the blockade radius of the control qubit. 

\textit{Controlling biases of error processes:}
We discussed above how the evolution of the target atom depends on whether the control condition is satisfied or not. As a result, the probability that the target atom will suffer an error depends on the state of the control atom. Here, we calculate the ratio $\eta_{\epsilon}$ between the error probability of the target atom when the control condition is satisfied to that when the control condition is violated.

The error probability when the control condition is not satisfied is given by the  population of the target Rydberg state
\begin{align}
    \epsilon_{v} \equiv \pi \frac{\Delta}{\Omega_p^2}\gamma_R\left(\frac{\Omega_p}{\Omega_c}\right)^2 = \tau_g  \gamma_R  x^2,
\end{align}
where the factor of $\pi$ is because we are implementing a NOT operation on the target qubit. Here, we neglected the diabatic corrections associated with the probe pulse, which are proportional to $x^6$ \cite{muller2009mesoscopic}.

On the other hand, when the control condition is violated, there are multiple contributions to the error probability
\begin{align}
    \epsilon_{s} = \tau_g   \left[\left(\frac{\Omega_p}{\Delta}\right)^2\gamma_P + \gamma_R + \left(\frac{\Omega_p \Omega_c}{\Delta J}\right)^2 \gamma_R\right],
    \label{eq:error_sat}
\end{align}
where the first term in the brackets is the radiative decay probability from the intermediate state $\ket{P}$ of the target atom. The second and the third terms in the brackets correspond to the errors due to the perturbative occupation of the Rydberg state of the control and the target atoms, respectively. In the following, we take $(\Omega_p/\Delta)^2\gamma_P \leftarrow \gamma_R$ to fix the free parameters $\Omega_p/\Delta$ and ensure that the two error sources contribute to the error probability the same way. In Eq. ($\ref{eq:error_sat}$), we assume $\tau_g$ is chosen such that the occupation of the target Rydberg state at the end of the protocol is negligible. As a result, because the ratio $\eta_{\epsilon}\equiv \frac{\epsilon_{v}}{\epsilon_{s}}$ depends on $x^2$ when $\left(\frac{\Omega_p \Omega_c}{\Delta J}\right)^2 \leq 1$, it can be lowered $\propto \Omega_c^{-2}$ as long as 
\begin{align}
    \Omega_c < J \frac{\Delta}{\Omega_p} = J \sqrt{\frac{\gamma_{P}}{\gamma_R}}.
    \label{eq:Condition}
\end{align}

The above result is the main justification for the Error Bounded Gate Count (EBGC) for a single-qubit-controlled Pauli operation described in the Section $\ref{sec:EBGC}$, which neglects the error-probability conditioned on the violation of the control condition. Then given the control condition $\ket{0_C}$ and the input state of the control register $\ket{\psi_c}$, the single-control k-target unitaries $CU_1 \cdots U_k$ can be implemented using $(2+k)/3|\bra{0_c}\psi_c\rangle|^2$ error-bounded gates and a depth of 3. Next, we discuss the conditions under which the error-probability conditioned on the violation of the control condition can be neglected for an implementation of $\bar{U}$.

\textit{Error probability for $\bar{U}$ using OHE address states}: In order to demonstrate the advantage of single-qubit controlled Pauli operators, we consider their action on a control register prepared in an $N$ qubit One-Hot encoding state. Then each single-qubit controlled Pauli operation conditioned on the state of the $j^{\rm th}$ control atom is equivalent to a controlled Pauli operation conditioned on a One-Hot encoding bitstring $\ket{{\rm ohe},j}$. 
Assuming that $\epsilon_v N \ll 1 $, total error associated with $N$ single-qubit controlled Paulis acting on the $N$-qubit One-Hot encoding state is 
    \begin{align}
        \epsilon_{\rm tot} = \left[\epsilon_{s} + \epsilon_v (N-1)\right] = \epsilon_{s}\left[1 + \eta_{\epsilon}(N-1)\right].
    \end{align}
    The expression for $\epsilon_{\rm tot}$ follows from the fact that the probability that the control condition is satisfied for any one of the $N$ controlled Pauli operations add up to 1. When the condition in Eq.~($\ref{eq:Condition}$) is satisfied, $\epsilon_{s}$ does not scale with $\Omega_c$, while $\eta_{\epsilon}\propto \Omega_{c}^{-2}$. Hence, if we chose $\Omega_{c}~=~O(\sqrt{N})$, $\epsilon_{\rm tot}$ remains constant. As a result, in this regime, the unitary $\bar{U}$ can be implemented with a constant error probability, although the gate complexity is $O(N)$ and the implementation of the protocol is error-robust.
    
    Surprisingly, there is also a regime biased errors still help realize an error-robust implementation, although
    the condition Eq. ($\ref{eq:Condition}$) is no longer satisfied. 
    In this strong-drive limit, the total error is not constant, but scales as $\epsilon_{\rm tot} = O(\sqrt{N})$. To see this, we simply note that in the regime  $\Omega_{c}\gg J\sqrt{\gamma_P/\gamma_R}$, we have $\eta_{\epsilon}\propto \Omega_{c}^{-4}$, while $\epsilon_{s}\propto \Omega_c^{2}$. Hence, picking $\Omega_c = O(N^{1/4})$, we obtain a total error probability scaling as $O(\sqrt{N})$. This scaling is error-robust, since the error probability scales quadratically slower than the gate complexity of the protocol. Notice that the property that the error reduction scaling quadratically with the drive strength is preserved. 
    
    \textit{Suppression using reported parameters}: Whether the Rabi frequency $\Omega_c$ can be increased at will depends both on the intensity of the laser field and the ability to single out the desired Rydberg states during the excitation between the hyperfine and Rydberg manifolds. While a more detailed discussion of the internal structure of Rydberg atoms is beyond the scope of our work, we can simply use the reported values for (i) the Rabi frequencies $\Omega_c$ achieved as well as (ii) the lifetimes $2\pi/\gamma_{P(R)}$ from the literature.
    
    To determine the maximum suppression of errors, use the reported values for the lifetimes $\tau_{R}=146\, \mu s$ for the Rydberg state $\ket{n=70,J=1/2,m_j=-1/2}$ and $\tau_{P}= 115 \, ns$ for the intermediate state $\ket{n=6\,P_{3/2},F=3,m=-3}$ \cite{omran2019generation}. Hence, we set $\frac{\Omega_p}{\Delta} \leftarrow \sqrt{\frac{\gamma_R}{\gamma_P}}=\approx 1/36$. On the other hand, we use the Rabi frequency $\Omega_R \approx 2\pi \times 120$ MHz reported by Ref. \cite{omran2019generation} for a transition between the intermediate state $\ket{P}$ and the $\ket{R}$. Thus, if we would like to have a 100 fold suppression of errors if the control condition is not satisfied (i.e., $\eta_{\epsilon} = 1/100$), then we need to set $\Omega_P = \Omega_R/10 = 2\pi \times 12 $ MHz. To determine the error probability for 100 gates, we calculate the error probability 
    $$\epsilon_s = 1- \exp{\left(- 2\pi * \frac{\Delta }{\Omega_P^2 \tau_R}\right)} \approx 2\pi *\frac{36}{\Omega_P \tau_R} \approx \frac{1}{50}. $$
    As a result, up to 100 gates conditioned on a One-Hot Encoding state can in principle be achieved with below $5\%$ error probability.


\textit{The size of the ancillary register}:
The results of this section allows one to implement the walk operator $W$ [see Eq. ($\ref{eq:QubitWalk}$)] with constant error probability if the ancillary control register is prepared in a One-Hot Encoding state. However, if we only use One-Hot Encoding states, then the cost of the error-robustness is an ancillary register of size $N$ for a block-encoding operator that can be decomposed into $N$ Pauli strings. In this case the size of the ancillary register makes the implementation unfeasible. In order to address this issue, we propose to use $k$-Hot Encoding states realized by a tensor product of $k$ One-Hot Encoding states. In Section $\ref{sec:DesignPrinc}$, we show that the error-probability associated with the action of $N$ single qubit controlled operations on a $k$-Hot encoded state scales as $O(k)$. Before we do so, we introduce two additional Rydberg gates which enables us to use the $k$HE states in our protocols.

\subsubsection{Utilizing $\ket{k{\rm he}}$ states: $CX^{(R)}$ and C$V_{\rm OHE}$ gates}
\label{sec:CVOHE}

If we want to realize an error-robust implementation of QSP protocols using $k$-Hot Encoding states, we have to overcome two challenges. First, to implement the relevant $\bar{U}$, we need an error-robust implementation of $k$-qubit controlled unitaries. Second, we need to be able to prepare $\ket{k{\rm he}}$ through an error-robust implementation of $V$. Here, we introduce two additional gates that will be instrumental in meeting these challenges. 

\textit{$CX^{(R)}$}: 
In order to implement $k$-qubit controlled Pauli operations using their single-qubit controlled counterparts, we use a controlled transfer of a logical states of the target atom to the Rydberg state conditionally on the state of the control atom. We denote this gate as $CX^{(R)}$.
To this end, we use an additional hyperfine state $\ket{r_t}$ [see Fig. $\ref{fig:CZ}\, c)$], and apply the EIT-based blockade gate where the probe lasers on the target atom induce transitions between $\ket{1_t}$ and $\ket{r_t}$. Then, the population in $\ket{r_t}$ can be transferred to the Rydberg state $\ket{R_t}$ using an additional $\pi$ pulse. As a result, the depth of the implementation is 4. The EBGC depends on the input state of both the control and the target registers. Given the input state $\ket{\psi_c} = \sqrt{1-|\alpha|^2}\ket{0_c} + \alpha \ket{1_c}$ and $\ket{\psi_t}=\sqrt{1-|\beta|^2}\ket{0_t} +\beta \ket{1_t}$, the EBGC of the controlled unitary is $2/3|\alpha|^2(1+|\beta|^2)$ for a single target qubit. The calculation of the EBGC for larger number of target atoms is straightforward.

\textit{$CV_{\rm OHE}$}:
We prepare the $k$HE address state using a controlled version of $V_{\rm OHE}$. Unlike the situation with the tensor products of Pauli operators, a controlled version of the $V_{\rm OHE}$ gate is challenging because $V_{\rm OHE}$ utilizes interactions between the Rydberg states amongst all atoms. Thus, we need to introduce a new mechanism to implement $CV_{\rm OHE}$. Our strategy is to use the One-Hot encoding state Rydberg state $\ket{\rm{ohe}^{(R)}}$ [see Eq. ($\ref{eq:ProjEvo}$)] as the intermediate state $\ket{P}$ discussed in Section $\ref{sec:EBGC}$. We are allowed to make such a substitution because the strong and long-range dipolar interactions between Rydberg states constrain the system of $n$ atoms to only a a two-dimensional subspace. 

During the implementation of the $CV_{\rm OHE}$ gate, the dynamics of the target register can be described by a 5 level system depicted in Fig. $\ref{fig:Levels}$. Besides the initial state $\ket{0}^{\otimes n}$, our scheme uses four One-Hot Encoding states which are distinguished by the state which specifies the type of excitation present. We denote these states as $\ket{{\rm ohe}^{(\eta)}}$, where $\ket{\eta} \in \{ \ket{1}, \ket{R}, \ket{r}, \ket{R_{p}}\}$ denoting different single atom states. Crucially, Moreover, starting from the state $\ket{\rm{ohe}^{(\eta_{\rm{in}})}}$ and transferring each qubit to a their corresponding state $\eta_{\rm{fin}}$ results in $\ket{\rm{ohe}^{(\eta_{\rm{fin}})}}$. 
The newly introduced Rydberg state $\ket{R_p}$ has two important properties.  First, it is only accessible from $\ket{R}$ via a microwave transition (see Fig. $\ref{fig:Levels}$) \cite{maxwell2013storage,paredes2014all}. Secondly, 
the angular momentum quantum numbers of $\ket{R}$ and $\ket{R_p}$ are different, such that the two states experience different energy shifts due to dipolar interactions. Thus, for our intents and purposes, we can assume that it is possible to have an energy shift on $\ket{R_p}$ while the energy of $\ket{R}$ stays constant. 

The gate protocol is based on an EIT scheme where the two states in the hyperfine subspace are $\ket{0}^{\otimes n}$ and $\ket{\rm{ohe}^{(r)}}$, and the intermediate state of the EIT scheme is the One-Hot encoding Rydberg state $\ket{\rm{ohe}^{(R)}}$. Finally, the state which controls whether the EIT condition is satisfied is $\ket{\rm{ohe}^{(R_p)}}$.
The first step of $CV_{\rm OHE}$ is to implement a transition between $\ket{0}^{\otimes n_a}$ and $\ket{\rm{ohe}^{(r)}}$ controlled by the energy shift of $\ket{\rm{ohe}^{(R_p)}}$. In the second step, $\ket{\rm{ohe}^{(r)}}$ is transferred to the $\ket{\rm{ohe}^{(R)}}$ state by a tensor product of single-qubit rotations.
The depth of the implementation is 4 and the EBGC given the state of the control register $\ket{\psi_c}=\sqrt{1-|\beta|^2}\ket{0}+\beta\ket{1}$ is $5/3|\beta|^2$. The prefactor 5 is a result of taking into account both the radiative and non-adiabatic errors into account.
We emphasize that the One-Hot encoding keeps EBGC small during the transfer between $\ket{\rm ohe^{r}}$ and $\ket{\rm ohe}$.

\setbox0\vbox{
\subsection{Advantages of the blockade based gates for implementing OHE and $C_{x_0}U_1\cdots U_k$ gates}
\label{sec:Advantages}

In this subsection, we clarify the advantage of implementing the multi-qubit gates $V_{\rm OHE}$ and $C_{x_0} U_1\cdots U_{k}$ using the blockade mechanism native to the Rydberg atom platform. In particular, we answer the question of whether the implementation of these gates via the \textit{blockade} mechanism that utilizes always-on two-atom interactions is advantageous compared to an implementation which controls arbitrary two-atom interactions in a time-dependent way. In the following, we call the latter implementation method Two-Qubit Control (TQC). 

In our discussion of the runtime and the errors associated with the TQC and blockade strategies, we do not consider the specific error model relevant for either implementation, and assume that each gate operation introduces a constant error independent of the input states. 
Without loss of generality \cite{Vatan2004}, we consider systems where the two-qubit interactions are diagonal in the computational basis, namely those generated by a mutually commuting subset of Pauli operators, $\sigma^{i}_{z}\sigma^{j}_{z}$, describing the interaction between the $i^{\rm th }$ and $j^{\rm th}$ qubits.

In the case of the TQC method, we allow the control of two-qubit interactions between any two qubits at a time. Moreover, we assume that multiple interaction terms can be turned on simultaneously. 
To compare the runtime of the two implementation strategies, 
we consider a unit of time given by the maximum strength of two-qubit interactions $J_{\rm max}$. 
Regarding errors, we assume that the time-dependent control of effective interaction strength $J$ leads to non-adiabatic errors which scale as $O[(J/J_{\rm max})^{2}]$ for each pair of qubits involved. This adiabaticity assumption puts the errors introduced by the blockade and TQC strategies on equal footing by demanding that the error scaling of the gates implemented using the two strategies to be identical. Therefore, in the following, we will be only considering scaling of the time complexity and errors of the implementations with respect to the number of qubits involved. In making these assumption on adiabaticity, we do not consider shortcuts to adiabaticity and their applications to implementing two-qubit gates \cite{Petersen2010,choi2014optimal,brown2016co}. Yet, to the best of our knowledge, although the shortcuts to adiabaticity can in principle reduce the error, the protocols which aim to counteract the non-adiabatic errors by designing custom engineered drive pulses typically require expensive numerical optimization routines that become unfeasible in the scaling limit relevant to our discussion. Moreover, it is important to remind the reader that our definition of the TQC method is motivated mainly by the purpose of accentuating the advantages of the blockade gates and that the current experimental platforms do not have the resources required to control two-qubit interactions between any two qubits in a time-dependent way.

We start our comparison by considering the implementation of the OHE gate in a system of $M$ qubits. As discussed in Section $\ref{sec:AmpEncGate}$, the OHE gate can be implemented using the blockade mechanism in time $1/\Omega_0$ as in Section $\ref{sec:AmpEncGate}$. The errors due to non-adiabatic process scale independently of $M$.
We can compare this performance to that of the TQC implementation of the OHE gate. To the best of our knowledge, the lowest depth implementation of the OHE gate has circuit depth $O(\log(M))$ and requires $O(M)$ two-qubit gates \cite{cruz2019efficient}. Therefore, the total runtime of the implementation based on tunable two-qubit interactions scales as $O(\log{(M)})$ while the total error scales as $O(M)$. As a result, the optimal scaling both the runtime and the errors of the blockade implementation drastically improves that of a TQC implementation of the same gate.

Next, we consider the implementation of the $C_{x_0}U_1\cdots U_k$ gates. The blockade-based implementation of $C_{x_0}U_1\cdots U_k$, has a constant runtime, and the total error due to non-adiabatic and radiative process scales as $O(m + k)$, where $m$ is the number of qubits in the control register. In the case that of TQC, on the other hand, the depth of implementing $C_{x_0}U_1\cdots U_k$ gate is the same as that of an $m$-bit Toffoli gate.  Specifically, an $m$-bit Toffoli gate can be used to entangle the state of the control register to the $\ket{1}_{\rm anc}$ or $\ket{0}_{\rm anc}$ of the ancilla qubit depending on whether or not it satifies the control condition. Then the unitary $U_1 \cdots U_k$ can be implemented conditionally on the single additional ancilla in constant time with the errors scaling as $O(k)$. To our knowledge, the lowest depth implementation of the $m$-bit Toffoli gate without ancillae requires a circuit of depth $O(m)$ and the errors scale as $O(m^2)$ \cite{Saeedi2013} \footnote{We note that theoretical lower bound for the number of CNOT gates required for the $m$-bit Toffoli gate is $\Omega(m)$ \cite{shende2008cnot}}. Hence, the runtime of $C_{x_0}U_1\cdots U_k$ gate implementation is $O(m)$, and the total error scales as $O(m^2+k)$. Again, the blockade based implementation drastically improves the scaling of the runtime and errors for the TQC implementation of the $C_{x_0}U_1 \cdots U_k$ gates.

As a result, the blockade-based implemetation of set of multi-qubit gates that we propose as building blocks for QSP-based algorithms provide drastic advantages in terms of runtime and errors with respect implementations based on TQC. We emphasize again that this advantage is present despite the fact that the TQC method requires a level of control over two-qubit interactions that has not been realized in current experimental platforms.  }

\section{Error-Robust implementation of LCU-based block-encoding unitary}
\label{sec:DesignPrinc}
In this section, we describe protocols for implementing LCU-based QSP walk operators using $k$HE ancillary address states, using the Rydberg atom gates described in Section $\ref{sec:MainStrat}$. The resulting implementation is has an EBGC scaling as $O(k^2)$ when the block-encoded operator $A$ is a linear combination of $k$-local Pauli strings acting on $n_{\rm site}$ qubits. The size ancilla register, on the other hand grows only linearly with $k$.

In Section $\ref{sec:CustomV}$, we present the $k$-Hot state preparation unitary based on C$V_{
\rm OHE}$ gates (see Section $\ref{sec:CVOHE}$), which efficiently prepares ancillary states that are customized for an error-robust implementation of LCU-based block-encoding. In Section $\ref{sec:ParalU}$, we discuss the implementation of $\bar{U}$ that complement the state preparation protocol in $\ref{sec:CustomV}$.


\subsection{Implementation of state preparation unitary $V$}
\label{sec:CustomV}
We demonstrate that it is possible to reduce the EBGC of state preparation to a constant even when the number of atoms in the address register is increased polylogarithmically with respect to the number of addresses in the LCU protocol.
To this end, we construct a protocol that only consists of $V_{\rm OHE}$ and its single-qubit-controlled counterpart $CV_{\rm OHE}$. 


As a first step, we describe a state preparation protocol which uses 2 ancilla registers $a_1$ and $a_2$, and prepares the following 2-Hot encoded state
\begin{align}
    \ket{\Psi_{\rm 2HE}} = \sum_{l=1}^{L}\beta^{(1)}_l \ket{{\rm ohe},l}\otimes \left(\sum_{i=1}^{n_l} \beta_{i}^{(2;l)}  \ket{{\rm ohe},i}\right),
    \label{eq:OHEancilla}
\end{align}
where $\beta_{i}^{n;l}$ is the coefficient of the state $\ket{{\rm ohe},i}$ of the $n^{\rm th}$ ancillary register $a_n$, conditioned on the $(n-1)^{\rm st}$ ancilla register being in the state $\ket{{\rm ohe},l}$.
To prepare $\ket{\Psi_{\rm 2HE}}$, we use two ancilla registers $a_{1}$ and $a_2$ consisting of $n_{a_1}$ and $n_{a_2}$ qubits, respectively. The state preparation unitary can then be implemented by first applying $V_{{\rm OHE},a_1}$ on the first ancilla register, followed by an application of $V^{(l)}_{{\rm OHE},a_2}$ on the second ancilla register conditional on the $l^{\rm th}$ qubit in $a_1$ being in state $\ket{1}$ (we denote this operation by $C_{1;a_{1,l}}V^{(l)}_{{\rm OHE};a_2}$).
The state-preparation protocol requires $n_{\rm tot} = n_{a_1}+n_{a_2}$ ancillary qubits and as an EBGC of only $\frac{1}{3}\left(3 + 5\sum_{l=1}^{L} |\beta^{(1)}_{l}|^2\right) = 8/3$. The depth of the protocol is $2 + 4 n_{a_1}$ (see Section $\ref{sec:CVOHE}$ for EBGC calculation for single-qubit-controlled $V_{\rm OHE}$).

\begin{figure*}
    \centering
    \includegraphics[width=0.8\textwidth]{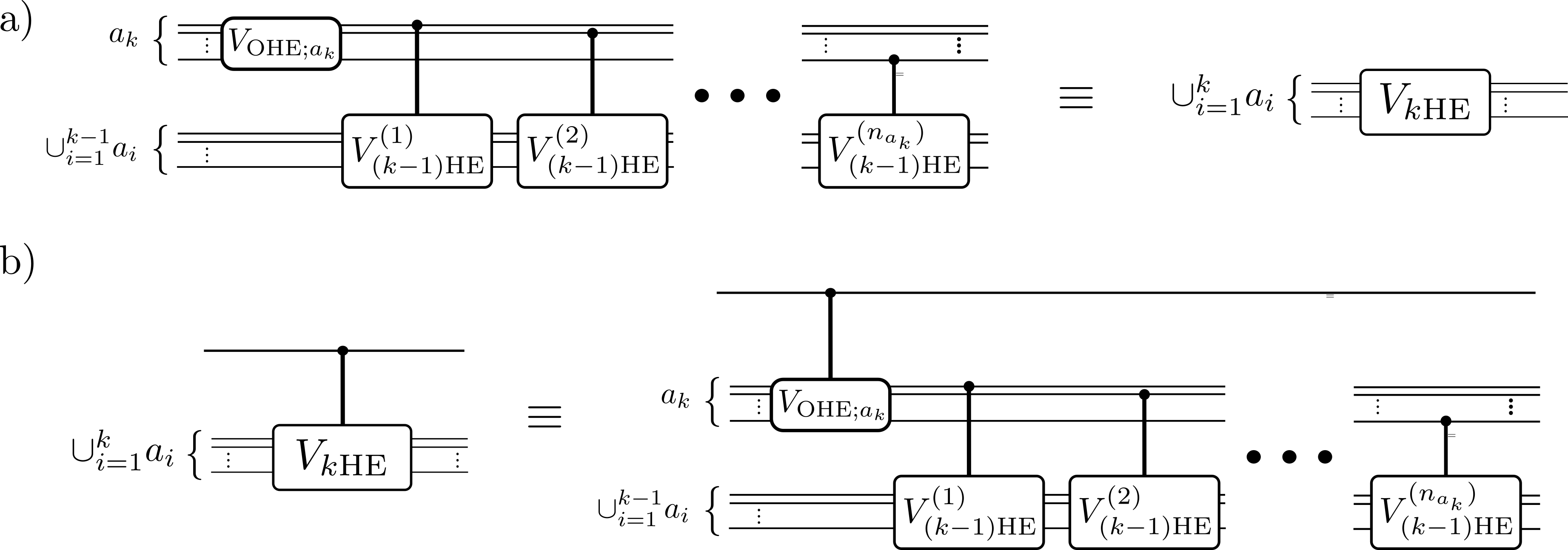}
    \caption{a) The recursion relation for constructing the $k$-Hot Encoding state preparation unitary. b) The controlled version of the $k$-Hot Encoding state preparation unitary. That at each step of the state preparation only controlled one-Hot encoding gates $V_{\rm OHE}$ are used, the controlled state preparation requires a single additional C$V_{\rm OHE}$.}
    \label{fig:StatePrep}
\end{figure*}

The preparation of the 2HE clearly exhibits a space-time trade-off. When we prepare a state with $N$ address components with $n_{a_1}=1$, the protocol takes constant time, but the number of ancillae scales as $O(N)$. Increasing $n_{a_1}$ by $L$, results in a protocol that takes $O(L)$ time but the number of ancillae is $O(N/L)$. The space-time trade-off can be made more advantageous for smaller ancillary registers if we encode the addresses in $k$HE states. In particular, the protocol for the preparation of the 2HE state above can be concatenated over 
$k$ ancillary registers [see Fig. $\ref{fig:StatePrep}$], with size $n_j$. The size of the ancillary address register grows linearly with $k$, while the number of address states grow as $\prod_{j=1}^{k} n_j$. If we set $n_j = n_{\rm site}$ for a system register of size $n_{\rm site}$, the concatenated protocol requires $O(k n_{\rm site})$ ancillae and $O(n_{\rm site}^{k-1})$ time. More importantly for the discussion of error-robustness the EBGC of the state preparation of $k$HE states is
\begin{align}
    \nonumber &\frac{1}{3}\Bigg[3+ \sum_{l_{k-1}=1}^{n_{k-1}}|\beta^{(k-1)}_{l_{k}}|^2\\
    \nonumber &\times\Bigg(5 + \sum_{l_{k-2}=1}^{n_{k-2}}|\beta^{(k-2 ; l_{k-1} )}_{l_{k-2}}|^2\left(5 +  \stackrel{k-3\, \mathrm{times}}{\cdots}\right)\Bigg)\Bigg]\\
    &= \frac{3+ 5(k-1)}{3} = O(k).
\end{align}

The prepared state is a product of $k$ One-Hot encoded states, each associated with a different ancillary register
\begin{align}
    \nonumber &\ket{\Psi_{k{\rm HE}}} =\\ &\sum_{l_1=1}^{n_1} \sum_{l_2=1}^{n_{2}} \cdots \sum_{l_k=1}^{n_k} \beta^{(1)}_{l_1}\beta_{l_2}^{(2;l_1)}\cdots \beta_{l_{k}}^{(k;l_{k-1})} \bigotimes_{j=1}^{k}\ket{{\rm ohe}, l_j}. 
    \label{eq:kHE}
\end{align}
We emphasize that our protocol allows one to adjust the amplitude associated with each $k$-Hot computational basis state, for instance by using a regression tree decomposition of the sorted list of coefficients $\{\alpha_i\}$ \cite{james2013introduction}.

Lastly, notice that the protocol for preparing a $k$HE state shares the same characteristic of $V_{\rm OHE}$ [see Eq.~($\ref{eq:ImplementReflection}$)] in that we can implement a reflection operator $\mathbf{I}-2\Pi_{0}$ by simply changing the phases of the laser drives that implement each $CV_{\rm OHE}$ gate.

\subsection{Implementation of $\bar{U}_{k{\rm HE}}$}
\label{sec:ParalU}

When the control conditions are encoded in $k$-Hot Encoding basis states, $\bar{U}_{k{\rm HE}}$ can be implemented as a sequence of $k$-qubit controlled Pauli operations. Here, we give a protocol for an error-robust implementation of $k$-qubit controlled unitaries made out of their biased-error single-qubit controlled counterparts. We show that the $k$-qubit controlled unitary preserves the biased error characteristics of its single-qubit controlled counterpart. However, the error probability when the control condition is satisfied scales as $O(k)$.

\begin{table}[]
    \centering
    \begin{tabular}{c|c|c}
         &  $\tilde{V}_{k{\rm HE}}$ &  $\bar{U}_{k{\rm HE}}$\\ \hline
        EBGC &  $\left( 5k-2\right)/3$ &  $O\left(k ^2\right)$ \\  \hline
         ${\rm depth}$ & $2(1+2L)$ & $9L$ \\\hline
    \end{tabular}
    \caption{The EBGCs and the depth associated with the protocols introduced in Section $\ref{sec:DesignPrinc}$. 
    $L$ is the number of atoms in the $a_1$ register for $\tilde{V}_{k{\rm HE}}$ and the number of Pauli operators that can be implemented in parallel for $\bar{U}_{k{\rm HE}}$. 
    The controlled versions of the LCU protocol can be implemented with an additional $2/3$ error-bounded gates and 2 steps. }
    \label{tab:gatecount2}
\end{table}


In the case when we have only 2 ancillary registers (as in Section $\ref{sec:CustomV}$)
a two-qubit controlled Pauli operation can be implemented by using two single-qubit controlled Pauli operations. First, we apply a unitary that excites the second control atom from the $\ket{1}$ state to the Rydberg manifold conditionally on the state of the first control atom (we denote this operation as $C_{1}X_{2}^{(R)}$). Next, the Pauli operation is implemented on the target register. If the second control atom is in the Rydberg state, then a Pauli operation $C_{R,2}P_{t}^{(l)}$, which is implemented on the target qubit conditioned on the second control atom being in its Rydberg state. If the second control atom is not excited to the Rydberg state, then the target qubit remains in the dark state due to the EIT effect (see Section $\ref{sec:CmUk}$). Crucially, the two-qubit controlled unitary only induces errors when both control atoms satisfy the control condition because if the first control atom is not excited to its Rydberg state no other atom is excited to the Rydberg state. By repeating this protocol using $k$ control atoms, we obtain a $k$-qubit controlled Pauli operation which induces errors only if all $k$ bits of the control condition is satisfied. In the case that the $k$-qubit control condition is satisfied, then the EBGC of the implementation scales as $O(k)$.

The unitary $U_{k{\rm HE}}$ can be implemented as a series of $k$-qubit controlled Pauli operations. When $k=2$, $U_{2{\rm HE}}$ can be implemented by the following protocol.
For each layer ancillary qubit $l$ in the first ancillary register $a_1$
\begin{enumerate}
    \item Apply $C_{a_{1l}} X^{(R)}_{a_{2,1}}\cdots X^{(R)}_{a_{2,n_l}}$ to excite the qubits in the second ancilla register to the Rydberg state conditionally on the state of the $l^{\rm th}$ qubit in $a_1$ being in state $\ket{1}_{a_{1,l}}$.
    \item Apply $\{C_{R;a_{2,i}}P_{i}^{(l)}\}$ in parallel
    \item Apply $C_{a_{1,l}} X^{(R)}_{a_{2,1}}\cdots  X^{(R)}_{a_{2,n_l}}$.
\end{enumerate}
The implementation depth of the above protocol is $(2*4+1)n_{a_1}=9n_{a_1}$. We note that the second step requires depth 1 as the control register is already excited to the Rydberg manifold. 
The EBGC is 
\begin{align}
    &1/3\sum_{l=1}^{L}|\beta^{(1)}_l|^2\left\{8 + \sum_{i=1}^{n_{a_2}}\left[|\beta_{i}^{(2;l)}|^2 \rm{supp}(P_{i}^{(l)})\right]\right\} \\
    \nonumber &= O(\max_{i,l} {\rm supp}(P_i^{(l)}))=O(k),
    \label{eq:GateLCU}
\end{align} 
where in order to obtain the last equality, we assumed that the Pauli strings $\{P_i^{l}\}$ are $k$-local.
We emphasize that the EBGC that does not scale with $n_{\rm site}$ or $N$, but only depends on the maximum support of the multi-qubit Pauli operators in the decomposition of the signal operator. Note also how we take advantage of the Rydberg state to avoid introducing new ancillae in the implementation of a two-qubit controlled unitary \cite{nielsen2002quantum}.

\begin{figure*}
    \centering
    \includegraphics[width=\textwidth]{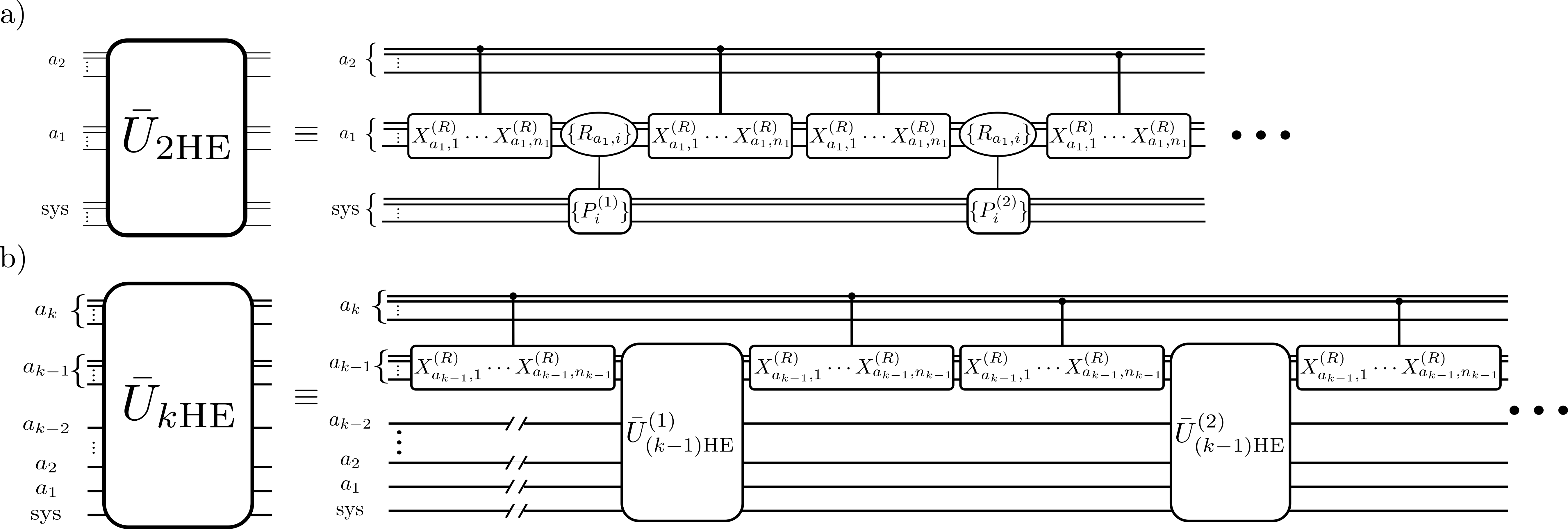}
    \caption{The circuit diagrams for implementing the unitaries: (a) $\bar{U}_{2{\rm HE}}$  and (b) $\bar{U}_{k{\rm HE}}$. The ancillary registers are denoted as $a_i$. In $a)$, the conditional Pauli operations with the curly brackets indicate a parallelized application. The same building block is repeated $n_{a_2}$ times to complete the application of $\bar{U}_{2{\rm HE}}$. In $b)$, we depict the concatenation of unitaries which result in the application of $\bar{U}_{k{\rm HE}}$. The building block is to be repeated $n_{a_{(k-1)}}$ times. Although the circuit suggests that the atoms in $a_l$ remain in the Rydberg manifold as we apply $\bar{U}_{l-1{\rm HE}}$, the atoms occupying the Rydberg state should be de-excited to the long-lived $\ket{r}$ state, to achieve an error-robust implementation.}
    \label{fig:barUkHE}
\end{figure*}

The above scheme can be extended to the case of $k$ ancillary registers usign $k$-qubit controlled operations and its EBGC is increased by an additional factor of $O(k)$. In Fig. $\ref{fig:barUkHE}$ $b)$, we depict the circuit identity which recursively implement $\bar{U}_{k{\rm HE}}$. 
Considering a scheme where the atoms in their Rydberg states in the $l^{\rm th}$ register are transferred to the $\ket{r}$ state when they are not needed, EBGC of $\bar{U}_{k{\rm HE}}$ conditioned on a $k$HE state is 
\begin{align}
    \nonumber &\sum_{l_k=1}^{n_{a_{k}}} |\beta^{(k)}_{l_k}|^2\left( 8 + \sum_{l_{k-1}=1}^{n_{a_{k-1}}}|\beta_{l_{k-1}}^{(k-1;l_k)}|^2\left(8+ \stackrel{k-2\, \mathrm{times}}{\cdots} \right)\right)\\
    &= O(k^2),
\end{align}
where we again consider $k$-local Pauli operators.

\textit{ Controlled-$W$ gates}:
\label{sec:CW}
In order to implement QSP protocols where the processing step of each iteration contribute only a constant error probability to the EBGC, we need to implement a single-qubit controlled version of
the walk operators $W$ in Eq.~($\ref{eq:QubitWalk}$). This can be easily implemented by conditioning the first step of the $k$HE state-preparation unitary. The controlled walk operator has a additional EBGC of $2/3$, and the depth of the protocol is increased by 2.

As a result, using the multi-qubit gates described in this section, the QSP walk operator [see Eq. (\ref{eq:QubitWalk})] block-encoding an operator that is a linear combination of $k$-local Pauli operators can be implemented with a total EBGC that scales as $O(k^2)$

\begin{figure*}
    \centering
    \includegraphics[width=\textwidth]{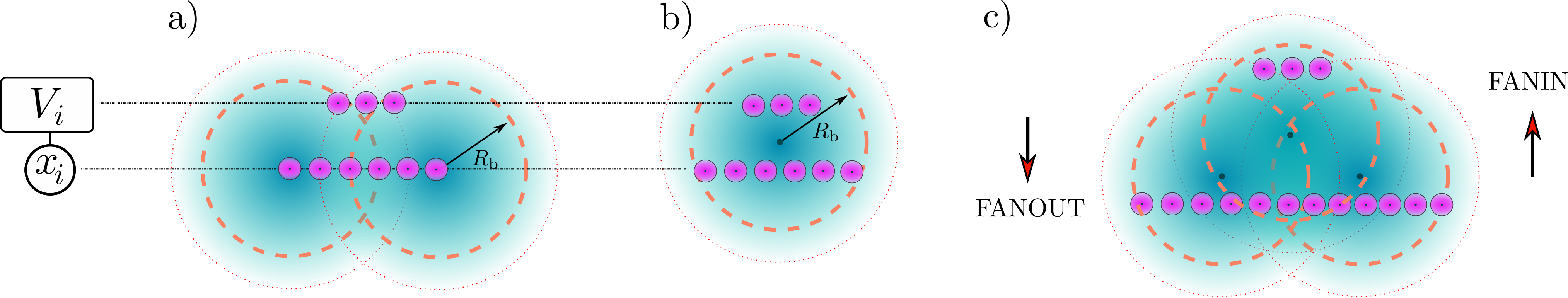}
    \caption{The problem of scalability and its resolution through FANIN and FANOUT protocols a) Due to the finite length scale of the Rydberg blockade, the conditional unitary gates cannot be applied when the size of the target or control register is larger than the maximum blockade volume $V_{b,{\rm max}}$. b) The communication between registers whose size exceeds $V_{b,{\rm max}}$ can be facilitated by  
    an additional network of ancillae, and c) the implementation of FANIN ($\uparrow$) and FANOUT ($\downarrow$) protocols. }
    \label{fig:Connector}
\end{figure*}

\section{Scalable implementation of LCU on the Rydberg atom platform}
\label{sec:scalable}

So far we have considered the situation where the largest blockade radius attainable is infinite. In this section, we consider the more realistic situation where the maximum range $R_{b,\rm max}$ of blockade interactions is finite. In a typical experiment the range of the resonant dipole interactions that result in F\"{o}rster processes do not exceed $30\,\mu$m, while the separation of the Rydberg atoms trapped by holographic optical tweezers is around $2\,\mu$m \cite{barredo2018synthetic}. 
Hence, the scalability of the protocols introduced in the last two sections is restricted ultimately by $R_{b,{\rm max}}$ because they assume a blockade radius larger than the system size. To engineer scalable protocols, we divide the system and ancilla qubits into a total of $n_{\rm sub}$ modules whose sizes are determined by $R_{b,{\rm max}}$. The main challenge in designing a scalable implementation of QSP protocols on the Rydberg atom platform is to make sure that the different subsystems can communicate efficiently.

Remarkably, the scalable protocols for implementing LCU-based QSP walk operators 
only require the $n_{\rm sub}$ subsystems to communicate a single qubit of information between themselves. This information can be communicated either by what we call ``connector" ancillae which serve as wires connecting different modules, or by physically transporting the ancillae appropriately using optical tweezers~\cite{bluvstein2021quantum}. The incoming information is processed and then output by a gadget we refer to as the telecommunication port, which introduces only three ancilla qubits per subsystem. 

Here, we describe explicit protocols to realize a modular and distributed implementation of the QSP walk operator constructed out of multi-qubit gates $V_{\rm OHE}$ and $C_{x_0}U_1 \cdots U_k$. The main contributions of this section is the  demonstration of a scalable LCU protocol which maintains an error-robust implementation, with an EBGC scaling $O(n_{\rm sub})$. Hence, when the EBGC is valid, the implementation of the LCU-based QSP walk operator has an error probability that does not scale with the number of Pauli operations in Eq. ($\ref{eq:BlockEncOp}$) and thus has an error-robust implementation. The analysis below demonstrates that the EBGC scaling is dominated by the implementation of the state-preparation step.


\subsection{Telecommunication ports and the implementation of FANIN and FANOUT protocols}

\label{sec:Connector}

In the absence of additional ancillae, blockade interactions cannot be used to entangle registers larger than the blockade volume $V_{b,{\rm max}} \propto R_{b,{\rm max}}^d$ in $d$ dimensions.
We depict the geometric constraints resulting from a finite $R_{b,{\rm max}}$ in Fig. $\ref{fig:Connector}$.
Similarly, the finite blockade radius does not allow the implementation of the $V_{\rm OHE}$ gate when the qubits in the relevant register occupy a volume larger than the blockade volume. The solution to this problem requires the ability (i) to broadcast the information regarding a single subsystem to many others (1-to-many communication), and (ii) to bring the relevant information of many subsystems to one particular subsystem (many-to-1 communication). We satisfy these requirements by utilizing FANOUT and state transfer protocols. Because both of these protocols are implemented through single-qubit controlled unitaries introduced in Section $\ref{sec:CmUk}$,
the resulting implementations are subject to EBGC.

In the following we consider a protocol involving where the ancillary target registers with $n_{\rm sub}^{(t)}$ subsystems. On the other hand, each one of the $k$ control registers encoding $k$HE addresses is divided into $n_{\rm sub}^{(c)}$ subsystems. We denote the $i^{\rm th}$ subsystem of the target register as $s_{i}^{(t)}$, and the $i^{\rm th}$ subsystem of the $k^{\rm th}$ control register as $s_{i,k}^{(c)}$.

\begin{figure}
    \centering
    \includegraphics[width = 0.45\textwidth ]{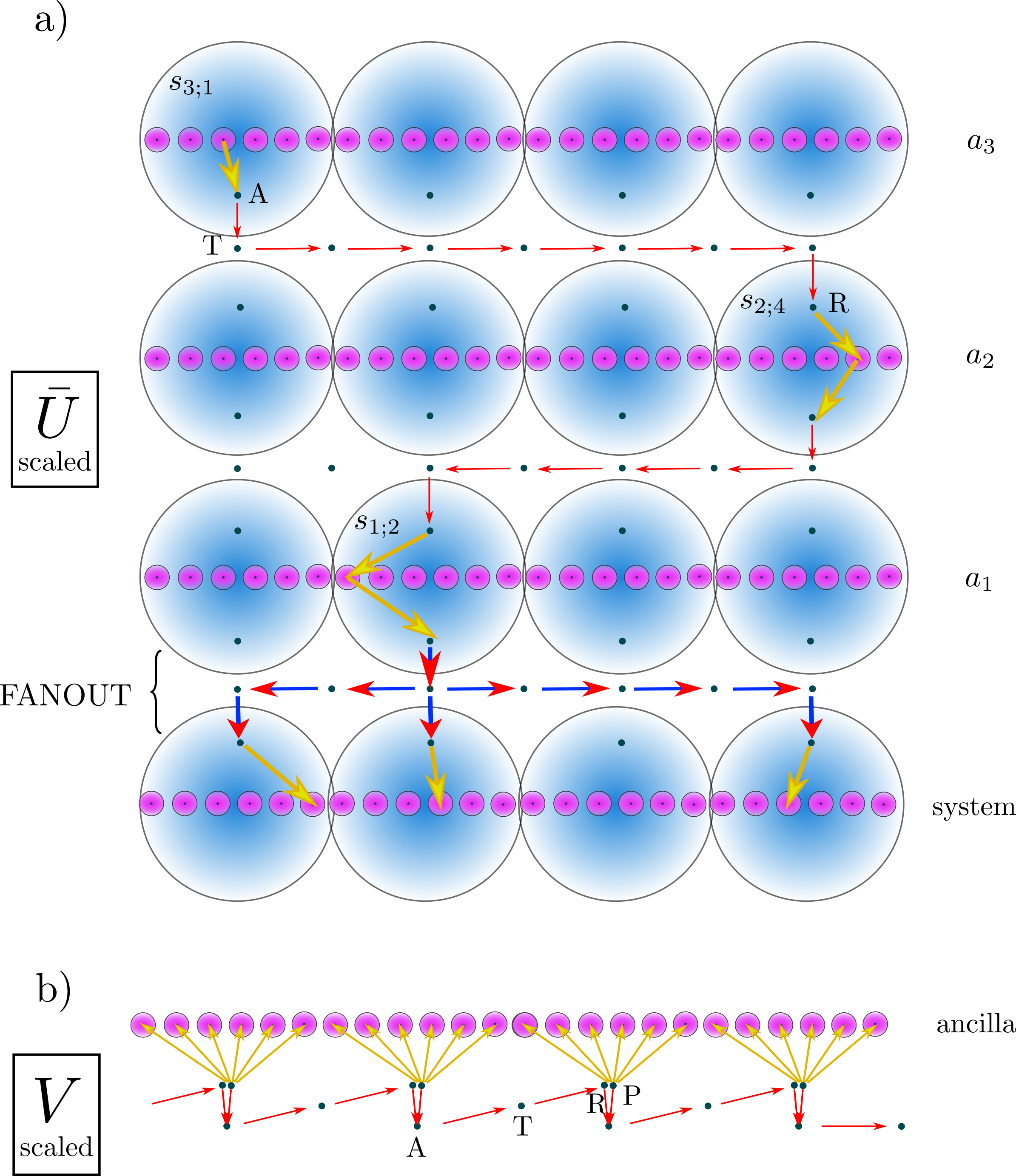}
    \caption{ The schemes for a scalable implementation of unitaries $C_m U_1 \cdots U_k$ and $V_{\rm OHE}$ in $a)$ and $b)$, respectively. The direction of the arrows convey the flow of the information of whether the control condition is satisfied. The yellow arrows connect the ancillary address and the system register to the network of telecommunication ancillae, and the red arrows depict the routing of the condition satisfaction information. The labels $A$, $T$, $R$, and $Q$ on the ancillary network stand for antenna, transmission, receiver, and processor ancillae described in the text, respectively. }
    \label{fig:Scaling}
\end{figure}

\textit{Telecommunication ports}: For each subsystem~$s_i$, we also introduce a telecommunication port (see Fig. \ref{fig:Scaling}) consisting of 3 ancilla qubits referred as: antenna ($A_i$), receiver ($R_i$), and processor ($Q_i$). For simplicity, we also assume that a set of connector ancillae $\{T_{ij}\}$ connecting $A_i$ to $R_j$. The role of $A_i$ and $R_j$ is to facilitate the communication of whether a control condition is satisfied or violated between $s_i$ and $s_{j\neq i}$. The processor ancilla $Q_i$, is only necessary for the scalable version of the One-Hot Encoding state preparation unitary, and is used to load the required amplitude information into each subsystem (see Section $\ref{sec:VOHEScale}$).

\begin{figure}
    \centering
    \includegraphics[width=0.45 \textwidth]{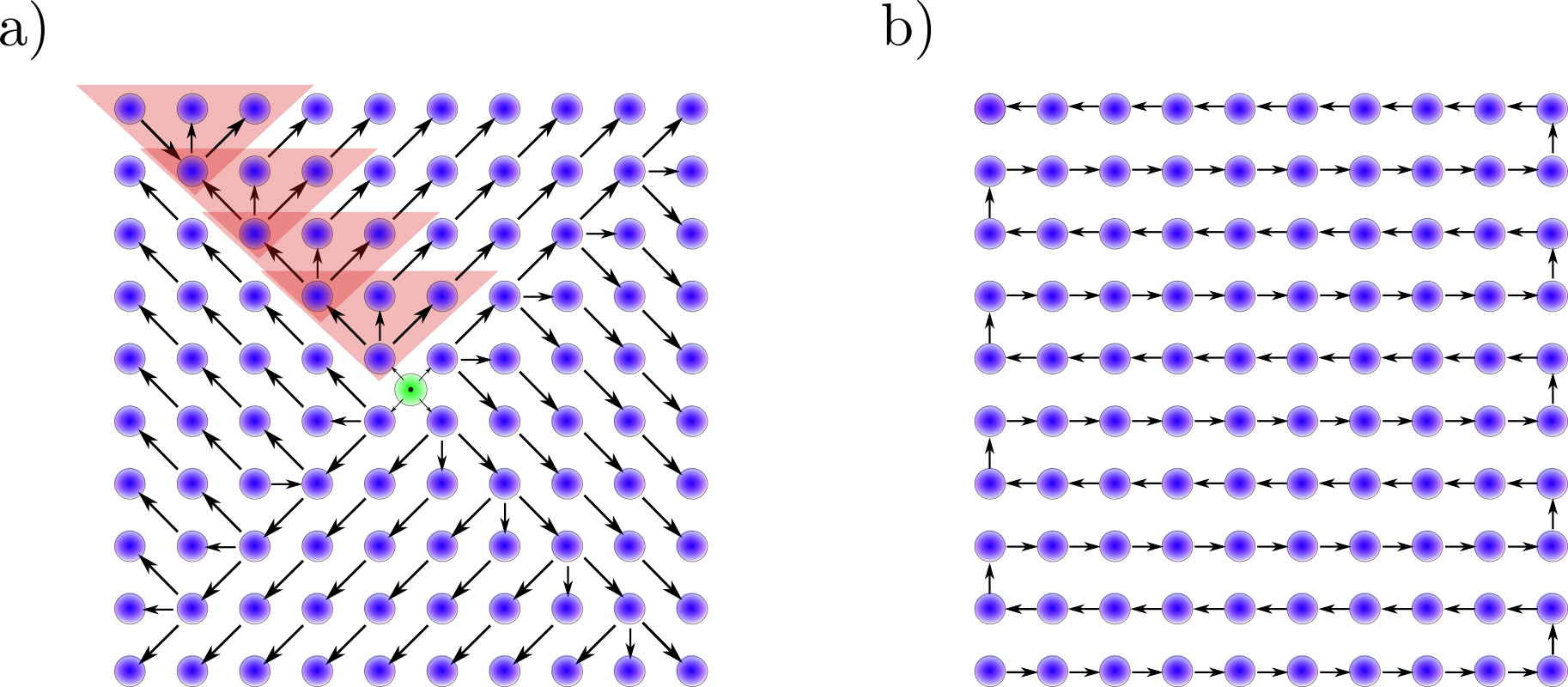}
    \caption{The schematic for the scalable protocols implementing $C_{x_0}U_1\cdots U_k$ (a) and $V_{\rm OHE}$ (b) gates, on a distributed network. The blue circles depict the telecommunication port of each subsystem. In (a) pattern of the arrows show the FANOUT of the state of the central atom (in green) to the target modules. The red triangles depict the regions where the information from one module is fanned-out to more than a single target module. The FANOUT protocol has optimal depth given the locality of the problem. In (b) we depict the path that the information regarding whether a unitary $V_{\rm OHE}^{(i)}$ is applied on subsystem $s_i$. The depth of the implementation scales linearly with $n_{\rm sub}$, which is optimal in one-dimensions, but suboptimal otherwise.}
    \label{fig:2Dlayout}
\end{figure}

\textit{FANOUT and single-qubit state transfer}:
The FANOUT protocol broadcasts the state of a single qubit to the receiver ancillae of many subsystems \cite{deutsch1989quantum}.
It can be implemented using single control multi-target  $\mathrm{CNOT}_k$ gates assuming that the target qubits are all initiated in the $\ket{0}$ state. Considering the 2D layout depicted in Fig. $\ref{fig:2Dlayout}$, the state of the central ``source" atom $A_c$ (in green) can be broadcasted using a parallelized implementation of CNOT and ${\rm CNOT}_3$ gates in accordance with the arrows connecting the subsystems in Fig. $\ref{fig:2Dlayout}$. This scheme, and its extension to $d$ dimensions implements the FANOUT gate using $O(n_{\rm sub})$ gates and in $O(n_{\rm sub}^{1/d})$ steps, which is optimal for local systems. Because  we are using single-qubit-controlled unitaries at each step, this protocol has an error-robust implementation. In particular, the EBGC for broadcasting a state $\ket{\psi_c}$ to $n_{\rm sub}$ subsystems scale as $|\langle 1_c \ket{\psi_c}|^2 O(n_{\rm sub})$. As a result, when the Pauli strings implemented are $k$-local with $k<n_{\rm sub}$, the EBGC for implementing $\bar{U}$ in Eq. ($\ref{eq:Ubar}$) scales a $O(k n_{\rm sub}^{1/d})$. 

The error-robust implementation of state transfer is similar to the FANOUT protocol. Starting from a source qubit in state $\ket{\psi_c}$, and all target qubits initialized to $\ket{0}$, each step of the state transfer is implemented by two CNOT gates, where the second CNOT gate has the control and target qubits swapped. This state transfer protocol preserves the biased error model and the induced error is $O(m|1_c\langle \ket{\psi_c}|^2)$ for a state transfer of $m$ steps. We emphasize that the error-robust implementation of the FANOUT and state-transfer protocols is possible because the state of all qubits is known at each step of the evolution. 


\subsection{Scalable implementation of $C_{x_0}U_1\cdots U_k$ and $V_{\rm OHE}$ }
\label{sec:EntLongDist2}

In this section, we utilize the FANOUT and state transfer protocols discussed above to construct the modular and distributed versions of the $C_{x_0}U_1\cdots U_k$ and $V_{\rm OHE}$ gates. We assume that the control conditions are given by $k$HE bistrings, and all telecommunication ports are initiated to $\ket{0}$. Given these constructions, the implementation of modular and distributed versions of LCU-based QSP walk operators, and their controlled versions are straightforward. 

\subsubsection{$C_{x_0}U_1\cdots U_k$}
\label{sec:CmUkScale}

A modular and distributed version of $C_{x_0}U_1\cdots U_k$ can be implemented by first applying a $k$-qubit controlled NOT operation on an additional ancilla $A_c$ conditionally on the $k$-bit control condition $x_0$ being satisfied. 
Then, the state of $A_c$ can be broadcasted to the $O(k)$ receiver ancillae $\{R^{(t)}_i\}$ associated with the subsystems of the target register, using a FANOUT protocol as depicted in Fig. $\ref{fig:2Dlayout}$. Lastly, (denoting the length of the Pauli string on the $i^{\rm th}$ subsystem of the target register as $k_i$), the controlled unitary $C_{R^{(t)}_i}U_{1}\cdots U_{k_i}$ is applied to the $i^{\rm th}$ subsystem of the target register and the control and FANOUT circuits are uncomputed. Next, we discuss how the FANOUT and the state transfer protocols introduced in the previous subsection allows each step of $C_{x_0}U_1\cdots U_{k}$ to be implemented in an error-robust manner. 

First, in order to apply a $k$-qubit controlled unitary over $O(k)$ subsystems of the $k$HE control register, we make use the state transfer protocol described in the last subsection. To describe our protocol, we first determine the subsystems of the control register that are relevant for the desired $k$-qubit control condition. Then, because each of the $k$ control registers is in a OHE state, the control condition $x_0$ can be checked via one subsystem per control register. We denote the relevant subsystem of the $i^{\rm th}$ ($i\in [k]$) control register as $s_{i;j(i,x_0)}$ (notice that $j$ is a function of the control condition $x_0$). The state transfer protocol is used to transfer whether the control condition is satisfied in subsystem $s_{i;j(i,x_0)}$ to the receiver ancilla $R_{i-1,j(i-1,x_0)}$ of  $s_{i-1;j(i-1,x_0)}$  (See Fig.~$\ref{fig:Scaling}$). Repeating this state transfer protocol starting from the $k^{\rm th}$ control register to the $1^{\rm st}$ control register, we ensure that the $A_c$ is excited to the Rydberg state only if the $k$-bit control condition is satisfied. The $k$-qubit controlled NOT operation does not induce errors when the first control register violates the control condition, in which case all ancillae used for state transfer steps remain in the $\ket{0}$ state. Hence, the $C_{x_0}$NOT$_c$ gate has an EBGC of $O(k n_{\rm sub}^{(c)1/d}|\beta^{1}_{l}|^2)$, where $n_{\rm sub}^{(c)}$ is the number of subsystems per One-Hot-Encoding ancillary register.

Next, to implement the conditional $k$-local Pauli operation acting on $O(k)$ subsystems in the target register, we use the FANOUT protocol to transfer the state of $A_c$ to the $O(k)$ subsystems of the target register. Because the FANOUT protocol preserves the error bias, the associated EBGC scales as $O(kn_{\rm sub}^{(t)1/d}|\beta^{(1)}_{l}|^2)$, with $n_{\rm sub}^{(t)}$ is the total number of subsystems in the target register.

Finally the required Pauli operations are applied on each subsystem of the control register conditionally on the state of the corresponding receiver ancillae. Again, the biased error model is preserved, and the EBGC associated with the implementation of Pauli operations is only $O(k |\beta^{(1)}_{l}|^2)$, . 
As a result, the total error probability of implementing a $k$-qubit controlled $k$-local Pauli operation scales as $O(k(n_{\rm sub}^{(c)1/d} +  n_{\rm sub}^{(t)1/d})  |\beta^{1}_{l}|^2)$. 
On the other hand, the implementation of $C_{x_0}U_1\cdots U_k$ has a gate complexity that scales as $O\left(k (n_{\rm sub}^{(c)1/d} +  n_{\rm sub}^{(t)1/d})\right)$, and our implementation is error-robust. The time complexity of the protocol is $O(n_{\rm sub}^{1/d})$.

We can implement $\bar{U}_{k{\rm HE}}$ in a scalable way by applying a sequence of $N$ $k$-qubit controlled Pauli operations conditioned on a $k$-HE control register. The EBGC of this implementation is $O(k(n_{\rm sub}^{(c)1/d} +  n_{\rm sub}^{(t)1/d}))$, and is independent of the number $N$ of $k$-qubit controlled Pauli operations. 


 
\subsubsection{$V_{\rm OHE}$}
\label{sec:VOHEScale}
The unitary $V_{\rm OHE}$ acting on $n_{\rm sub}$ subsystems of size $n_j$ achieves the following transformation
\begin{align}
    \nonumber &V_{\rm OHE} \prod_{j}^{n_{\rm sub}} \ket{0}^{
    \otimes n_j} =  \\
    \nonumber&= \sum_{j=1}^{n_{\rm sub}} \beta_j\sum_{i=1}^{n_j} \alpha_{i}^{(j)} \ket{0}^{\otimes n_0}\cdots\ket{0}^{\otimes n_{j-1}}\otimes \ket{\rm{ohe},i}_j\otimes\\
    &\quad \ket{0}^{\otimes n_{j+1}}\cdots \ket{0}^{\otimes n_{\rm sub}},
    \label{eq:Vscale}
\end{align}
where $\ket{\rm{ohe},i}_j$ is a One-Hot Encoding computational basis state of the  subsystem $s_j$, and $\alpha_i^{(j)}$ are the OHE amplitudes of the $j^{\rm th}$ subsystem. Unitarity entails $\sum_{j}^{n_{\rm sub}}|\beta_j|^{2} = \sum_{i=1}^{n_j}|\alpha_{i}^{(j)}|^{2}=1$. 

In order to implement $V_{\rm OHE}$ in a scalable manner, we need to design a protocol where the unitary $V_{\rm OHE}^{(j)}$ is applied only (i) with a probability amplitude $\beta_j$ and (ii) if all the atoms in the register are in the $\ket{0}$ state.
To overcome these challenges, we 
introduce the processor ancillae $Q_j$ for each subsystem. Then $V_{\rm OHE}$ can be implemented over $n_{\rm sub}$ subsystems using single-qubit controlled version of $V_{\rm OHE}^{(j)}$ conditioned on $Q_j$ being in the $\ket{1}_{Q_j}$ state. The single-qubit controlled $V_{\rm OHE}^{(j)}$ allows us to (i) impose the probability amplitude $\beta_{j}$ to each $V_{\rm OHE}^{(j)}$, and (ii) determine whether $V_{\rm OHE}^{(j)}$ is implemented without using a multi-qubit controlled measurement. 


The protocol that we propose for implementing the distributed version of $V_{\rm OHE}$ is the following. First, set all antenna ancillae to the state $\ket{0}$. Then, for each subsystem $s_{i}$ with $i\geq 1$ [arranged along a line as in Fig. $\ref{fig:2Dlayout}$ b)], 
\begin{enumerate}
    \item (Query whether $V^{(j)}_{\rm OHE}$ applied for $1<j<i$.) Transfer the state of $A_{i-1}$ to $R_{i}$. Uncompute the ancillary qubits $\{T_{i-1,i}\}$. 
    \item (Input coefficient $\beta_i$) Apply a single-qubit rotation to $Q_i$ along the $x$-axis by an angle $\theta_{i}\equiv \arcsin{(\beta_i)}$
    \item Apply $C_{01;R_iQ_i}V^{(i)}_{\rm OHE}$
    \item (if  $V^{(l)}_{\rm OHE}$ for $l\leq i$ is applied, turn on antenna~) Apply $C_{\neq 00;R_iQ_i}X_{A_i}$. 
    \item Repeat for the $i+1$-st subsystem, until $i=n_{\rm sub}$.
\end{enumerate}
The protocol completes in $O(n_{\rm sub})$ steps, which is optimal in one-dimensions, but suboptimal in higher dimensions. 
However, the condition that each subsystem should receive the information from every other subsystem is a constraint that likely makes an implementation with circuit depth $O(n_{\rm sub}^{1/d})$ impossible. 
The EBGC of the above protocol is $O(n_{\rm sub})$ due to the gates applied to on the telecommunication port of each subsystem.
Lastly, note that as discussed in Section $\ref{sec:LCU}$, the ancillary qubits do not need to be uncomputed for a successful implementation of the LCU-based block encoding.

\section{Implementation of Optimal Hamiltonian simulation protocols on the Rydberg atom platform}
\label{sec:HamSimImpl}

In the previous section, we demonstrated that the implementation of the algorithmic primitives of the QSP framework in Rydberg atoms is error-robust and scalable. Next, we focus on Hamiltonian simulation as a particular application of QSP. We first give an overview of different approaches to the Hamiltonian simulation problem, including (i) Hamiltonian simulation algorithms based on product formulas, (ii) QSP-based optimal Hamiltonian simulation of generic Hamiltonians, and (iii) optimal simulation algorithm of Ref. \cite{haah2021quantum} for geometrically local Hamiltonians (which we oversimplistically refer to as block-decimated QSP). In Section $\ref{sec:Results}$, we compare these three approaches by comparing the EBGC counts and circuit depths for implementations on the Rydberg atom platform. The details of each Hamiltonian simulation algorithm, as well as the explicit calculations of EBGC and circuit depth are presented in Appendix \textcolor{red}{$\ref{sec:GateCount}$}

\subsection{Hamiltonian simulation}
\label{sec:HamSim}
The use of physical quantum systems to simulate quantum dynamics has a rich tradition.
The task of quantum Hamiltonian simulation is simply stated: given any initial state $\ket{\psi_0}$ of $n$ qubits, a Hamiltonian $H$ and evolution time $t$, construct a sequence of quantum gates, which approximates the final state $\ket{\psi_{f}}=e^{-iHt}\ket{\psi_0}$. 
In Ref. \cite{lloyd1996universal}, Lloyd provided the first demonstration that this task is feasible. The strategy of what is now known as product formulas (PF) \cite{nielsen2002quantum} is to make use of the algebraic structure of the local terms in the expansion $H = \sum_{l=1}^{L} H_{l}$ through the Baker-Campbell-Haussdorf identity
\begin{align*}
&\exp{(-iHt)} =\\
&=\left(e^{-iH_1t/r}e^{-iH_2t/r} \cdots e^{-iH_Nt/r}\right)^r + O\left(\frac{(L\Lambda t)^2}{r}\right),
\end{align*}
where $\Lambda = \max_{l} |H_l|$, and $r$ is the number of time slices used in the approximation. For a fixed error tolerance $\epsilon$, and a geometrically local Hamiltonian for which $O(L) = O(n)$ the number of time slices required is quadratic in the simulated space-time volume $r=O((n t)^2/\epsilon)$. Since each time slice has $O(n)$ operations, the total gate complexity of the PF algorithm is $O(n^3t^2/\epsilon)$. Higher order PF exist \cite{suzuki1991general}, and at order $2k$ the dependence of the the gate complexity on the system size improves to $O(5^{2k}n^2 t/\epsilon^{1/2k}))$, although in the limit of large $k$ the prefactor becomes prohibitive. Recently Ref. \cite{childs2019nearly} showed that in the case of a one-dimensional system with nearest-neighbor interactions, the gate complexity can be reduced by a factor of $n$ using an integral representation of the Trotterization error, and the resulting algorithm has a gate complexity of $O((nt)^{1+1/2k}/\epsilon^{1/2k})$ which scales almost linearly in the simulated space-time volume. 

The PF algorithm of Ref. \cite{childs2019nearly} for the 1D system with nearest-neighbor interactions analytically demonstrates the validity of arguments put forward by Jordan, Lee, and Preskill \cite{Jordan2012} which claimed that the simulation of quantum dynamics generated by geometrically local Hamiltonians requires a gate complexity at least linear in the simulated space-time volume. 
The more general question: ``Can the same gate complexity be obtained for any time-dependent local Hamiltonian?" was answered affirmatively by Haah, Hastings, Kothari, and Low \cite{haah2021quantum}, and the gate complexity of the algorithm, $O(nt {\rm polylog}(nt/\epsilon))$ was proved to be optimal even for simulating only local observables. 
Optimal Hamiltonian simulation algorithm of Ref. \cite{haah2021quantum} makes use of two facts. The first is known as Lieb-Robinson bounds \cite{lieb1972finite,hastings2010locality}, which constrain how information spreads in local Hamiltonian systems. The second is that the novel algorithmic frameworks of LCU and QSP enables optimal Hamiltonian simulation in a small subsystem with respect to all parameters (i.e, with polynomial cost in the system size while achieving a polylogarithmic dependence on the error threshold)\cite{low2019hamiltonian}.

For generic Hamiltonians, \cite{low2019hamiltonian} showed that the QSP-based simulation algorithm has optimal query complexity. However, whether the gate complexity of the algorithm is optimal depends on the gate complexity of the QSP walk operator. Considering the implementation of the walk operator described in our work, the gate complexity of implementing the walk operator scales as $O(N)=O(n^{k})$ for a $k$-local Hamiltonian, resulting in a total gate complexity of $O(N^2t)$, which is not optimal even for geometrically local Hamiltonians. On the other hand, our results show that it is possible to implement the walk operator for $k$-local Hamiltonians with constant EBGC. Together with the optimal query complexity of the QSP-based Hamiltonian simulation, we conjecture that the scaling of the error probability is optimal for $k$-local Hamiltonians.


\begin{figure*}
    \centering
    \includegraphics[width = 0.9\textwidth]{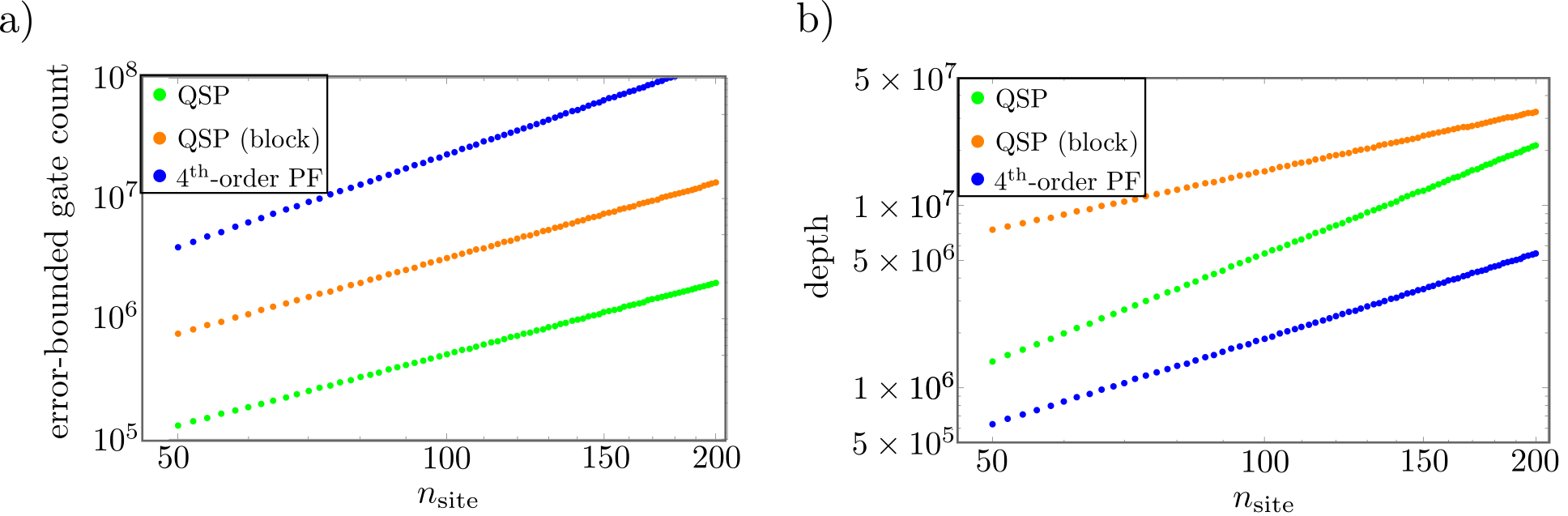}
    \caption{a) The EBGCs and b) the circuit depth of Hamiltonian simulation algorithms (i) based on the fourth-order product formula (blue, Ref. \cite{childs2018toward}), (ii) QSP-based local Hamiltonian simulation (orange , Ref. \cite{haah2021quantum}) which uses block-decimation, and (iii) QSP-based Hamiltonian simulation (green, Ref. \cite{low2019hamiltonian}). We assume that the QSP-based Hamiltonian simulation of Ref. \cite{low2019hamiltonian} can be implemented without the scalable protocols discussed in Section $\ref{sec:scalable}$. While from the point of view of depth complexity, the fourth-order product formula is superior in all system sizes considered, the QSP-based Hamiltonian simulation of Ref. \cite{low2019hamiltonian} exhibits the lowest implementation overhead in terms of EBGCs, achieving more than an order of magnitude reduction compared to the fourth order product formula, while simultaneously exhibiting better asymptotic scaling. }
    \label{fig:TheData}
\end{figure*}

\subsection{Results}
\label{sec:Results}
We calculate the EBGC and circuit depth of three Hamiltonian simulation algorithms for the one-dimensional disordered transverse field Ising model: (i) fourth-order product formula, (ii) block-decimated QSP of Ref. \cite{haah2021quantum}, and (iii) QSP-based simulation algorithm of Ref. \cite{low2019hamiltonian}. The results are displayed in Fig.~$\ref{fig:TheData}$.

Most strikingly, the QSP-based Hamiltonian simulation of Ref. \cite{low2019hamiltonian} has an EBGC (green in Fig. $\ref{fig:TheData}$ a) ) that is more than an order of magnitude smaller than that of the simulation based on the fourth-order product formula (in blue), for $n_{\rm site}=50$.
The optimal local Hamiltonian simulation algorithm of Ref. \cite{haah2021quantum} has an increased overhead (in orange) due to the block-decimation, which results in a substantial overhead in the query complexity in comparison to the algorithm of Ref. \cite{low2019hamiltonian}. Note that since the scaling of EBGC for both QSP-based Hamiltonian simulations are optimal, there is no system size for which the fourth order product formula is more robust to errors than its QSP-based counterparts.  

From the point of view of the circuit depth, the fourth order product formula results in the shortest circuit depth Hamiltonian simulation for NISQ devices, although the optimal local Hamiltonian simulation of Ref. \cite{haah2021quantum} has a better scaling. Note that the time complexity of the QSP-based Hamiltonian simulation of Ref. \cite{low2019hamiltonian} shows suboptimal scaling as it does not take advantage of the geometric locality to parallelize the simulation.

As a result, the choice of using product formula vs. QSP-based Hamiltonian simulation on the Rydberg atom platform depends on the errors relevant for the implementation. If the lifetime of the logical states are the main contributor to the decoherence, then using the product-formula-based Hamiltonian simulation is the most advantageous. On the other hand, if the logical states are long-lived and the majority of errors are introduced during gate operations, and it is possible to implement controlled unitaries with biased errors, then QSP-based Hamiltonian simulation has a clear advantage.

\section{Conclusions and Outlook}
\label{sec:conclusions}

\textit{Conclusions:}
Our work has two main messages. First, it is possible to design implementations of a wide range of quantum protocols where the error-probability scales slower than the gate complexity, by co-tailoring the relevant error-model and the compilation method. Second, the Rydberg atom platform is exceptionally well-suited for realizing such error-robust implementations in a hardware efficient manner.

To design error-robust implementations, we took the structure of two general compilation methods, LCU-based block-encoding unitaries and QSP, as a guide to determining a structured error model, as formalized in the EBGC we introduce in Section $\ref{sec:EBGC}$. Besides being very general and having near-optimal query complexity, the iterative QSP protocols allowed us to focus solely on an error-robust implementation of a walk operator. On the other hand, the most decisive property of the LCU method was that it allowed the address states (see Section $\ref{sec:LCU}$) to be chosen from $k$-Hot Encoding bitstrings, which drastically reduced the ancillae requirements.

Two observations on the Rydberg atom platform were crucial in achieving an error-robust implementation. First is the availability of biased-error single-qubit controlled Pauli operations using Rydberg atoms. Physically, the protocol for these gates minimizes the laser power that is absorbed by the Rydberg atoms during the implementation, conditionally their initial state. As a result, 
the gate induced errors are only relevant when the control condition is satisfied. The second observation is the availability of a One-Hot Encoding (OHE) state preparation using Rydberg atoms. Concatenation of OHE state preparation unitaries result in a $k$-Hot Encoding state, which can be used to encode many address states for the LCU method without drastically increasing the size of the control register.  The versatility of the Rydberg atoms was also crucial in demonstrating that the error-robust implementations of QSP protocols are scalable ( see Section $\ref{sec:scalable}$). 

We also highlighted the efficiency of our proposed implementation of Hamiltonian simulation in terms of the error-bounded gate complexity, by comparing it to a state-of-the-art implementation of product-formula-based Hamiltonian simulation algorithm.

We determined the error-robustness of our implementations based on the Error Bounded Gate Count (EBGC) introduced in Section $\ref{sec:EBGC}$. The accuracy of EBGC relies crucially on whether one can implement a single-qubit controlled Pauli with a strong suppression of the error probability conditionally on the state of the control register. We showed that such a suppression is possible at the expense of increased laser drive amplitude. Most importantly, the suppression of the error rate is proportional to the \textit{intensity} of the drive amplitude, rather than the \textit{amplitude} as is the case of conventional implementations of multi-qubit gates \cite{jaksch2000fast,Levine2019}. The quadratic improvement of error-robustness with respect to the laser amplitude sets our proposed implementation apart from other gate protocols in Rydberg atoms. Lastly, we showed in Section $\ref{sec:MainStrat}$ that using currently available Rabi frequencies, it is possible to suppress the gate induced error rate up to a factor of hundred.

\textit{Outlook}: Given the generality of QSP and LCU frameworks, and the recent successes of the Rydberg atom platform \cite{semeghini2021probing,bluvstein2021quantum}, we foresee many promising avenues of research that originate from our work. First, our techniques can be extended to implementing algorithms based on Quantum Singular Value Transformation (QSVT) in a straightforward manner. A crucial question in this direction is whether the properties of Rydberg atoms provide other substantial advantages in realizing specific QSVT-based algorithms. In particular, whether the versatility of Rydberg atoms allow for efficient and error-robust implementations of the phase rotations of the form $\exp{i\phi (2\Pi - \mathbf{I})}$ \cite{gilyen2019quantum} is an open question. Second, our work demonstrates that block-encoding unitaries for geometrically local operators have error-robust implementations using shallow circuits. It is therefore important to understand the power of QSP protocols which process geometrically local signal operators. 

Our work also provides opportunities to explore important questions for the realization of fault-tolerant quantum computation using Rydberg atoms. In particular, a demonstration of the compatibility of the error-robust implementation presented and the fault-tolerant architecture proposed by Ref. \cite{cong2021hardware} would greatly ease the resource requirements for realizing FTQC on the Rydberg atom platform.
Moreover, our results motivate the search for new and more versatile biased-error gate protocols which can further improve error-robustness with reduced classical resource requirements. Lastly, the question of whether sparse encodings (e.g., the $k$-Hot Encoding address states) are useful in the context of quantum algorithmic frameworks, such as variational quantum algorithms, seems to be a widely open.

\begin{acknowledgments}

S.Z. and S.S. thank I. Chuang, M.D. Lukin, D. Bluvstein, I. Cong, S. Ebadi, and A. Keesling for insightful comments and helpful discussions. S.Z. acknowledges the financial support from the Swiss National Science Foundation through the Early Postdoc.Mobility grant (P2EZP2\_184320) and from the Army Research Office, ARO MURI (grant no W911NF1910517). S.S. and S.Z. acknowledges NTT Research for their financial and technical support.

\end{acknowledgments}

\appendix
\section{Quantum Signal Processing}
\label{app:QSP}
We begin our discussion with the quantum control theory of a single qubit. The pioneering work of Ref. \cite{low2017quantum} asked the following question. Given two unitary operations on a single-qubit $G(\phi)=e^{i\phi \sigma_z}$, and $R(\theta)=-ie^{i\pi/4 \sigma_z}e^{i\theta \sigma_x}e^{-i\pi/4 \sigma_z}$, with Pauli operators $\sigma_{i=x,y,z}$, what single-qubit transformations can we design by the following sequence of gates
\begin{align}
    G(\phi_k)R(\theta)G(\phi_{k-1})R(\theta)\cdots G(\phi_1)R(\theta)G(\phi_0)
    \label{eq:Alternating}
\end{align}
alternating between $G(\phi)$ and $R(\theta)$ while keeping $\theta$ constant and varying $\phi_i$ between each iteration.

The concept of signal processing is established by considering $cos(\theta)=x$ as the signal encoded in the signal unitary 
\begin{align}
    R(\theta) =
    \left(\begin{array}{cc} x & \sqrt{1-x^2} \\ \sqrt{1-x^2} & - x \end{array}\right),
    \label{eq:Ref}
\end{align}
which is to be processed by the control angles $\{\phi_i\}$. We emphasize that the single-qubit rotation $R(\theta)$ can be interpreted as a block encoding of the signal $x$ since $\bra{0}R(\theta)\ket{0}=x$. Ref. \cite{low2019hamiltonian} showed that the first diagonal matrix element of the unitary resulting from a $k-$fold iteration of $G(\phi_i)R(\theta)$ can be designed to be \textit{any} degree $k$ complex-valued fixed-parity polynomial $P(x)$ via a judicious choice of the angles $\phi_i$. Formally,
\begin{align}
    \nonumber &\mathcal{U}_s \equiv\left[\prod_{i=1}^{k} G(\phi_i)R(\theta)\right]G(\phi_0) \\
    &= \left(\begin{array}{cc} P(x) & iQ(x)\sqrt{1-x^2} \\ iQ^*(x)\sqrt{1-x^2} & P(x)^* \end{array}\right),
    \label{eq:QSP1Qubit}
\end{align}
where $Q(x)\in\mathbf{C}$ is a degree $k-1$ polynomial whose parity is opposite to that of $P(x)$. Unitarity introduces the constraint   $|P(x)|^2+(1-x^2)|Q(x)|^2=1$ for $x\in[0,1]$.

While the above scheme seems to block-encode only fixed-parity polynomial $P(x)$, it is straightforward to block-encode an arbitrary-parity polynomial if we notice
\begin{align}
    \nonumber &\left(\begin{array}{cc} P(x) & iQ(x)\sqrt{1-x^2} \\ iQ^*(x)\sqrt{1-x^2} & P(x)^* \end{array}\right) \\
    &= A(x)\mathbf{1} + iB(x) \sigma_{z} +i C(x) \sigma_{x} +i D(x) \sigma_{y},
\end{align}
where all coefficients are polynomials of fixed parity, with $A(x)$ and $B(x)$ having degree $k$, while $C(x)$ and $D(x)$ having degree $k-1$ (see Ref. \cite{low2016methodology} for a full characterization). Hence, we can obtain block-encodings of arbitrary parity polynomials by a simple rotation of the qubit.
To summarize, interweaving single-qubit rotations $G(\phi_i)$ and $R(\theta)$ allows one to construct a block-encoding of an arbitrary-parity polynomial of a block-encoded signal $x$ given a suitable set of phases $\{\phi_i\}$. We note that determining the desired set of phases $\{\phi_i\}$ is not a trivial task. For instance, see Ref. \cite{MartynChuang2021} for concrete procedures for various examples and its appendix for numerically optimized phase angles. 


Ref. \cite{low2017quantum} further showed that the signal processing of scalar $x$ can be extended to processing of multi-dimensional operators using only a single additional ancilla qubit, which we will call the ``exit" ancilla in the following. Intuitively, by applying a conditional block-encoding of the signal operator $H$, we can elevate the eigenvalues $\lambda_i$ of $A$ (e.g., $A\ket{\lambda} =\lambda \ket{\lambda}$) to rotation angles $\theta_i$ for the exit ancilla.

Formally, given that the block-encoding unitary is Hermitian $U^2=\mathbf{1}$, we can introduce an iterate $W \equiv (2\left(\ket{0}\bra{0}\right)^{\otimes n_a} - \mathbf{1})U $, which can be written as a direct sum over $SU(2)$ invariant subspaces associated with each eigenvalue of $H$
\begin{align}
    W =  \bigoplus_{\lambda} \left(\begin{array}{cc}
        \lambda & -\sqrt{1-\lambda^2}  \\
         \sqrt{1-\lambda^2} & \lambda 
    \end{array}\right)_{\lambda},
    \label{eq:W}
\end{align}
where the subscript $\lambda$ means that the matrix representation is written in the following basis 
\begin{align}
    \ket{G_{\lambda}} = \ket{0}^{\otimes n_{a}}\ket{\lambda} \quad\quad \ket{G^{\perp}_{\lambda}} = \frac{\lambda \ket{G_{\lambda}}  - U\ket{G_{\lambda}} }{\sqrt{1-\lambda^2}}.
\end{align}
Hence, the eigenvectors of $W$ are given by 
\begin{align}
    \ket{G_{\lambda_{\pm}}} = \frac{1}{\sqrt{2}}\left(\ket{G_{\lambda}} \pm \ket{G^{\perp}_{\lambda}}\right),
\end{align}
with associated eigenvalues $e^{\pm i\theta_{\lambda}}$, where  $\theta_{\lambda}~\equiv~\arccos{(\lambda)}$. 

Ref. \cite{low2017quantum} showed that using a controlled version of $W$ with the exit-ancilla as the control, 
it is possible to implement the  unitary,  
\begin{align}
    U_{\phi} = \sum_{\lambda,\eta=\pm} R_{\phi}(\theta_{\lambda}) \otimes \ket{G_{\lambda \eta}}\bra{G_{\lambda\eta}},
\end{align}
which rotates the exit-ancilla along a fixed axis on the $x-y$ plane as determined by $\phi$ and by an angle determined by the phased of the eigenvalue $e^{i\theta_{\lambda}}$. 
The decomposition of $U_{\phi}$ in terms of a controlled version of $W$ and single-qubit rotations of the exit-ancilla is the following:
\begin{align}
    \nonumber U_{\phi} &= (e^{-i\phi \sigma^{(\rm{ex})}_z/2}\otimes \mathbf{1})U_0(e^{-i\phi \sigma^{(\rm{ex})}_z/2}\otimes \mathbf{1})\\
    \nonumber U_{0} &\equiv \ket{+}_{\rm ex}\bra{+}\otimes \mathbf{1} + \ket{-}_{\rm ex}\bra{-}\otimes W\\
    &=\sum_{\lambda,\eta=\pm} e^{i\eta\theta_{\lambda}/2}R_{\rm ex}(\eta\theta_{\lambda})\otimes \ket{G_{\lambda \eta}}\bra{G_{\lambda\eta}},
\end{align}
and $\sigma^{(\rm{ex})}_z$ is a Pauli operator acting on the exit ancilla.

Noting that the eigenvectors of $W$ satisfy 
\begin{align}
    \bra{0^{\otimes n_a}}\otimes \mathbf{1}_{\rm sys})\ket{G_{\lambda \pm}} = \sqrt{\frac{1}{2}}\ket{\lambda},
\end{align}
an arbitrary-parity polynomial of a Hermitian signal operator can be block-encoded. That is, 
\begin{align}
    &\bra{+}_{\rm ex}\bra{0}^{\otimes n_a} \prod_{j=1}^{k/2}U_{\phi_{2j}}U^{\dagger}_{\phi_{2j+1}+\pi} \ket{0}^{\otimes n_a}\ket{+}_{\rm ex} \\
    &= \sum_{\lambda}\tilde{P}(\lambda) \ket{\lambda}\bra{\lambda},
\end{align}
where $\tilde{P}(x)$ is an arbitrary parity polynomial of degree $k$. As a result, QSP provides an indispensable tool for processing block-encoded signal operators. Most importantly, for the following discussion on Hamiltonian simulation we would like to implement $\tilde{P}(\lambda) \approx e^{i\lambda t}$. 

\section{QSP-based optimal Hamiltonian simulation }
\label{app:QSPSim}
The Hamiltonian simulation algorithm based on the QSP framework uses the block-encoding of the Hamiltonian $H$ to construct a polynomial approximation of $P(H)\approx e^{-iHt}$. Ref. \cite{low2019hamiltonian} proved that this method results in an optimal query complexity, which is
\begin{align}
 k^* = O\left(\alpha t + \frac{\log(1/\epsilon))}{\log\log{(1/\epsilon)}}\right),
\end{align}
where we define $\alpha\equiv ||H||=O(N)$, for a Hamiltonian composed of $N$ Pauli strings. The linear scaling of $k^{*}$ with respect to the spectral norm of the Hamiltonian is due to the unitarity block-encoding utilized by QSP.  We also emphasize that the scaling of the query complexity with respect to error tolerance $\epsilon$ is exponentially improved compared to the simulation algorithms based on Trotterization \cite{childs2018toward}. Moreover, it is possible show that the number of queries $k$ can be bound by the following inequality \cite{gilyen2019quantum} 
\begin{align}
    k^* \leq e^{q}\alpha t + \frac{\ln{(1/\epsilon)}}{q} \quad \forall q \in \mathbf{R}.
    \label{eq:kcalc}
\end{align}

The query complexity of the QSP-based Hamiltonian simulation can be related to the time and error-bounded gate complexities, given a specific implementation of the query. We emphasize that even if the time required to implement the query is constant, the time complexity of Hamiltonian simulation of a system of size $n_{\rm site}$ for a time $t=n_{\rm site}$ results in an implementation time of $O(n_{\rm site}^2)$, which is suboptimal. On the other hand, implementing the query (i.e., the walk operator) with constant EBGC implies an implementation optimal with respect to errors. In Section~$\ref{sec:GateCount}$, we demonstrate that this optimal error complexity is achieved for our proposal for $k$-local Hamiltonians. 
 

\section{Optimal Quantum Hamiltonian simulation of local Hamiltonians}
\label{sec:OptimalHaah}

Since Ref. \cite{Jordan2012} argued that the optimal gate complexity of an algorithm simulating local Hamiltonian simulations should scale roughly as the simulated space-time volume, there has been a renewed interest in finding rigorous tight bounds for optimal simulation of local Hamiltonians \cite{childs2019nearly,haah2021quantum}. In 2018, an algorithm by Haah et al. \cite{haah2021quantum}, constructed an algorithm with a gate complexity linear in the simulated $d+1$ dimensional space-time volume $O(n t {\rm polylog}( n^{1/d} t/\epsilon))$, where $\epsilon$ is the total error tolerance of Hamiltonian simulation, and proved the optimality of this bound. In this section, we give a brief review of this work to motivate the constructions in Appendix $\ref{sec:GateCount}$.

\begin{figure*}
    \centering
    \includegraphics[width=\textwidth]{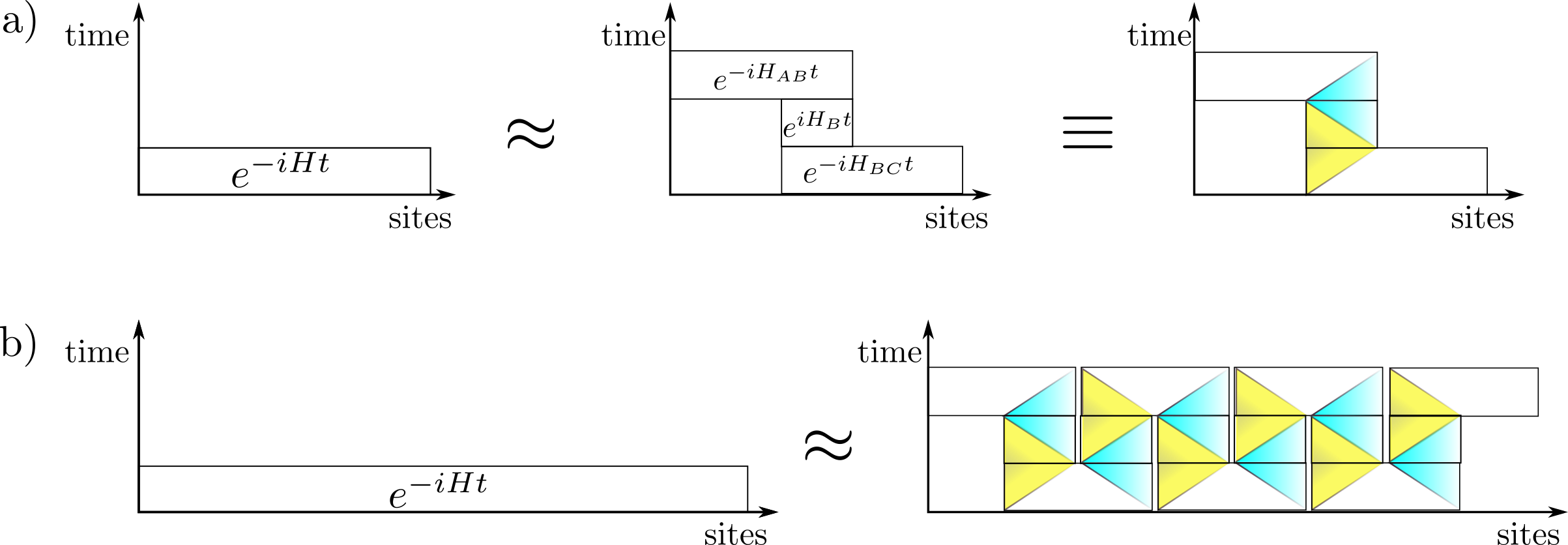}
    \caption{Overview of the algorithm in Ref. \cite{haah2021quantum}. (a) the first approximation depicts the principle relation between the initial and decimated circuits for Hamiltonian simulation. On the other hand, the equivalence relation depicts the spread and reversal of errors originating from the left (yellow) and right (blue) edges of each block. At the first time step, the error due to the left edge of a block simulating $e^{-iH_{BC}t}$ spreads within the light-cone depicted in yellow. At the second step, the evolution operator $e^{iH_Bt}$ fixes this error but it also introduces new errors due to the right boundary of the block. The final evolution by  $e^{-iH_{AB}t}$ reverses the error introduced by $e^{iH_Bt}$, and the approximation error is bounded by the exponentially small error outside of the light cone given the intermediate block is large enough to contain the light-cone. (b) Demonstration of how the block decimation operator can be repeated in space and the pattern of errors induced.}
    \label{fig:HaahOverview}
\end{figure*}

From the bird's eye view, the algorithm in Ref. \cite{haah2021quantum} uses a block-decimation of the time-evolution unitary $e^{-iHt}$ which guarantees that the error due to the decomposition is bound by the Lieb-Robinson (LR) theorems \cite{lieb1972finite,hastings2010locality}. LR theorems formalize the intuition that for local Hamiltonians, the maximum speed that information can travel is a well-defined constant, called the Lieb-Robinson velocity $v$. The overall structure of the algorithm in Ref. \cite{haah2021quantum} is depicted in Fig. $\ref{fig:HaahOverview}$ for the case of a one-dimensional system, which we consider for sake of simplicity.
Any block-decimation of the evolution operator introduces local errors at each boundary. These errors can be described as the emission of Hamiltonian terms $H_{bd}$ at the boundaries of neighboring blocks. Upon time evolution by a local Hamiltonian, it is possible to ensure that these errors propagate only within the associated light-cones (depicted as yellow and blue triangles in Fig. $\ref{fig:HaahOverview}$) up to exponentially small corrections. The knowledge of the local spread of such errors allows on to design a spin-echo-like algorithm which reverses the spread of the errors, leaving behind only the exponentially small corrections. Formally, one can show that for a given a block decimation of the sites $X = A \bigcup B \bigcup C$, a constant $\mu = O(1)$, Lieb-Robinson velocity $v$, and $ v t \ll l = dist(B)$
\begin{align}
    |U_t - e^{-iH_{A\bigcup B}t}e^{iH_{B}t}e^{-iH_{B\bigcup C}t}|< O(|H_{bd}|\exp{(-\mu l)} ),
    \label{eq:HaahDecomp}
\end{align}
where $H_X$ denotes the sum of Hamiltonian terms which have a support on a region $X$. Longer times can be simulated by applying the same decimated evolution operator repeatedly $t$ times and obtain an error linearly increasing with $t$.

Given the above discussion, the choice of the smallest dimensions of each block, denoted $l$ in the space axis and $t_{\square}$ in the time axis, is determined by (i) the tolerance $\epsilon_{\rm LR}$ for errors resulting from the block-decimation, (ii) the tolerance $\epsilon_{\square}$ for errors due to the QSP-based approximate Hamiltonian simulation of each block, and (iii) the number of blocks $m = O(nt/lt_{\square})$. In particular, given a total error tolerance $\epsilon$, we would like the error associated with each block to be $\epsilon_{\square}+\epsilon_{\rm LR} = O(\epsilon/ m)$ \cite{childs2017lecture,nielsen2002quantum}. In the following, we assume $\epsilon_{\square}= \epsilon_{\rm LR}$. Given Eq. ($\ref{eq:HaahDecomp}$), we have $\epsilon_{\rm LR} = O(e^{-\mu l})$, and the spatial dimension of each block is  $l=O(\log{(nt/t_{\square}\epsilon)})$. 
We are then left with the challenge of realizing the Hamiltonian simulation of a system of size  $O(\log{\left(nt/t_{\square}\epsilon\right)})$, with error tolerance $\epsilon_{\square} = O(\epsilon \frac{lt_{\square}}{nT})$ using only $O(\mathrm{polylog}{\left(nt/t_{\square}\epsilon\right)})$ gates. 
Fortunately, the QSP-based Hamiltonian simulation algorithm discussed in Section $\ref{app:QSPSim}$ has a query complexity that scales logarithmically with $\epsilon_{\square}^{-1}$,
resulting in a the gate complexity that scales almost linearly with the simulated space-time volume 
\begin{align}
    O(nt \,{\rm polylog}(nt/t_{\square}\epsilon)).
\end{align}
In summary, the optimal quantum Hamiltonian simulation algorithm by Haah et al. uses the QSP-based query optimal Hamiltonian simulation algorithm in combination with a clever block-decimation of the space-time evolution to obtain a near optimal gate count for Hamiltonian simulation algorithm for local Hamiltonians. 

We emphasize that the aforementioned optimal gate count is not error-bounded in the sense discussed in Section $\ref{sec:subadditivity}$, as it does not take into account the specific error model of the implementation.
Moreover, the near optimal gate count of the algorithm in Ref. \cite{haah2021quantum} comes at a price of a high implementation overhead \cite{childs2018toward}, which makes it less efficient than the product-formula algorithms for Hamiltonian simulation on near-term intermediate scale quantum (NISQ) processors. In the next section, we calculate the overhead associated with the Rydberg platform implementations of both the product formula and the QSP-based Hamiltonian simulation protocols. While the overhead of the algorithm in Ref. \cite{haah2021quantum} makes it undesirable for NISQ devices, we find that the EBGC of QSP-based Hamiltonian simulation of Ref. \cite{low2019hamiltonian} has orders of magnitude of smaller overhead for gate errors compared to its competitors.

\section{Concrete circuits and gate counts}
\label{sec:GateCount}

Here, we use the results of the Section $\ref{sec:DesignPrinc}$ to analyze the resource requirements for the implementation of QSP-based and  Hamiltonian simulation algorithms on the Rydberg atom platform, and compare them to those needed to implement Hamiltonian simulation based on product formula. To this end, we briefly discuss the implementation of product-formula-based Hamiltonian simulation algorithms on the Rydberg platform and calculate the associated EBGCs.


\subsection{Hamiltonian simulation}

Here, we explicitly calculate the resources needed for implementing algorithms in the framework of QSP, using LCU-based block-encoding.
In particular, for a system of $n_{\rm site}$ qubits, we consider the number of error-bounded gates, the runtime, and the number of ancillas required to implement a (i) conditional version of LCU block-encoding, (ii) QSP-based Hamiltonian simulation, (iii) Optimal simulation of local Hamiltonians in Ref. \cite{haah2021quantum}, and finally (iv) Hamiltonian simulation using fourth order product formula. We note that all of calculations in this section, we assume that the blockade radius can be taken large enough such that the scalable protocols discussed in Section $\ref{sec:scalable}$ are not necessary.



\subsubsection{Implementing Haah's Optimal Hamiltonian Simulation (Ref. \cite{haah2021quantum})}

To facilitate the calculation of resources needed for Hamiltonian simulation, it is necessary to chose an explicit Hamiltonian to be simulated. 
Here, we chose a the one-dimensional disordered Heisenberg Hamiltonian $H_{\rm DH}$ as our target system
\begin{align}
    H_{\rm DH} = \sum^{n_{\rm site}}_{i<j} \sigma_i\cdot\sigma_j + \sum^{n_{\rm site}}_{i} h_{i}\sigma_i^{(z)}.
    \label{eq:DisHeisenberg}
\end{align}
The choice of the disordered Heisenberg Hamiltonian as our target allows us to directly compare the cost of our implementation of QSP-based Hamiltonian simulation to that implied by the previous empirical studies that use product formulas \cite{childs2018toward,childs2019nearly}. 


Our first task is to find the dimensions of each block in the  decimation given the parameters $n_{\rm site}$, $t$, and the error tolerance $\epsilon$.

\begin{figure}[htp]
    \centering
    \includegraphics[width=0.4\textwidth]{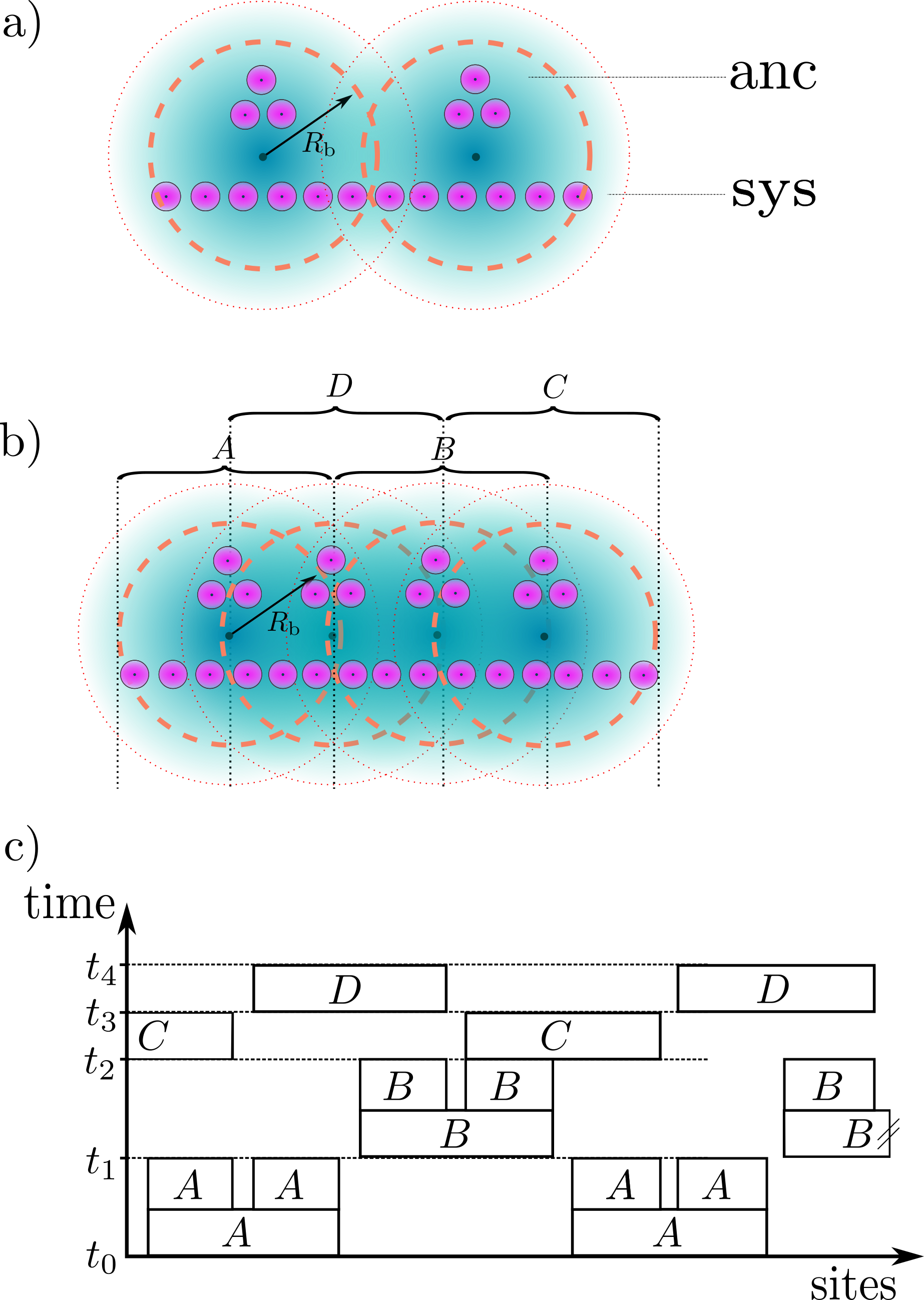}
    \caption{The overall schematic for the paralelized implementation the optimal Hamiltonian simulation of \cite{haah2021quantum} on the Rydberg platform. a) A depiction of the sources of cross-talk errors due to the algebraic decay of the interaction potential, which makes the definition of Rydberg blockade radius $R_b$ fuzzy. The two system atoms at the boundary between the two blockade volumes introduce errors due to unwanted blockade interactions. b) The experimental configuration for the parallelized application of the algorithm in Ref \cite{haah2021quantum}. We use twice as many overlapping simulation regions (i.e., $A\,,B\,,C\,$,and $D$) as depicted in Fig. $\ref{fig:HaahOverview}$ to reduce the errors due to the algebraic decay of the dipolar interactions. c) The schedule for applying the QSP-based Hamiltonian simulation algorithm on 4 different simulation regions to reduce the crosstalk errors. The crosstalk errors can be reduced arbitrarily at the expense of an increase in the circuit depth. 
    }
    \label{fig:Parallelize}
\end{figure}

In one-dimensions, the number of blocks is 
\begin{align}
    m  = 4 \left(\frac{2t n_{\rm site}}{t_{\square}l}\right),
    \label{eq:blocks}
\end{align}
where $l$ and $t_{\square}$ are the shortest dimensions of each block along space and time coordinates (see Fig. $\ref{fig:Parallelize}$), respectively. Hence, we have $m/2$ blocks of length $l$ and $m/2$ blocks of length $2l$, and the overall factor of $4$ in Eq. ($\ref{eq:blocks}$) is due to the normalization of the Hamiltonian for each site (i.e., $H_{i,i+1}\leq 1$). For a given spatial extent $l$ of each block, the parameter $t_{\square}$ can be determined by studying how the errors due the decimation scale as a function of time for a single block. This was done in Ref. \cite{haah2021quantum} and the following relation was found
\begin{align}
    0.175\left(\frac{7.9 t_{\square}}{l+0.95}\right)^{l+0.95} = \frac{\epsilon}{3m}\equiv \epsilon_{\rm LR}.
\end{align}
For the data presented here, we set $t=4n_{\rm site}$ and $ m\epsilon_{\square}= m\epsilon_{LR}=10^{-3}/2$.

Once the parameters $l$ and $t_{\square}$ are determined, we can also calculate the order $k_{\square}$ of the polynomial appoximation to the Hamiltonian evolution associated with each block using Eq. ($\ref{eq:kcalc}$).
For the smaller blocks of spatial size $l$, we get 
\begin{align}
    k_{\square}(l) = \min_{q} \left[ e^{q}t_{\square} l + \ln\left(\frac{\epsilon_{\square}}{3 m} \right)/q \right],
\end{align}
while for blocks of spatial size $2l$, we replace $l\rightarrow 2l$.

The simultaneous implementation of Hamiltonian simulation in each block may introduce unwanted cross-talk errors due to the algebraic decay of the dipolar interactions, even if the blockade radii of adjacent blocks do not overlap (see Fig. $\ref{fig:Parallelize}$ a). We circumvent this problem by doubling the implementation time required to simulate evolution for a time $t_{\square}$.
The scheme is depicted in Fig. $\ref{fig:Parallelize}$ for a one dimensional system. For each time step, we require that the blockade radii associated with different spatial blocks have negligible overlap. We also note that, in principle, the block-encoding and the QSP-based Hamiltonian simulation algorithms have the flexibility to implement the simulation of a Hamiltonians with different boundary terms are omitted at each step (see Fig. $\ref{fig:HaahOverview}$), as well as changing the overall sign of the Hamiltonian $H\rightarrow -H$.

Having determined $l$, $t_{\square}$, and $k_{\square}$, we next
calculate the depth and EBGCs for the CPHASE gate and LCU-based block-encoding using the techniques discussed in Section $\ref{sec:DesignPrinc}$ (see also Table $\ref{tab:gatecount2}$), and assuming that we are using the customized One-Hot encoding state-preparation protocol
\begin{align}
    \nonumber d_{\rm LCU} &= 2d_{V_{\rm OHE}} + d_{\bar{U}_{\rm OHE}}+1 \\
    &= 4 ( 1 + 2L ) + 9L +1 = 124\\
    n_{\rm LCU} &= 2 n_{\tilde{V}_{\rm OHE}}+ n_{\bar{U}_{\rm OHE}} =  \frac{16+10}{3} =\frac{26}{3} \\
    a_{\rm LCU} &= n_{\rm site}+L ;\quad d_{\rm CPHASE} = 3; \quad n_{\rm CPHASE} = \frac{4}{3},
    \label{eq:ParticularCounts}
\end{align}
where $d_{*}$, $n_*$, and $a_{*}$ denote the circuit depth, EBGC, and the number of ancillae needed for implementing $``*"$. In order to calculate the EBGC for $\bar{U}_{\rm OHE}$, we used Eq. ($\ref{eq:GateLCU}$), and set $L=2*3+1=7$ to implement Pauli terms associated with the 3 axes of interactions and the disordered field in Eq. $\ref{eq:DisHeisenberg}$. Lastly, here we assumed the implementation of an $L$-qubit controlled phase rotation of an additional phase ancilla, which introduces only a constant EBGC thanks to the One-Hot Encoding nature of the ancillary registers. 

Hence, implementing the conditional version of the iterate $W$ in Eq. ($\ref{eq:QubitWalk}$) costs depth $d_{CW} = 2*2 + d_{\rm CPHASE}+ d_{\rm LCU}$, where we added depth 2 for each controlled version of the unitaries. On the other hand, the EBGC of $CW$ is $n_{CW} = 2*\frac{2}{3}+n_{\rm CPHASE}+n_{\rm LCU}$.
As a result, the QSP-based Hamiltonian simulation algorithm of Ref. \cite{low2019hamiltonian} can be implemented using
\begin{align}
    \nonumber d_{QSP} &= k^* (d_{CW} + 1)\\
    \nonumber a_{QSP} &= a_{LCU} + 1\\
    n_{QSP} &= k^* (n_{CW} + 1),
\end{align}
where $k^*$ is calculated using Eq. ($\ref{eq:kcalc}$). Lastly, the computation resources to implement the local Hamiltonian simulation algorithm of Ref. \cite{haah2021quantum} are 
\begin{align*}
    \nonumber d_{\rm H1D} &=  2\frac{t}{t_{\square}}(k_{\square}(l)+2 k_{\square}(2l))(d_{CW} + 1) \\
    \nonumber a_{\rm H1D} &= \frac{n_{\rm site}}{2l} \left(4+2l+2\right) \\
    n_{\rm H1D} &= \frac{m}{2}k_{\square}(2l)\left[ n_{CW} + 1 \right].
\end{align*}
Notice that the overall factor of 2 in the depth of implementation comes from our method of reducing the cross-talk between the blocks (see Fig. $\ref{fig:Parallelize}$).
The number of ancillae is increased by an additional $O
\left(\frac{n_{\rm site}}{l}\right)$ ancillae compared to the requirements for QSP-based Hamiltonian simulation in order to facilitate the parallelization by the block-decimation.

\subsubsection{Comparison to Hamiltonian simulation with product formulas}

Here, we compare the resource requirements for our implementation of the QSP-based Hamiltonian simulation algorithm on the Rydberg platform to those of Hamiltonian simulation using product formulas \cite{childs2018toward,childs2019nearly}.

In order to have a fair comparison for the disordered Heisenberg model, we consider the ordering structure of the product formula proposed by Ref. \cite{childs2019nearly}. In particular, we focus on the types of Hamiltonians that can be written in the following form 
\begin{align}
    H=\sum_{i}^{n_{\rm site}-1} H_{i,i+1}, 
\end{align}
where the terms $H_{i,i+1}$ can be decomposed into Pauli operators which act non-trivially only on sites $i$ and $i+1$. Then the first order product formula has the following form
\begin{align}
   \nonumber e^{-iH\delta t} \approx \mathcal{\rho}_1(\delta t) &= \prod_{k=1}^{n/2-1} e^{-i\delta t H_{2k,2k+1}} \prod_{l=1}^{n/2} e^{-i\delta t H_{2k-1,2k}}\\
   &=e^{-i\delta t H_{\rm even}}e^{-i\delta t H_{\rm odd}}.
   \label{eq:ProdForm}
\end{align}
Moreover, the higher order product formulas can be constructed as the following \cite{childs2019nearly}
\begin{align}
    \mathcal{\rho}_2(\delta t)    &= e^{-i\frac{\delta t}{2} H_{\rm even}}e^{-i\delta t H_{\rm odd}} e^{-i\frac{\delta t}{2} H_{\rm even}}\\
    \mathcal{\rho}_{2k}(\delta t) &= \mathcal{\rho}_{2k-2}(p_k \delta t)^2 \mathcal{\rho}_{2k-2}((1-4p_k)\delta t ) \mathcal{\rho}_{2k-2}(p_k\delta t)^2,
    \label{eq:ProdForm2k}
\end{align}
where $p_k = 1/4-4^{\frac{1}{2k-1}}$. In the following, we will only focus on the 4$^{\rm th}$ order product formula algoritm as it results in the best conventional gate counts in Ref. \cite{childs2018toward}.
The errors induced by the $4^{\rm st}$ order product formula approximation to the evolution operator $e^{-iH_{H1D}\delta t}$ scale as $O(n(\delta t)^5)$ for small $\delta t$ \cite{childs2019nearly}. In order to simulate larger times, one conducts the simulation in $r = t/\delta t$ segments. Ref. \cite{childs2019nearly} utilized a numerical optimization algorithm which determined that the number of segments $r$ for an error threshold $\epsilon=10^{-3}$ and 4$^{\rm th}$-order product formula
\begin{align}
    r_{4} \approx 4 n_{\rm site}^{1.555}.
    \label{eq:evenodd}
\end{align}



In order to calculate the resource costs of implementing product formulas on the Rydberg atom platform, we consider the specific protocols proposed in Ref. \cite{weimer2010rydberg}. This proposal is based on an implementation of the exponential of a Pauli term in the Hamiltonian (i.e., $e^{-i \alpha_j P_j}$) using a single ancilla $a$. In particular, the scheme uses the following gate sequence
\begin{align}
    e^{-i\alpha_j P_j} = G^{\dagger} e^{i\alpha_j \sigma_{z}^{a} } G,
\end{align}
where 
\begin{align}
    G = e^{-i\pi/4 \sigma_x^{(a)}} U_{j} e^{i\pi/4 \sigma^{(a)}_x},
\end{align}
and $U_{j} = \ket{1}_{a}\bra{1}\otimes P_j + \left(\mathbf{I}-\ket{0}_{a}\bra{0}\right) \otimes \mathbf{I}$. Intuitively, the transformation $G$ maps the eigenstates of the Pauli operator $P_j$ with eigenvalues $\pm 1$ onto the $\ket{\frac{1\pm 1 }{2}}_{\rm anc}$ states of the ancilla qubit. The two eigenstates acquire phases with opposite signs using the single-qubit phase rotation $e^{i\alpha_j \sigma_{z}^{a} }$. The exponentiation requires 1 ancilla per two-qubit Pauli operator and can be implemented in depth $8+\alpha_j/\pi$. The EBGC of the exponentiation step is \begin{align}
    \frac{1}{3}\left[ 2\left(2+\frac{{\rm supp}(P_j)}{2}\right)+\frac{\alpha_j}{\pi}\right].
\end{align}
Thus, the exponentiation of each two-qubit Pauli takes  $2+\alpha_j/(3\pi)$ error-bounded gates.

The resources needed for the simulation of the 1D disordered Heisenberg model using $1^{\rm th}$ order product formula for a single segment \cite{childs2018toward} that implements a time evolution for $\delta t$ are the following 
\begin{align*}
d_{\rm PF} &= 2*3*(8+\delta t/\pi) + \delta t/\pi = 48 + 7\delta t/\pi\\
a_{\rm PF} &= n_{\rm site}/2\\
n_{\rm PF} &= n_{\rm site} \left[3\left(2+\frac{\delta t}{3\pi}\right)+ \frac{\delta t}{3\pi}\right] = n_{\rm site}\left(6+ \frac{4\delta t}{3\pi}\right),
\end{align*}
where we note the factors of 2 in the calculation of $d_{PF}$ arise from the serial application of evolution by $H_{\rm even}$ and $H_{\rm odd}$, and we assume that the local random field can be implemented using 1 single-qubit rotation without any need for ancillas. Because the ancillae are uncomputed after each step, $a_{PF}$ does not change with increasing $k$. Here, we compare the QSP-based Hamiltonian simulation to the 4$^{\rm th}$ order product formula according to Eq. ($\ref{eq:ProdForm2k}$), as it results in the lowest gate counts in Ref. \cite{childs2018toward}.
To calculate the resources for longer time evolution, the above expressions should be multiplied by $r_{4}$ in Eq.~$(\ref{eq:evenodd})$.

%


\begin{thebibliography}{83}%
\makeatletter
\providecommand \@ifxundefined [1]{%
 \@ifx{#1\undefined}
}%
\providecommand \@ifnum [1]{%
 \ifnum #1\expandafter \@firstoftwo
 \else \expandafter \@secondoftwo
 \fi
}%
\providecommand \@ifx [1]{%
 \ifx #1\expandafter \@firstoftwo
 \else \expandafter \@secondoftwo
 \fi
}%
\providecommand \natexlab [1]{#1}%
\providecommand \enquote  [1]{``#1''}%
\providecommand \bibnamefont  [1]{#1}%
\providecommand \bibfnamefont [1]{#1}%
\providecommand \citenamefont [1]{#1}%
\providecommand \href@noop [0]{\@secondoftwo}%
\providecommand \href [0]{\begingroup \@sanitize@url \@href}%
\providecommand \@href[1]{\@@startlink{#1}\@@href}%
\providecommand \@@href[1]{\endgroup#1\@@endlink}%
\providecommand \@sanitize@url [0]{\catcode `\\12\catcode `\$12\catcode
  `\&12\catcode `\#12\catcode `\^12\catcode `\_12\catcode `\%12\relax}%
\providecommand \@@startlink[1]{}%
\providecommand \@@endlink[0]{}%
\providecommand \url  [0]{\begingroup\@sanitize@url \@url }%
\providecommand \@url [1]{\endgroup\@href {#1}{\urlprefix }}%
\providecommand \urlprefix  [0]{URL }%
\providecommand \Eprint [0]{\href }%
\providecommand \doibase [0]{https://doi.org/}%
\providecommand \selectlanguage [0]{\@gobble}%
\providecommand \bibinfo  [0]{\@secondoftwo}%
\providecommand \bibfield  [0]{\@secondoftwo}%
\providecommand \translation [1]{[#1]}%
\providecommand \BibitemOpen [0]{}%
\providecommand \bibitemStop [0]{}%
\providecommand \bibitemNoStop [0]{.\EOS\space}%
\providecommand \EOS [0]{\spacefactor3000\relax}%
\providecommand \BibitemShut  [1]{\csname bibitem#1\endcsname}%
\let\auto@bib@innerbib\@empty
\bibitem [{\citenamefont {Brennen}\ \emph {et~al.}(1999)\citenamefont
  {Brennen}, \citenamefont {Caves}, \citenamefont {Jessen},\ and\ \citenamefont
  {Deutsch}}]{brennen1999quantum}%
  \BibitemOpen
  \bibfield  {author} {\bibinfo {author} {\bibfnamefont {G.~K.}\ \bibnamefont
  {Brennen}}, \bibinfo {author} {\bibfnamefont {C.~M.}\ \bibnamefont {Caves}},
  \bibinfo {author} {\bibfnamefont {P.~S.}\ \bibnamefont {Jessen}},\ and\
  \bibinfo {author} {\bibfnamefont {I.~H.}\ \bibnamefont {Deutsch}},\
  }\bibfield  {title} {\bibinfo {title} {Quantum logic gates in optical
  lattices},\ }\href@noop {} {\bibfield  {journal} {\bibinfo  {journal}
  {Physical Review Letters}\ }\textbf {\bibinfo {volume} {82}},\ \bibinfo
  {pages} {1060} (\bibinfo {year} {1999})}\BibitemShut {NoStop}%
\bibitem [{\citenamefont {Briegel}\ \emph {et~al.}(2000)\citenamefont
  {Briegel}, \citenamefont {Calarco}, \citenamefont {Jaksch}, \citenamefont
  {Cirac},\ and\ \citenamefont {Zoller}}]{briegel2000quantum}%
  \BibitemOpen
  \bibfield  {author} {\bibinfo {author} {\bibfnamefont {H.-J.}\ \bibnamefont
  {Briegel}}, \bibinfo {author} {\bibfnamefont {T.}~\bibnamefont {Calarco}},
  \bibinfo {author} {\bibfnamefont {D.}~\bibnamefont {Jaksch}}, \bibinfo
  {author} {\bibfnamefont {J.~I.}\ \bibnamefont {Cirac}},\ and\ \bibinfo
  {author} {\bibfnamefont {P.}~\bibnamefont {Zoller}},\ }\bibfield  {title}
  {\bibinfo {title} {Quantum computing with neutral atoms},\ }\href@noop {}
  {\bibfield  {journal} {\bibinfo  {journal} {Journal of modern optics}\
  }\textbf {\bibinfo {volume} {47}},\ \bibinfo {pages} {415} (\bibinfo {year}
  {2000})}\BibitemShut {NoStop}%
\bibitem [{\citenamefont {Jaksch}\ \emph {et~al.}(2000)\citenamefont {Jaksch},
  \citenamefont {Cirac}, \citenamefont {Zoller}, \citenamefont {Rolston},
  \citenamefont {C{\^o}t{\'e}},\ and\ \citenamefont {Lukin}}]{jaksch2000fast}%
  \BibitemOpen
  \bibfield  {author} {\bibinfo {author} {\bibfnamefont {D.}~\bibnamefont
  {Jaksch}}, \bibinfo {author} {\bibfnamefont {J.}~\bibnamefont {Cirac}},
  \bibinfo {author} {\bibfnamefont {P.}~\bibnamefont {Zoller}}, \bibinfo
  {author} {\bibfnamefont {S.}~\bibnamefont {Rolston}}, \bibinfo {author}
  {\bibfnamefont {R.}~\bibnamefont {C{\^o}t{\'e}}},\ and\ \bibinfo {author}
  {\bibfnamefont {M.}~\bibnamefont {Lukin}},\ }\bibfield  {title} {\bibinfo
  {title} {Fast quantum gates for neutral atoms},\ }\href@noop {} {\bibfield
  {journal} {\bibinfo  {journal} {Physical Review Letters}\ }\textbf {\bibinfo
  {volume} {85}},\ \bibinfo {pages} {2208} (\bibinfo {year}
  {2000})}\BibitemShut {NoStop}%
\bibitem [{\citenamefont {Brion}\ \emph {et~al.}(2007)\citenamefont {Brion},
  \citenamefont {M{\o}lmer},\ and\ \citenamefont {Saffman}}]{brion2007quantum}%
  \BibitemOpen
  \bibfield  {author} {\bibinfo {author} {\bibfnamefont {E.}~\bibnamefont
  {Brion}}, \bibinfo {author} {\bibfnamefont {K.}~\bibnamefont {M{\o}lmer}},\
  and\ \bibinfo {author} {\bibfnamefont {M.}~\bibnamefont {Saffman}},\
  }\bibfield  {title} {\bibinfo {title} {Quantum computing with collective
  ensembles of multilevel systems},\ }\href@noop {} {\bibfield  {journal}
  {\bibinfo  {journal} {Physical review letters}\ }\textbf {\bibinfo {volume}
  {99}},\ \bibinfo {pages} {260501} (\bibinfo {year} {2007})}\BibitemShut
  {NoStop}%
\bibitem [{\citenamefont {M{\o}lmer}\ \emph {et~al.}(2011)\citenamefont
  {M{\o}lmer}, \citenamefont {Isenhower},\ and\ \citenamefont
  {Saffman}}]{molmer2011efficient}%
  \BibitemOpen
  \bibfield  {author} {\bibinfo {author} {\bibfnamefont {K.}~\bibnamefont
  {M{\o}lmer}}, \bibinfo {author} {\bibfnamefont {L.}~\bibnamefont
  {Isenhower}},\ and\ \bibinfo {author} {\bibfnamefont {M.}~\bibnamefont
  {Saffman}},\ }\bibfield  {title} {\bibinfo {title} {Efficient grover search
  with rydberg blockade},\ }\href@noop {} {\bibfield  {journal} {\bibinfo
  {journal} {Journal of Physics B: Atomic, Molecular and Optical Physics}\
  }\textbf {\bibinfo {volume} {44}},\ \bibinfo {pages} {184016} (\bibinfo
  {year} {2011})}\BibitemShut {NoStop}%
\bibitem [{\citenamefont {Saffman}(2016)}]{saffman2016quantum}%
  \BibitemOpen
  \bibfield  {author} {\bibinfo {author} {\bibfnamefont {M.}~\bibnamefont
  {Saffman}},\ }\bibfield  {title} {\bibinfo {title} {Quantum computing with
  atomic qubits and rydberg interactions: progress and challenges},\
  }\href@noop {} {\bibfield  {journal} {\bibinfo  {journal} {Journal of Physics
  B: Atomic, Molecular and Optical Physics}\ }\textbf {\bibinfo {volume}
  {49}},\ \bibinfo {pages} {202001} (\bibinfo {year} {2016})}\BibitemShut
  {NoStop}%
\bibitem [{\citenamefont {Weiss}\ and\ \citenamefont
  {Saffman}(2017{\natexlab{a}})}]{Weiss2017}%
  \BibitemOpen
  \bibfield  {author} {\bibinfo {author} {\bibfnamefont {D.~S.}\ \bibnamefont
  {Weiss}}\ and\ \bibinfo {author} {\bibfnamefont {M.}~\bibnamefont
  {Saffman}},\ }\bibfield  {title} {\bibinfo {title} {Quantum computing with
  neutral atoms},\ }\href {https://doi.org/10.1063/PT.3.3626} {\bibfield
  {journal} {\bibinfo  {journal} {Physics Today}\ }\textbf {\bibinfo {volume}
  {70}},\ \bibinfo {pages} {44} (\bibinfo {year} {2017}{\natexlab{a}})},\
  \Eprint {https://arxiv.org/abs/https://doi.org/10.1063/PT.3.3626}
  {https://doi.org/10.1063/PT.3.3626} \BibitemShut {NoStop}%
\bibitem [{\citenamefont {Adams}\ \emph {et~al.}(2019)\citenamefont {Adams},
  \citenamefont {Pritchard},\ and\ \citenamefont {Shaffer}}]{adams2019rydberg}%
  \BibitemOpen
  \bibfield  {author} {\bibinfo {author} {\bibfnamefont {C.~S.}\ \bibnamefont
  {Adams}}, \bibinfo {author} {\bibfnamefont {J.~D.}\ \bibnamefont
  {Pritchard}},\ and\ \bibinfo {author} {\bibfnamefont {J.~P.}\ \bibnamefont
  {Shaffer}},\ }\bibfield  {title} {\bibinfo {title} {Rydberg atom quantum
  technologies},\ }\href@noop {} {\bibfield  {journal} {\bibinfo  {journal}
  {Journal of Physics B: Atomic, Molecular and Optical Physics}\ }\textbf
  {\bibinfo {volume} {53}},\ \bibinfo {pages} {012002} (\bibinfo {year}
  {2019})}\BibitemShut {NoStop}%
\bibitem [{\citenamefont {Henriet}\ \emph {et~al.}(2020)\citenamefont
  {Henriet}, \citenamefont {Beguin}, \citenamefont {Signoles}, \citenamefont
  {Lahaye}, \citenamefont {Browaeys}, \citenamefont {Reymond},\ and\
  \citenamefont {Jurczak}}]{henriet2020quantum}%
  \BibitemOpen
  \bibfield  {author} {\bibinfo {author} {\bibfnamefont {L.}~\bibnamefont
  {Henriet}}, \bibinfo {author} {\bibfnamefont {L.}~\bibnamefont {Beguin}},
  \bibinfo {author} {\bibfnamefont {A.}~\bibnamefont {Signoles}}, \bibinfo
  {author} {\bibfnamefont {T.}~\bibnamefont {Lahaye}}, \bibinfo {author}
  {\bibfnamefont {A.}~\bibnamefont {Browaeys}}, \bibinfo {author}
  {\bibfnamefont {G.-O.}\ \bibnamefont {Reymond}},\ and\ \bibinfo {author}
  {\bibfnamefont {C.}~\bibnamefont {Jurczak}},\ }\bibfield  {title} {\bibinfo
  {title} {Quantum computing with neutral atoms},\ }\href@noop {} {\bibfield
  {journal} {\bibinfo  {journal} {Quantum}\ }\textbf {\bibinfo {volume} {4}},\
  \bibinfo {pages} {327} (\bibinfo {year} {2020})}\BibitemShut {NoStop}%
\bibitem [{\citenamefont {Bernien}\ \emph {et~al.}(2017)\citenamefont
  {Bernien}, \citenamefont {Schwartz}, \citenamefont {Keesling}, \citenamefont
  {Levine}, \citenamefont {Omran}, \citenamefont {Pichler}, \citenamefont
  {Choi}, \citenamefont {Zibrov}, \citenamefont {Endres}, \citenamefont
  {Greiner} \emph {et~al.}}]{bernien2017probing}%
  \BibitemOpen
  \bibfield  {author} {\bibinfo {author} {\bibfnamefont {H.}~\bibnamefont
  {Bernien}}, \bibinfo {author} {\bibfnamefont {S.}~\bibnamefont {Schwartz}},
  \bibinfo {author} {\bibfnamefont {A.}~\bibnamefont {Keesling}}, \bibinfo
  {author} {\bibfnamefont {H.}~\bibnamefont {Levine}}, \bibinfo {author}
  {\bibfnamefont {A.}~\bibnamefont {Omran}}, \bibinfo {author} {\bibfnamefont
  {H.}~\bibnamefont {Pichler}}, \bibinfo {author} {\bibfnamefont
  {S.}~\bibnamefont {Choi}}, \bibinfo {author} {\bibfnamefont {A.~S.}\
  \bibnamefont {Zibrov}}, \bibinfo {author} {\bibfnamefont {M.}~\bibnamefont
  {Endres}}, \bibinfo {author} {\bibfnamefont {M.}~\bibnamefont {Greiner}},
  \emph {et~al.},\ }\bibfield  {title} {\bibinfo {title} {Probing many-body
  dynamics on a 51-atom quantum simulator},\ }\href@noop {} {\bibfield
  {journal} {\bibinfo  {journal} {Nature}\ }\textbf {\bibinfo {volume} {551}},\
  \bibinfo {pages} {579} (\bibinfo {year} {2017})}\BibitemShut {NoStop}%
\bibitem [{\citenamefont {Browaeys}\ and\ \citenamefont
  {Lahaye}(2020)}]{browaeys2020many}%
  \BibitemOpen
  \bibfield  {author} {\bibinfo {author} {\bibfnamefont {A.}~\bibnamefont
  {Browaeys}}\ and\ \bibinfo {author} {\bibfnamefont {T.}~\bibnamefont
  {Lahaye}},\ }\bibfield  {title} {\bibinfo {title} {Many-body physics with
  individually controlled rydberg atoms},\ }\href@noop {} {\bibfield  {journal}
  {\bibinfo  {journal} {Nature Physics}\ }\textbf {\bibinfo {volume} {16}},\
  \bibinfo {pages} {132} (\bibinfo {year} {2020})}\BibitemShut {NoStop}%
\bibitem [{\citenamefont {Omran}\ \emph {et~al.}(2019)\citenamefont {Omran},
  \citenamefont {Levine}, \citenamefont {Keesling}, \citenamefont {Semeghini},
  \citenamefont {Wang}, \citenamefont {Ebadi}, \citenamefont {Bernien},
  \citenamefont {Zibrov}, \citenamefont {Pichler}, \citenamefont {Choi} \emph
  {et~al.}}]{omran2019generation}%
  \BibitemOpen
  \bibfield  {author} {\bibinfo {author} {\bibfnamefont {A.}~\bibnamefont
  {Omran}}, \bibinfo {author} {\bibfnamefont {H.}~\bibnamefont {Levine}},
  \bibinfo {author} {\bibfnamefont {A.}~\bibnamefont {Keesling}}, \bibinfo
  {author} {\bibfnamefont {G.}~\bibnamefont {Semeghini}}, \bibinfo {author}
  {\bibfnamefont {T.~T.}\ \bibnamefont {Wang}}, \bibinfo {author}
  {\bibfnamefont {S.}~\bibnamefont {Ebadi}}, \bibinfo {author} {\bibfnamefont
  {H.}~\bibnamefont {Bernien}}, \bibinfo {author} {\bibfnamefont {A.~S.}\
  \bibnamefont {Zibrov}}, \bibinfo {author} {\bibfnamefont {H.}~\bibnamefont
  {Pichler}}, \bibinfo {author} {\bibfnamefont {S.}~\bibnamefont {Choi}}, \emph
  {et~al.},\ }\bibfield  {title} {\bibinfo {title} {Generation and manipulation
  of schr{\"o}dinger cat states in rydberg atom arrays},\ }\href@noop {}
  {\bibfield  {journal} {\bibinfo  {journal} {Science}\ }\textbf {\bibinfo
  {volume} {365}},\ \bibinfo {pages} {570} (\bibinfo {year}
  {2019})}\BibitemShut {NoStop}%
\bibitem [{\citenamefont {Samajdar}\ \emph {et~al.}(2021)\citenamefont
  {Samajdar}, \citenamefont {Ho}, \citenamefont {Pichler}, \citenamefont
  {Lukin},\ and\ \citenamefont {Sachdev}}]{Samajdar2021}%
  \BibitemOpen
  \bibfield  {author} {\bibinfo {author} {\bibfnamefont {R.}~\bibnamefont
  {Samajdar}}, \bibinfo {author} {\bibfnamefont {W.~W.}\ \bibnamefont {Ho}},
  \bibinfo {author} {\bibfnamefont {H.}~\bibnamefont {Pichler}}, \bibinfo
  {author} {\bibfnamefont {M.~D.}\ \bibnamefont {Lukin}},\ and\ \bibinfo
  {author} {\bibfnamefont {S.}~\bibnamefont {Sachdev}},\ }\bibfield  {title}
  {\bibinfo {title} {Quantum phases of rydberg atoms on a kagome lattice},\
  }\bibfield  {journal} {\bibinfo  {journal} {Proceedings of the National
  Academy of Sciences}\ }\textbf {\bibinfo {volume} {118}},\ \href
  {https://doi.org/10.1073/pnas.2015785118} {10.1073/pnas.2015785118} (\bibinfo
  {year} {2021}),\ \Eprint
  {https://arxiv.org/abs/https://www.pnas.org/content/118/4/e2015785118.full.pdf}
  {https://www.pnas.org/content/118/4/e2015785118.full.pdf} \BibitemShut
  {NoStop}%
\bibitem [{\citenamefont {Verresen}\ \emph {et~al.}(2020)\citenamefont
  {Verresen}, \citenamefont {Lukin},\ and\ \citenamefont
  {Vishwanath}}]{verresen2020prediction}%
  \BibitemOpen
  \bibfield  {author} {\bibinfo {author} {\bibfnamefont {R.}~\bibnamefont
  {Verresen}}, \bibinfo {author} {\bibfnamefont {M.~D.}\ \bibnamefont
  {Lukin}},\ and\ \bibinfo {author} {\bibfnamefont {A.}~\bibnamefont
  {Vishwanath}},\ }\bibfield  {title} {\bibinfo {title} {Prediction of toric
  code topological order from rydberg blockade},\ }\href@noop {} {\bibfield
  {journal} {\bibinfo  {journal} {arXiv preprint arXiv:2011.12310}\ } (\bibinfo
  {year} {2020})}\BibitemShut {NoStop}%
\bibitem [{\citenamefont {Semeghini}\ \emph {et~al.}(2021)\citenamefont
  {Semeghini}, \citenamefont {Levine}, \citenamefont {Keesling}, \citenamefont
  {Ebadi}, \citenamefont {Wang}, \citenamefont {Bluvstein}, \citenamefont
  {Verresen}, \citenamefont {Pichler}, \citenamefont {Kalinowski},
  \citenamefont {Samajdar} \emph {et~al.}}]{semeghini2021probing}%
  \BibitemOpen
  \bibfield  {author} {\bibinfo {author} {\bibfnamefont {G.}~\bibnamefont
  {Semeghini}}, \bibinfo {author} {\bibfnamefont {H.}~\bibnamefont {Levine}},
  \bibinfo {author} {\bibfnamefont {A.}~\bibnamefont {Keesling}}, \bibinfo
  {author} {\bibfnamefont {S.}~\bibnamefont {Ebadi}}, \bibinfo {author}
  {\bibfnamefont {T.~T.}\ \bibnamefont {Wang}}, \bibinfo {author}
  {\bibfnamefont {D.}~\bibnamefont {Bluvstein}}, \bibinfo {author}
  {\bibfnamefont {R.}~\bibnamefont {Verresen}}, \bibinfo {author}
  {\bibfnamefont {H.}~\bibnamefont {Pichler}}, \bibinfo {author} {\bibfnamefont
  {M.}~\bibnamefont {Kalinowski}}, \bibinfo {author} {\bibfnamefont
  {R.}~\bibnamefont {Samajdar}}, \emph {et~al.},\ }\bibfield  {title} {\bibinfo
  {title} {Probing topological spin liquids on a programmable quantum
  simulator},\ }\href@noop {} {\bibfield  {journal} {\bibinfo  {journal} {arXiv
  preprint arXiv:2104.04119}\ } (\bibinfo {year} {2021})}\BibitemShut {NoStop}%
\bibitem [{\citenamefont {Weiss}\ and\ \citenamefont
  {Saffman}(2017{\natexlab{b}})}]{weiss2017quantum}%
  \BibitemOpen
  \bibfield  {author} {\bibinfo {author} {\bibfnamefont {D.~S.}\ \bibnamefont
  {Weiss}}\ and\ \bibinfo {author} {\bibfnamefont {M.}~\bibnamefont
  {Saffman}},\ }\bibfield  {title} {\bibinfo {title} {Quantum computing with
  neutral atoms},\ }\href@noop {} {\bibfield  {journal} {\bibinfo  {journal}
  {Physics Today}\ }\textbf {\bibinfo {volume} {70}} (\bibinfo {year}
  {2017}{\natexlab{b}})}\BibitemShut {NoStop}%
\bibitem [{\citenamefont {Lukin}\ \emph {et~al.}(1999)\citenamefont {Lukin},
  \citenamefont {Yelin}, \citenamefont {Fleischhauer},\ and\ \citenamefont
  {Scully}}]{Lukin1998}%
  \BibitemOpen
  \bibfield  {author} {\bibinfo {author} {\bibfnamefont {M.~D.}\ \bibnamefont
  {Lukin}}, \bibinfo {author} {\bibfnamefont {S.~F.}\ \bibnamefont {Yelin}},
  \bibinfo {author} {\bibfnamefont {M.}~\bibnamefont {Fleischhauer}},\ and\
  \bibinfo {author} {\bibfnamefont {M.~O.}\ \bibnamefont {Scully}},\ }\bibfield
   {title} {\bibinfo {title} {Quantum interference effects induced by
  interacting dark resonances},\ }\href
  {https://doi.org/10.1103/PhysRevA.60.3225} {\bibfield  {journal} {\bibinfo
  {journal} {Phys. Rev. A}\ }\textbf {\bibinfo {volume} {60}},\ \bibinfo
  {pages} {3225} (\bibinfo {year} {1999})}\BibitemShut {NoStop}%
\bibitem [{\citenamefont {Lukin}\ \emph {et~al.}(2000)\citenamefont {Lukin},
  \citenamefont {Yelin},\ and\ \citenamefont {Fleischhauer}}]{Lukin2000}%
  \BibitemOpen
  \bibfield  {author} {\bibinfo {author} {\bibfnamefont {M.~D.}\ \bibnamefont
  {Lukin}}, \bibinfo {author} {\bibfnamefont {S.~F.}\ \bibnamefont {Yelin}},\
  and\ \bibinfo {author} {\bibfnamefont {M.}~\bibnamefont {Fleischhauer}},\
  }\bibfield  {title} {\bibinfo {title} {Entanglement of atomic ensembles by
  trapping correlated photon states},\ }\href
  {https://doi.org/10.1103/physrevlett.84.4232} {\bibfield  {journal} {\bibinfo
   {journal} {Physical Review Letters}\ }\textbf {\bibinfo {volume} {84}},\
  \bibinfo {pages} {4232} (\bibinfo {year} {2000})}\BibitemShut {NoStop}%
\bibitem [{\citenamefont {Bajcsy}\ \emph {et~al.}(2003)\citenamefont {Bajcsy},
  \citenamefont {Zibrov},\ and\ \citenamefont {Lukin}}]{bajcsy2003stationary}%
  \BibitemOpen
  \bibfield  {author} {\bibinfo {author} {\bibfnamefont {M.}~\bibnamefont
  {Bajcsy}}, \bibinfo {author} {\bibfnamefont {A.~S.}\ \bibnamefont {Zibrov}},\
  and\ \bibinfo {author} {\bibfnamefont {M.~D.}\ \bibnamefont {Lukin}},\
  }\bibfield  {title} {\bibinfo {title} {Stationary pulses of light in an
  atomic medium},\ }\href@noop {} {\bibfield  {journal} {\bibinfo  {journal}
  {Nature}\ }\textbf {\bibinfo {volume} {426}},\ \bibinfo {pages} {638}
  (\bibinfo {year} {2003})}\BibitemShut {NoStop}%
\bibitem [{\citenamefont {Choi}\ \emph {et~al.}(2008)\citenamefont {Choi},
  \citenamefont {Deng}, \citenamefont {Laurat},\ and\ \citenamefont
  {Kimble}}]{choi2008mapping}%
  \BibitemOpen
  \bibfield  {author} {\bibinfo {author} {\bibfnamefont {K.~S.}\ \bibnamefont
  {Choi}}, \bibinfo {author} {\bibfnamefont {H.}~\bibnamefont {Deng}}, \bibinfo
  {author} {\bibfnamefont {J.}~\bibnamefont {Laurat}},\ and\ \bibinfo {author}
  {\bibfnamefont {H.}~\bibnamefont {Kimble}},\ }\bibfield  {title} {\bibinfo
  {title} {Mapping photonic entanglement into and out of a quantum memory},\
  }\href@noop {} {\bibfield  {journal} {\bibinfo  {journal} {Nature}\ }\textbf
  {\bibinfo {volume} {452}},\ \bibinfo {pages} {67} (\bibinfo {year}
  {2008})}\BibitemShut {NoStop}%
\bibitem [{\citenamefont {Urban}\ \emph {et~al.}(2009)\citenamefont {Urban},
  \citenamefont {Johnson}, \citenamefont {Henage}, \citenamefont {Isenhower},
  \citenamefont {Yavuz}, \citenamefont {Walker},\ and\ \citenamefont
  {Saffman}}]{urban2009observation}%
  \BibitemOpen
  \bibfield  {author} {\bibinfo {author} {\bibfnamefont {E.}~\bibnamefont
  {Urban}}, \bibinfo {author} {\bibfnamefont {T.~A.}\ \bibnamefont {Johnson}},
  \bibinfo {author} {\bibfnamefont {T.}~\bibnamefont {Henage}}, \bibinfo
  {author} {\bibfnamefont {L.}~\bibnamefont {Isenhower}}, \bibinfo {author}
  {\bibfnamefont {D.}~\bibnamefont {Yavuz}}, \bibinfo {author} {\bibfnamefont
  {T.}~\bibnamefont {Walker}},\ and\ \bibinfo {author} {\bibfnamefont
  {M.}~\bibnamefont {Saffman}},\ }\bibfield  {title} {\bibinfo {title}
  {Observation of rydberg blockade between two atoms},\ }\href@noop {}
  {\bibfield  {journal} {\bibinfo  {journal} {Nature Physics}\ }\textbf
  {\bibinfo {volume} {5}},\ \bibinfo {pages} {110} (\bibinfo {year}
  {2009})}\BibitemShut {NoStop}%
\bibitem [{\citenamefont {Lukin}\ \emph {et~al.}(2001)\citenamefont {Lukin},
  \citenamefont {Fleischhauer}, \citenamefont {Cote}, \citenamefont {Duan},
  \citenamefont {Jaksch}, \citenamefont {Cirac},\ and\ \citenamefont
  {Zoller}}]{lukin2001dipole}%
  \BibitemOpen
  \bibfield  {author} {\bibinfo {author} {\bibfnamefont {M.~D.}\ \bibnamefont
  {Lukin}}, \bibinfo {author} {\bibfnamefont {M.}~\bibnamefont {Fleischhauer}},
  \bibinfo {author} {\bibfnamefont {R.}~\bibnamefont {Cote}}, \bibinfo {author}
  {\bibfnamefont {L.}~\bibnamefont {Duan}}, \bibinfo {author} {\bibfnamefont
  {D.}~\bibnamefont {Jaksch}}, \bibinfo {author} {\bibfnamefont {J.~I.}\
  \bibnamefont {Cirac}},\ and\ \bibinfo {author} {\bibfnamefont
  {P.}~\bibnamefont {Zoller}},\ }\bibfield  {title} {\bibinfo {title} {Dipole
  blockade and quantum information processing in mesoscopic atomic ensembles},\
  }\href@noop {} {\bibfield  {journal} {\bibinfo  {journal} {Physical review
  letters}\ }\textbf {\bibinfo {volume} {87}},\ \bibinfo {pages} {037901}
  (\bibinfo {year} {2001})}\BibitemShut {NoStop}%
\bibitem [{\citenamefont {Madjarov}\ \emph {et~al.}(2020)\citenamefont
  {Madjarov}, \citenamefont {Covey}, \citenamefont {Shaw}, \citenamefont
  {Choi}, \citenamefont {Kale}, \citenamefont {Cooper}, \citenamefont
  {Pichler}, \citenamefont {Schkolnik}, \citenamefont {Williams},\ and\
  \citenamefont {Endres}}]{madjarov2020high}%
  \BibitemOpen
  \bibfield  {author} {\bibinfo {author} {\bibfnamefont {I.~S.}\ \bibnamefont
  {Madjarov}}, \bibinfo {author} {\bibfnamefont {J.~P.}\ \bibnamefont {Covey}},
  \bibinfo {author} {\bibfnamefont {A.~L.}\ \bibnamefont {Shaw}}, \bibinfo
  {author} {\bibfnamefont {J.}~\bibnamefont {Choi}}, \bibinfo {author}
  {\bibfnamefont {A.}~\bibnamefont {Kale}}, \bibinfo {author} {\bibfnamefont
  {A.}~\bibnamefont {Cooper}}, \bibinfo {author} {\bibfnamefont
  {H.}~\bibnamefont {Pichler}}, \bibinfo {author} {\bibfnamefont
  {V.}~\bibnamefont {Schkolnik}}, \bibinfo {author} {\bibfnamefont {J.~R.}\
  \bibnamefont {Williams}},\ and\ \bibinfo {author} {\bibfnamefont
  {M.}~\bibnamefont {Endres}},\ }\bibfield  {title} {\bibinfo {title}
  {High-fidelity entanglement and detection of alkaline-earth rydberg atoms},\
  }\href@noop {} {\bibfield  {journal} {\bibinfo  {journal} {Nature Physics}\
  }\textbf {\bibinfo {volume} {16}},\ \bibinfo {pages} {857} (\bibinfo {year}
  {2020})}\BibitemShut {NoStop}%
\bibitem [{\citenamefont {Wilson}\ \emph {et~al.}(2022)\citenamefont {Wilson},
  \citenamefont {Saskin}, \citenamefont {Meng}, \citenamefont {Ma},
  \citenamefont {Dilip}, \citenamefont {Burgers},\ and\ \citenamefont
  {Thompson}}]{Wilson2022}%
  \BibitemOpen
  \bibfield  {author} {\bibinfo {author} {\bibfnamefont {J.~T.}\ \bibnamefont
  {Wilson}}, \bibinfo {author} {\bibfnamefont {S.}~\bibnamefont {Saskin}},
  \bibinfo {author} {\bibfnamefont {Y.}~\bibnamefont {Meng}}, \bibinfo {author}
  {\bibfnamefont {S.}~\bibnamefont {Ma}}, \bibinfo {author} {\bibfnamefont
  {R.}~\bibnamefont {Dilip}}, \bibinfo {author} {\bibfnamefont {A.~P.}\
  \bibnamefont {Burgers}},\ and\ \bibinfo {author} {\bibfnamefont {J.~D.}\
  \bibnamefont {Thompson}},\ }\bibfield  {title} {\bibinfo {title} {Trapping
  alkaline earth rydberg atoms optical tweezer arrays},\ }\href
  {https://doi.org/10.1103/PhysRevLett.128.033201} {\bibfield  {journal}
  {\bibinfo  {journal} {Phys. Rev. Lett.}\ }\textbf {\bibinfo {volume} {128}},\
  \bibinfo {pages} {033201} (\bibinfo {year} {2022})}\BibitemShut {NoStop}%
\bibitem [{\citenamefont {Ma}\ \emph {et~al.}(2022)\citenamefont {Ma},
  \citenamefont {Burgers}, \citenamefont {Liu}, \citenamefont {Wilson},
  \citenamefont {Zhang},\ and\ \citenamefont {Thompson}}]{ma2022universal}%
  \BibitemOpen
  \bibfield  {author} {\bibinfo {author} {\bibfnamefont {S.}~\bibnamefont
  {Ma}}, \bibinfo {author} {\bibfnamefont {A.~P.}\ \bibnamefont {Burgers}},
  \bibinfo {author} {\bibfnamefont {G.}~\bibnamefont {Liu}}, \bibinfo {author}
  {\bibfnamefont {J.}~\bibnamefont {Wilson}}, \bibinfo {author} {\bibfnamefont
  {B.}~\bibnamefont {Zhang}},\ and\ \bibinfo {author} {\bibfnamefont {J.~D.}\
  \bibnamefont {Thompson}},\ }\bibfield  {title} {\bibinfo {title} {Universal
  gate operations on nuclear spin qubits in an optical tweezer array of yb 171
  atoms},\ }\href@noop {} {\bibfield  {journal} {\bibinfo  {journal} {Physical
  Review X}\ }\textbf {\bibinfo {volume} {12}},\ \bibinfo {pages} {021028}
  (\bibinfo {year} {2022})}\BibitemShut {NoStop}%
\bibitem [{\citenamefont {H{\"a}ffner}\ \emph {et~al.}(2008)\citenamefont
  {H{\"a}ffner}, \citenamefont {Roos},\ and\ \citenamefont
  {Blatt}}]{haffner2008quantum}%
  \BibitemOpen
  \bibfield  {author} {\bibinfo {author} {\bibfnamefont {H.}~\bibnamefont
  {H{\"a}ffner}}, \bibinfo {author} {\bibfnamefont {C.~F.}\ \bibnamefont
  {Roos}},\ and\ \bibinfo {author} {\bibfnamefont {R.}~\bibnamefont {Blatt}},\
  }\bibfield  {title} {\bibinfo {title} {Quantum computing with trapped ions},\
  }\href@noop {} {\bibfield  {journal} {\bibinfo  {journal} {Physics reports}\
  }\textbf {\bibinfo {volume} {469}},\ \bibinfo {pages} {155} (\bibinfo {year}
  {2008})}\BibitemShut {NoStop}%
\bibitem [{\citenamefont {Home}\ \emph {et~al.}(2009)\citenamefont {Home},
  \citenamefont {Hanneke}, \citenamefont {Jost}, \citenamefont {Amini},
  \citenamefont {Leibfried},\ and\ \citenamefont
  {Wineland}}]{home2009complete}%
  \BibitemOpen
  \bibfield  {author} {\bibinfo {author} {\bibfnamefont {J.~P.}\ \bibnamefont
  {Home}}, \bibinfo {author} {\bibfnamefont {D.}~\bibnamefont {Hanneke}},
  \bibinfo {author} {\bibfnamefont {J.~D.}\ \bibnamefont {Jost}}, \bibinfo
  {author} {\bibfnamefont {J.~M.}\ \bibnamefont {Amini}}, \bibinfo {author}
  {\bibfnamefont {D.}~\bibnamefont {Leibfried}},\ and\ \bibinfo {author}
  {\bibfnamefont {D.~J.}\ \bibnamefont {Wineland}},\ }\bibfield  {title}
  {\bibinfo {title} {Complete methods set for scalable ion trap quantum
  information processing},\ }\href@noop {} {\bibfield  {journal} {\bibinfo
  {journal} {Science}\ }\textbf {\bibinfo {volume} {325}},\ \bibinfo {pages}
  {1227} (\bibinfo {year} {2009})}\BibitemShut {NoStop}%
\bibitem [{\citenamefont {Wallraff}\ \emph {et~al.}(2004)\citenamefont
  {Wallraff}, \citenamefont {Schuster}, \citenamefont {Blais}, \citenamefont
  {Frunzio}, \citenamefont {Huang}, \citenamefont {Majer}, \citenamefont
  {Kumar}, \citenamefont {Girvin},\ and\ \citenamefont
  {Schoelkopf}}]{wallraff2004strong}%
  \BibitemOpen
  \bibfield  {author} {\bibinfo {author} {\bibfnamefont {A.}~\bibnamefont
  {Wallraff}}, \bibinfo {author} {\bibfnamefont {D.~I.}\ \bibnamefont
  {Schuster}}, \bibinfo {author} {\bibfnamefont {A.}~\bibnamefont {Blais}},
  \bibinfo {author} {\bibfnamefont {L.}~\bibnamefont {Frunzio}}, \bibinfo
  {author} {\bibfnamefont {R.-S.}\ \bibnamefont {Huang}}, \bibinfo {author}
  {\bibfnamefont {J.}~\bibnamefont {Majer}}, \bibinfo {author} {\bibfnamefont
  {S.}~\bibnamefont {Kumar}}, \bibinfo {author} {\bibfnamefont {S.~M.}\
  \bibnamefont {Girvin}},\ and\ \bibinfo {author} {\bibfnamefont {R.~J.}\
  \bibnamefont {Schoelkopf}},\ }\bibfield  {title} {\bibinfo {title} {Strong
  coupling of a single photon to a superconducting qubit using circuit quantum
  electrodynamics},\ }\href@noop {} {\bibfield  {journal} {\bibinfo  {journal}
  {Nature}\ }\textbf {\bibinfo {volume} {431}},\ \bibinfo {pages} {162}
  (\bibinfo {year} {2004})}\BibitemShut {NoStop}%
\bibitem [{\citenamefont {Blais}\ \emph {et~al.}(2021)\citenamefont {Blais},
  \citenamefont {Grimsmo}, \citenamefont {Girvin},\ and\ \citenamefont
  {Wallraff}}]{blais2021circuit}%
  \BibitemOpen
  \bibfield  {author} {\bibinfo {author} {\bibfnamefont {A.}~\bibnamefont
  {Blais}}, \bibinfo {author} {\bibfnamefont {A.~L.}\ \bibnamefont {Grimsmo}},
  \bibinfo {author} {\bibfnamefont {S.~M.}\ \bibnamefont {Girvin}},\ and\
  \bibinfo {author} {\bibfnamefont {A.}~\bibnamefont {Wallraff}},\ }\bibfield
  {title} {\bibinfo {title} {Circuit quantum electrodynamics},\ }\href@noop {}
  {\bibfield  {journal} {\bibinfo  {journal} {Reviews of Modern Physics}\
  }\textbf {\bibinfo {volume} {93}},\ \bibinfo {pages} {025005} (\bibinfo
  {year} {2021})}\BibitemShut {NoStop}%
\bibitem [{\citenamefont {Bluvstein}\ \emph {et~al.}(2021)\citenamefont
  {Bluvstein}, \citenamefont {Levine}, \citenamefont {Semeghini}, \citenamefont
  {Wang}, \citenamefont {Ebadi}, \citenamefont {Kalinowski}, \citenamefont
  {Keesling}, \citenamefont {Maskara}, \citenamefont {Pichler}, \citenamefont
  {Greiner} \emph {et~al.}}]{bluvstein2021quantum}%
  \BibitemOpen
  \bibfield  {author} {\bibinfo {author} {\bibfnamefont {D.}~\bibnamefont
  {Bluvstein}}, \bibinfo {author} {\bibfnamefont {H.}~\bibnamefont {Levine}},
  \bibinfo {author} {\bibfnamefont {G.}~\bibnamefont {Semeghini}}, \bibinfo
  {author} {\bibfnamefont {T.~T.}\ \bibnamefont {Wang}}, \bibinfo {author}
  {\bibfnamefont {S.}~\bibnamefont {Ebadi}}, \bibinfo {author} {\bibfnamefont
  {M.}~\bibnamefont {Kalinowski}}, \bibinfo {author} {\bibfnamefont
  {A.}~\bibnamefont {Keesling}}, \bibinfo {author} {\bibfnamefont
  {N.}~\bibnamefont {Maskara}}, \bibinfo {author} {\bibfnamefont
  {H.}~\bibnamefont {Pichler}}, \bibinfo {author} {\bibfnamefont
  {M.}~\bibnamefont {Greiner}}, \emph {et~al.},\ }\bibfield  {title} {\bibinfo
  {title} {A quantum processor based on coherent transport of entangled atom
  arrays},\ }\href@noop {} {\bibfield  {journal} {\bibinfo  {journal} {arXiv
  preprint arXiv:2112.03923}\ } (\bibinfo {year} {2021})}\BibitemShut {NoStop}%
\bibitem [{\citenamefont {Weimer}\ \emph {et~al.}(2010)\citenamefont {Weimer},
  \citenamefont {M{\"u}ller}, \citenamefont {Lesanovsky}, \citenamefont
  {Zoller},\ and\ \citenamefont {B{\"u}chler}}]{weimer2010rydberg}%
  \BibitemOpen
  \bibfield  {author} {\bibinfo {author} {\bibfnamefont {H.}~\bibnamefont
  {Weimer}}, \bibinfo {author} {\bibfnamefont {M.}~\bibnamefont {M{\"u}ller}},
  \bibinfo {author} {\bibfnamefont {I.}~\bibnamefont {Lesanovsky}}, \bibinfo
  {author} {\bibfnamefont {P.}~\bibnamefont {Zoller}},\ and\ \bibinfo {author}
  {\bibfnamefont {H.~P.}\ \bibnamefont {B{\"u}chler}},\ }\bibfield  {title}
  {\bibinfo {title} {A rydberg quantum simulator},\ }\href@noop {} {\bibfield
  {journal} {\bibinfo  {journal} {Nature Physics}\ }\textbf {\bibinfo {volume}
  {6}},\ \bibinfo {pages} {382} (\bibinfo {year} {2010})}\BibitemShut {NoStop}%
\bibitem [{\citenamefont {Wilson}\ \emph {et~al.}(2019)\citenamefont {Wilson},
  \citenamefont {Saskin}, \citenamefont {Meng}, \citenamefont {Ma},
  \citenamefont {Dilip}, \citenamefont {Burgers},\ and\ \citenamefont
  {Thompson}}]{wilson2019trapped}%
  \BibitemOpen
  \bibfield  {author} {\bibinfo {author} {\bibfnamefont {J.}~\bibnamefont
  {Wilson}}, \bibinfo {author} {\bibfnamefont {S.}~\bibnamefont {Saskin}},
  \bibinfo {author} {\bibfnamefont {Y.}~\bibnamefont {Meng}}, \bibinfo {author}
  {\bibfnamefont {S.}~\bibnamefont {Ma}}, \bibinfo {author} {\bibfnamefont
  {R.}~\bibnamefont {Dilip}}, \bibinfo {author} {\bibfnamefont
  {A.}~\bibnamefont {Burgers}},\ and\ \bibinfo {author} {\bibfnamefont
  {J.}~\bibnamefont {Thompson}},\ }\bibfield  {title} {\bibinfo {title}
  {Trapped arrays of alkaline earth rydberg atoms in optical tweezers},\
  }\href@noop {} {\bibfield  {journal} {\bibinfo  {journal} {arXiv preprint
  arXiv:1912.08754}\ } (\bibinfo {year} {2019})}\BibitemShut {NoStop}%
\bibitem [{\citenamefont {Barnes}\ \emph {et~al.}(2022)\citenamefont {Barnes},
  \citenamefont {Battaglino}, \citenamefont {Bloom}, \citenamefont {Cassella},
  \citenamefont {Coxe}, \citenamefont {Crisosto}, \citenamefont {King},
  \citenamefont {Kondov}, \citenamefont {Kotru}, \citenamefont {Larsen} \emph
  {et~al.}}]{barnes2022assembly}%
  \BibitemOpen
  \bibfield  {author} {\bibinfo {author} {\bibfnamefont {K.}~\bibnamefont
  {Barnes}}, \bibinfo {author} {\bibfnamefont {P.}~\bibnamefont {Battaglino}},
  \bibinfo {author} {\bibfnamefont {B.~J.}\ \bibnamefont {Bloom}}, \bibinfo
  {author} {\bibfnamefont {K.}~\bibnamefont {Cassella}}, \bibinfo {author}
  {\bibfnamefont {R.}~\bibnamefont {Coxe}}, \bibinfo {author} {\bibfnamefont
  {N.}~\bibnamefont {Crisosto}}, \bibinfo {author} {\bibfnamefont {J.~P.}\
  \bibnamefont {King}}, \bibinfo {author} {\bibfnamefont {S.~S.}\ \bibnamefont
  {Kondov}}, \bibinfo {author} {\bibfnamefont {K.}~\bibnamefont {Kotru}},
  \bibinfo {author} {\bibfnamefont {S.~C.}\ \bibnamefont {Larsen}}, \emph
  {et~al.},\ }\bibfield  {title} {\bibinfo {title} {Assembly and coherent
  control of a register of nuclear spin qubits},\ }\href@noop {} {\bibfield
  {journal} {\bibinfo  {journal} {Nature Communications}\ }\textbf {\bibinfo
  {volume} {13}},\ \bibinfo {pages} {1} (\bibinfo {year} {2022})}\BibitemShut
  {NoStop}%
\bibitem [{\citenamefont {Cong}\ \emph {et~al.}(2021)\citenamefont {Cong},
  \citenamefont {Wang}, \citenamefont {Levine}, \citenamefont {Keesling},\ and\
  \citenamefont {Lukin}}]{cong2021hardware}%
  \BibitemOpen
  \bibfield  {author} {\bibinfo {author} {\bibfnamefont {I.}~\bibnamefont
  {Cong}}, \bibinfo {author} {\bibfnamefont {S.-T.}\ \bibnamefont {Wang}},
  \bibinfo {author} {\bibfnamefont {H.}~\bibnamefont {Levine}}, \bibinfo
  {author} {\bibfnamefont {A.}~\bibnamefont {Keesling}},\ and\ \bibinfo
  {author} {\bibfnamefont {M.~D.}\ \bibnamefont {Lukin}},\ }\bibfield  {title}
  {\bibinfo {title} {Hardware-efficient, fault-tolerant quantum computation
  with rydberg atoms},\ }\href@noop {} {\bibfield  {journal} {\bibinfo
  {journal} {arXiv preprint arXiv:2105.13501}\ } (\bibinfo {year}
  {2021})}\BibitemShut {NoStop}%
\bibitem [{\citenamefont {Zhou}\ \emph {et~al.}(2020)\citenamefont {Zhou},
  \citenamefont {Stoudenmire},\ and\ \citenamefont {Waintal}}]{Zhou2020}%
  \BibitemOpen
  \bibfield  {author} {\bibinfo {author} {\bibfnamefont {Y.}~\bibnamefont
  {Zhou}}, \bibinfo {author} {\bibfnamefont {E.~M.}\ \bibnamefont
  {Stoudenmire}},\ and\ \bibinfo {author} {\bibfnamefont {X.}~\bibnamefont
  {Waintal}},\ }\bibfield  {title} {\bibinfo {title} {What limits the
  simulation of quantum computers?},\ }\href
  {https://doi.org/10.1103/PhysRevX.10.041038} {\bibfield  {journal} {\bibinfo
  {journal} {Phys. Rev. X}\ }\textbf {\bibinfo {volume} {10}},\ \bibinfo
  {pages} {041038} (\bibinfo {year} {2020})}\BibitemShut {NoStop}%
\bibitem [{\citenamefont {Oh}\ \emph {et~al.}(2021)\citenamefont {Oh},
  \citenamefont {Noh}, \citenamefont {Fefferman},\ and\ \citenamefont
  {Jiang}}]{Oh2021}%
  \BibitemOpen
  \bibfield  {author} {\bibinfo {author} {\bibfnamefont {C.}~\bibnamefont
  {Oh}}, \bibinfo {author} {\bibfnamefont {K.}~\bibnamefont {Noh}}, \bibinfo
  {author} {\bibfnamefont {B.}~\bibnamefont {Fefferman}},\ and\ \bibinfo
  {author} {\bibfnamefont {L.}~\bibnamefont {Jiang}},\ }\bibfield  {title}
  {\bibinfo {title} {Classical simulation of lossy boson sampling using matrix
  product operators},\ }\href {https://doi.org/10.1103/PhysRevA.104.022407}
  {\bibfield  {journal} {\bibinfo  {journal} {Phys. Rev. A}\ }\textbf {\bibinfo
  {volume} {104}},\ \bibinfo {pages} {022407} (\bibinfo {year}
  {2021})}\BibitemShut {NoStop}%
\bibitem [{\citenamefont {Pan}\ \emph {et~al.}(2022)\citenamefont {Pan},
  \citenamefont {Chen},\ and\ \citenamefont {Zhang}}]{pan2022solving}%
  \BibitemOpen
  \bibfield  {author} {\bibinfo {author} {\bibfnamefont {F.}~\bibnamefont
  {Pan}}, \bibinfo {author} {\bibfnamefont {K.}~\bibnamefont {Chen}},\ and\
  \bibinfo {author} {\bibfnamefont {P.}~\bibnamefont {Zhang}},\ }\bibfield
  {title} {\bibinfo {title} {Solving the sampling problem of the sycamore
  quantum circuits},\ }\href@noop {} {\bibfield  {journal} {\bibinfo  {journal}
  {Physical Review Letters}\ }\textbf {\bibinfo {volume} {129}},\ \bibinfo
  {pages} {090502} (\bibinfo {year} {2022})}\BibitemShut {NoStop}%
\bibitem [{\citenamefont {Aharonov}\ and\ \citenamefont
  {Ben-Or}(1997)}]{aharonov1997fault}%
  \BibitemOpen
  \bibfield  {author} {\bibinfo {author} {\bibfnamefont {D.}~\bibnamefont
  {Aharonov}}\ and\ \bibinfo {author} {\bibfnamefont {M.}~\bibnamefont
  {Ben-Or}},\ }\bibfield  {title} {\bibinfo {title} {Fault-tolerant quantum
  computation with constant error},\ }in\ \href@noop {} {\emph {\bibinfo
  {booktitle} {Proceedings of the twenty-ninth annual ACM symposium on Theory
  of computing}}}\ (\bibinfo {year} {1997})\ pp.\ \bibinfo {pages}
  {176--188}\BibitemShut {NoStop}%
\bibitem [{\citenamefont {Preskill}(1998)}]{preskill1998reliable}%
  \BibitemOpen
  \bibfield  {author} {\bibinfo {author} {\bibfnamefont {J.}~\bibnamefont
  {Preskill}},\ }\bibfield  {title} {\bibinfo {title} {Reliable quantum
  computers},\ }\href@noop {} {\bibfield  {journal} {\bibinfo  {journal}
  {Proceedings of the Royal Society of London. Series A: Mathematical, Physical
  and Engineering Sciences}\ }\textbf {\bibinfo {volume} {454}},\ \bibinfo
  {pages} {385} (\bibinfo {year} {1998})}\BibitemShut {NoStop}%
\bibitem [{\citenamefont {Knill}(2005)}]{knill2005quantum}%
  \BibitemOpen
  \bibfield  {author} {\bibinfo {author} {\bibfnamefont {E.}~\bibnamefont
  {Knill}},\ }\bibfield  {title} {\bibinfo {title} {Quantum computing with
  realistically noisy devices},\ }\href@noop {} {\bibfield  {journal} {\bibinfo
   {journal} {Nature}\ }\textbf {\bibinfo {volume} {434}},\ \bibinfo {pages}
  {39} (\bibinfo {year} {2005})}\BibitemShut {NoStop}%
\bibitem [{\citenamefont {Nielsen}\ and\ \citenamefont
  {Chuang}(2002)}]{nielsen2002quantum}%
  \BibitemOpen
  \bibfield  {author} {\bibinfo {author} {\bibfnamefont {M.~A.}\ \bibnamefont
  {Nielsen}}\ and\ \bibinfo {author} {\bibfnamefont {I.}~\bibnamefont
  {Chuang}},\ }\href@noop {} {\emph {\bibinfo {title} {Quantum computation and
  quantum information}}}\ (\bibinfo  {publisher} {American Association of
  Physics Teachers},\ \bibinfo {year} {2002})\BibitemShut {NoStop}%
\bibitem [{\citenamefont {Low}(2017)}]{low2017quantum}%
  \BibitemOpen
  \bibfield  {author} {\bibinfo {author} {\bibfnamefont {G.~H.}\ \bibnamefont
  {Low}},\ }\emph {\bibinfo {title} {Quantum signal processing by single-qubit
  dynamics}},\ \href@noop {} {Ph.D. thesis},\ \bibinfo  {school} {Massachusetts
  Institute of Technology} (\bibinfo {year} {2017})\BibitemShut {NoStop}%
\bibitem [{\citenamefont {Gily{\'e}n}\ \emph {et~al.}(2019)\citenamefont
  {Gily{\'e}n}, \citenamefont {Su}, \citenamefont {Low},\ and\ \citenamefont
  {Wiebe}}]{gilyen2019quantum}%
  \BibitemOpen
  \bibfield  {author} {\bibinfo {author} {\bibfnamefont {A.}~\bibnamefont
  {Gily{\'e}n}}, \bibinfo {author} {\bibfnamefont {Y.}~\bibnamefont {Su}},
  \bibinfo {author} {\bibfnamefont {G.~H.}\ \bibnamefont {Low}},\ and\ \bibinfo
  {author} {\bibfnamefont {N.}~\bibnamefont {Wiebe}},\ }\bibfield  {title}
  {\bibinfo {title} {Quantum singular value transformation and beyond:
  exponential improvements for quantum matrix arithmetics},\ }in\ \href@noop {}
  {\emph {\bibinfo {booktitle} {Proceedings of the 51st Annual ACM SIGACT
  Symposium on Theory of Computing}}}\ (\bibinfo {year} {2019})\ pp.\ \bibinfo
  {pages} {193--204}\BibitemShut {NoStop}%
\bibitem [{\citenamefont {Martyn}\ \emph {et~al.}(2021)\citenamefont {Martyn},
  \citenamefont {Rossi}, \citenamefont {Tan},\ and\ \citenamefont
  {Chuang}}]{MartynChuang2021}%
  \BibitemOpen
  \bibfield  {author} {\bibinfo {author} {\bibfnamefont {J.~M.}\ \bibnamefont
  {Martyn}}, \bibinfo {author} {\bibfnamefont {Z.~M.}\ \bibnamefont {Rossi}},
  \bibinfo {author} {\bibfnamefont {A.~K.}\ \bibnamefont {Tan}},\ and\ \bibinfo
  {author} {\bibfnamefont {I.~L.}\ \bibnamefont {Chuang}},\ }\bibfield  {title}
  {\bibinfo {title} {Grand unification of quantum algorithms},\ }\href
  {https://doi.org/10.1103/PRXQuantum.2.040203} {\bibfield  {journal} {\bibinfo
   {journal} {PRX Quantum}\ }\textbf {\bibinfo {volume} {2}},\ \bibinfo {pages}
  {040203} (\bibinfo {year} {2021})}\BibitemShut {NoStop}%
\bibitem [{\citenamefont {Chakraborty}\ \emph {et~al.}(2019)\citenamefont
  {Chakraborty}, \citenamefont {Gily{\'e}n},\ and\ \citenamefont
  {Jeffery}}]{Chakraborty2019}%
  \BibitemOpen
  \bibfield  {author} {\bibinfo {author} {\bibfnamefont {S.}~\bibnamefont
  {Chakraborty}}, \bibinfo {author} {\bibfnamefont {A.}~\bibnamefont
  {Gily{\'e}n}},\ and\ \bibinfo {author} {\bibfnamefont {S.}~\bibnamefont
  {Jeffery}},\ }\bibfield  {title} {\bibinfo {title} {{The Power of
  Block-Encoded Matrix Powers: Improved Regression Techniques via Faster
  Hamiltonian Simulation}},\ }in\ \href
  {https://doi.org/10.4230/LIPIcs.ICALP.2019.33} {\emph {\bibinfo {booktitle}
  {46th International Colloquium on Automata, Languages, and Programming (ICALP
  2019)}}},\ \bibinfo {series} {Leibniz International Proceedings in
  Informatics (LIPIcs)}, Vol.\ \bibinfo {volume} {132},\ \bibinfo {editor}
  {edited by\ \bibinfo {editor} {\bibfnamefont {C.}~\bibnamefont {Baier}},
  \bibinfo {editor} {\bibfnamefont {I.}~\bibnamefont {Chatzigiannakis}},
  \bibinfo {editor} {\bibfnamefont {P.}~\bibnamefont {Flocchini}},\ and\
  \bibinfo {editor} {\bibfnamefont {S.}~\bibnamefont {Leonardi}}}\ (\bibinfo
  {publisher} {Schloss Dagstuhl--Leibniz-Zentrum fuer Informatik},\ \bibinfo
  {address} {Dagstuhl, Germany},\ \bibinfo {year} {2019})\ pp.\ \bibinfo
  {pages} {33:1--33:14}\BibitemShut {NoStop}%
\bibitem [{\citenamefont {Levine}\ \emph {et~al.}(2019)\citenamefont {Levine},
  \citenamefont {Keesling}, \citenamefont {Semeghini}, \citenamefont {Omran},
  \citenamefont {Wang}, \citenamefont {Ebadi}, \citenamefont {Bernien},
  \citenamefont {Greiner}, \citenamefont {Vuleti\ifmmode~\acute{c}\else
  \'{c}\fi{}}, \citenamefont {Pichler},\ and\ \citenamefont
  {Lukin}}]{Levine2019}%
  \BibitemOpen
  \bibfield  {author} {\bibinfo {author} {\bibfnamefont {H.}~\bibnamefont
  {Levine}}, \bibinfo {author} {\bibfnamefont {A.}~\bibnamefont {Keesling}},
  \bibinfo {author} {\bibfnamefont {G.}~\bibnamefont {Semeghini}}, \bibinfo
  {author} {\bibfnamefont {A.}~\bibnamefont {Omran}}, \bibinfo {author}
  {\bibfnamefont {T.~T.}\ \bibnamefont {Wang}}, \bibinfo {author}
  {\bibfnamefont {S.}~\bibnamefont {Ebadi}}, \bibinfo {author} {\bibfnamefont
  {H.}~\bibnamefont {Bernien}}, \bibinfo {author} {\bibfnamefont
  {M.}~\bibnamefont {Greiner}}, \bibinfo {author} {\bibfnamefont
  {V.}~\bibnamefont {Vuleti\ifmmode~\acute{c}\else \'{c}\fi{}}}, \bibinfo
  {author} {\bibfnamefont {H.}~\bibnamefont {Pichler}},\ and\ \bibinfo {author}
  {\bibfnamefont {M.~D.}\ \bibnamefont {Lukin}},\ }\bibfield  {title} {\bibinfo
  {title} {Parallel implementation of high-fidelity multiqubit gates with
  neutral atoms},\ }\href {https://doi.org/10.1103/PhysRevLett.123.170503}
  {\bibfield  {journal} {\bibinfo  {journal} {Phys. Rev. Lett.}\ }\textbf
  {\bibinfo {volume} {123}},\ \bibinfo {pages} {170503} (\bibinfo {year}
  {2019})}\BibitemShut {NoStop}%
\bibitem [{\citenamefont {Childs}\ and\ \citenamefont
  {Wiebe}(2012)}]{childs2012hamiltonian}%
  \BibitemOpen
  \bibfield  {author} {\bibinfo {author} {\bibfnamefont {A.~M.}\ \bibnamefont
  {Childs}}\ and\ \bibinfo {author} {\bibfnamefont {N.}~\bibnamefont {Wiebe}},\
  }\bibfield  {title} {\bibinfo {title} {Hamiltonian simulation using linear
  combinations of unitary operations},\ }\href@noop {} {\bibfield  {journal}
  {\bibinfo  {journal} {arXiv preprint arXiv:1202.5822}\ } (\bibinfo {year}
  {2012})}\BibitemShut {NoStop}%
\bibitem [{\citenamefont {Childs}\ \emph {et~al.}(2018)\citenamefont {Childs},
  \citenamefont {Maslov}, \citenamefont {Nam}, \citenamefont {Ross},\ and\
  \citenamefont {Su}}]{childs2018toward}%
  \BibitemOpen
  \bibfield  {author} {\bibinfo {author} {\bibfnamefont {A.~M.}\ \bibnamefont
  {Childs}}, \bibinfo {author} {\bibfnamefont {D.}~\bibnamefont {Maslov}},
  \bibinfo {author} {\bibfnamefont {Y.}~\bibnamefont {Nam}}, \bibinfo {author}
  {\bibfnamefont {N.~J.}\ \bibnamefont {Ross}},\ and\ \bibinfo {author}
  {\bibfnamefont {Y.}~\bibnamefont {Su}},\ }\bibfield  {title} {\bibinfo
  {title} {Toward the first quantum simulation with quantum speedup},\
  }\href@noop {} {\bibfield  {journal} {\bibinfo  {journal} {Proceedings of the
  National Academy of Sciences}\ }\textbf {\bibinfo {volume} {115}},\ \bibinfo
  {pages} {9456} (\bibinfo {year} {2018})}\BibitemShut {NoStop}%
\bibitem [{\citenamefont {Lloyd}(1996)}]{lloyd1996universal}%
  \BibitemOpen
  \bibfield  {author} {\bibinfo {author} {\bibfnamefont {S.}~\bibnamefont
  {Lloyd}},\ }\bibfield  {title} {\bibinfo {title} {Universal quantum
  simulators},\ }\href@noop {} {\bibfield  {journal} {\bibinfo  {journal}
  {Science}\ ,\ \bibinfo {pages} {1073}} (\bibinfo {year} {1996})}\BibitemShut
  {NoStop}%
\bibitem [{\citenamefont {Harrow}\ \emph {et~al.}(2009)\citenamefont {Harrow},
  \citenamefont {Hassidim},\ and\ \citenamefont {Lloyd}}]{harrow2009quantum}%
  \BibitemOpen
  \bibfield  {author} {\bibinfo {author} {\bibfnamefont {A.~W.}\ \bibnamefont
  {Harrow}}, \bibinfo {author} {\bibfnamefont {A.}~\bibnamefont {Hassidim}},\
  and\ \bibinfo {author} {\bibfnamefont {S.}~\bibnamefont {Lloyd}},\ }\bibfield
   {title} {\bibinfo {title} {Quantum algorithm for linear systems of
  equations},\ }\href@noop {} {\bibfield  {journal} {\bibinfo  {journal}
  {Physical review letters}\ }\textbf {\bibinfo {volume} {103}},\ \bibinfo
  {pages} {150502} (\bibinfo {year} {2009})}\BibitemShut {NoStop}%
\bibitem [{Note1()}]{Note1}%
  \BibitemOpen
  \bibinfo {note} {See Ref. \cite {gilyen2019quantum,MartynChuang2021} for
  further examples}\BibitemShut {NoStop}%
\bibitem [{Note2()}]{Note2}%
  \BibitemOpen
  \bibinfo {note} {In contrast, compiling time-dependent Hamiltonian simulation
  is difficult because then multiple functional transformations $f_t(\cdot )$
  and their inputs $A_t$ need to be explicitly specified.}\BibitemShut {Stop}%
\bibitem [{\citenamefont {M{\"u}ller}\ \emph {et~al.}(2009)\citenamefont
  {M{\"u}ller}, \citenamefont {Lesanovsky}, \citenamefont {Weimer},
  \citenamefont {B{\"u}chler},\ and\ \citenamefont
  {Zoller}}]{muller2009mesoscopic}%
  \BibitemOpen
  \bibfield  {author} {\bibinfo {author} {\bibfnamefont {M.}~\bibnamefont
  {M{\"u}ller}}, \bibinfo {author} {\bibfnamefont {I.}~\bibnamefont
  {Lesanovsky}}, \bibinfo {author} {\bibfnamefont {H.}~\bibnamefont {Weimer}},
  \bibinfo {author} {\bibfnamefont {H.}~\bibnamefont {B{\"u}chler}},\ and\
  \bibinfo {author} {\bibfnamefont {P.}~\bibnamefont {Zoller}},\ }\bibfield
  {title} {\bibinfo {title} {Mesoscopic rydberg gate based on
  electromagnetically induced transparency},\ }\href@noop {} {\bibfield
  {journal} {\bibinfo  {journal} {Physical Review Letters}\ }\textbf {\bibinfo
  {volume} {102}},\ \bibinfo {pages} {170502} (\bibinfo {year}
  {2009})}\BibitemShut {NoStop}%
\bibitem [{\citenamefont {Boller}\ \emph {et~al.}(1991)\citenamefont {Boller},
  \citenamefont {Imamo{\u{g}}lu},\ and\ \citenamefont
  {Harris}}]{boller1991observation}%
  \BibitemOpen
  \bibfield  {author} {\bibinfo {author} {\bibfnamefont {K.-J.}\ \bibnamefont
  {Boller}}, \bibinfo {author} {\bibfnamefont {A.}~\bibnamefont
  {Imamo{\u{g}}lu}},\ and\ \bibinfo {author} {\bibfnamefont {S.~E.}\
  \bibnamefont {Harris}},\ }\bibfield  {title} {\bibinfo {title} {Observation
  of electromagnetically induced transparency},\ }\href@noop {} {\bibfield
  {journal} {\bibinfo  {journal} {Physical Review Letters}\ }\textbf {\bibinfo
  {volume} {66}},\ \bibinfo {pages} {2593} (\bibinfo {year}
  {1991})}\BibitemShut {NoStop}%
\bibitem [{\citenamefont {Lukin}\ and\ \citenamefont
  {Imamo{\u{g}}lu}(2000)}]{lukin2000nonlinear}%
  \BibitemOpen
  \bibfield  {author} {\bibinfo {author} {\bibfnamefont {M.}~\bibnamefont
  {Lukin}}\ and\ \bibinfo {author} {\bibfnamefont {A.}~\bibnamefont
  {Imamo{\u{g}}lu}},\ }\bibfield  {title} {\bibinfo {title} {Nonlinear optics
  and quantum entanglement of ultraslow single photons},\ }\href@noop {}
  {\bibfield  {journal} {\bibinfo  {journal} {Physical Review Letters}\
  }\textbf {\bibinfo {volume} {84}},\ \bibinfo {pages} {1419} (\bibinfo {year}
  {2000})}\BibitemShut {NoStop}%
\bibitem [{\citenamefont {Rebentrost}\ \emph {et~al.}(2014)\citenamefont
  {Rebentrost}, \citenamefont {Mohseni},\ and\ \citenamefont
  {Lloyd}}]{Rebentrost2014}%
  \BibitemOpen
  \bibfield  {author} {\bibinfo {author} {\bibfnamefont {P.}~\bibnamefont
  {Rebentrost}}, \bibinfo {author} {\bibfnamefont {M.}~\bibnamefont
  {Mohseni}},\ and\ \bibinfo {author} {\bibfnamefont {S.}~\bibnamefont
  {Lloyd}},\ }\bibfield  {title} {\bibinfo {title} {Quantum support vector
  machine for big data classification},\ }\href
  {https://doi.org/10.1103/PhysRevLett.113.130503} {\bibfield  {journal}
  {\bibinfo  {journal} {Phys. Rev. Lett.}\ }\textbf {\bibinfo {volume} {113}},\
  \bibinfo {pages} {130503} (\bibinfo {year} {2014})}\BibitemShut {NoStop}%
\bibitem [{\citenamefont {Low}\ and\ \citenamefont
  {Chuang}(2019)}]{low2019hamiltonian}%
  \BibitemOpen
  \bibfield  {author} {\bibinfo {author} {\bibfnamefont {G.~H.}\ \bibnamefont
  {Low}}\ and\ \bibinfo {author} {\bibfnamefont {I.~L.}\ \bibnamefont
  {Chuang}},\ }\bibfield  {title} {\bibinfo {title} {Hamiltonian simulation by
  qubitization},\ }\href@noop {} {\bibfield  {journal} {\bibinfo  {journal}
  {Quantum}\ }\textbf {\bibinfo {volume} {3}},\ \bibinfo {pages} {163}
  (\bibinfo {year} {2019})}\BibitemShut {NoStop}%
\bibitem [{\citenamefont {Haah}\ \emph {et~al.}(2021)\citenamefont {Haah},
  \citenamefont {Hastings}, \citenamefont {Kothari},\ and\ \citenamefont
  {Low}}]{haah2021quantum}%
  \BibitemOpen
  \bibfield  {author} {\bibinfo {author} {\bibfnamefont {J.}~\bibnamefont
  {Haah}}, \bibinfo {author} {\bibfnamefont {M.~B.}\ \bibnamefont {Hastings}},
  \bibinfo {author} {\bibfnamefont {R.}~\bibnamefont {Kothari}},\ and\ \bibinfo
  {author} {\bibfnamefont {G.~H.}\ \bibnamefont {Low}},\ }\bibfield  {title}
  {\bibinfo {title} {Quantum algorithm for simulating real time evolution of
  lattice hamiltonians},\ }\href@noop {} {\bibfield  {journal} {\bibinfo
  {journal} {SIAM Journal on Computing}\ ,\ \bibinfo {pages} {FOCS18}}
  (\bibinfo {year} {2021})}\BibitemShut {NoStop}%
\bibitem [{\citenamefont {Childs}(2017)}]{childs2017lecture}%
  \BibitemOpen
  \bibfield  {author} {\bibinfo {author} {\bibfnamefont {A.~M.}\ \bibnamefont
  {Childs}},\ }\bibfield  {title} {\bibinfo {title} {Lecture notes on quantum
  algorithms},\ }\href@noop {} {\bibfield  {journal} {\bibinfo  {journal}
  {Lecture notes at University of Maryland}\ } (\bibinfo {year}
  {2017})}\BibitemShut {NoStop}%
\bibitem [{\citenamefont {Petrosyan}\ \emph {et~al.}(2017)\citenamefont
  {Petrosyan}, \citenamefont {Motzoi}, \citenamefont {Saffman},\ and\
  \citenamefont {M\o{}lmer}}]{Petrosyan2017}%
  \BibitemOpen
  \bibfield  {author} {\bibinfo {author} {\bibfnamefont {D.}~\bibnamefont
  {Petrosyan}}, \bibinfo {author} {\bibfnamefont {F.}~\bibnamefont {Motzoi}},
  \bibinfo {author} {\bibfnamefont {M.}~\bibnamefont {Saffman}},\ and\ \bibinfo
  {author} {\bibfnamefont {K.}~\bibnamefont {M\o{}lmer}},\ }\bibfield  {title}
  {\bibinfo {title} {High-fidelity rydberg quantum gate via a two-atom dark
  state},\ }\href {https://doi.org/10.1103/PhysRevA.96.042306} {\bibfield
  {journal} {\bibinfo  {journal} {Phys. Rev. A}\ }\textbf {\bibinfo {volume}
  {96}},\ \bibinfo {pages} {042306} (\bibinfo {year} {2017})}\BibitemShut
  {NoStop}%
\bibitem [{\citenamefont {L{\"o}w}\ \emph {et~al.}(2012)\citenamefont
  {L{\"o}w}, \citenamefont {Weimer}, \citenamefont {Nipper}, \citenamefont
  {Balewski}, \citenamefont {Butscher}, \citenamefont {B{\"u}chler},\ and\
  \citenamefont {Pfau}}]{Low2012}%
  \BibitemOpen
  \bibfield  {author} {\bibinfo {author} {\bibfnamefont {R.}~\bibnamefont
  {L{\"o}w}}, \bibinfo {author} {\bibfnamefont {H.}~\bibnamefont {Weimer}},
  \bibinfo {author} {\bibfnamefont {J.}~\bibnamefont {Nipper}}, \bibinfo
  {author} {\bibfnamefont {J.~B.}\ \bibnamefont {Balewski}}, \bibinfo {author}
  {\bibfnamefont {B.}~\bibnamefont {Butscher}}, \bibinfo {author}
  {\bibfnamefont {H.~P.}\ \bibnamefont {B{\"u}chler}},\ and\ \bibinfo {author}
  {\bibfnamefont {T.}~\bibnamefont {Pfau}},\ }\bibfield  {title} {\bibinfo
  {title} {An experimental and theoretical guide to strongly interacting
  rydberg gases},\ }\href {https://doi.org/10.1088/0953-4075/45/11/113001}
  {\bibfield  {journal} {\bibinfo  {journal} {Journal of Physics B: Atomic,
  Molecular and Optical Physics}\ }\textbf {\bibinfo {volume} {45}},\ \bibinfo
  {pages} {113001} (\bibinfo {year} {2012})}\BibitemShut {NoStop}%
\bibitem [{\citenamefont {Yavuz}\ \emph {et~al.}(2006)\citenamefont {Yavuz},
  \citenamefont {Kulatunga}, \citenamefont {Urban}, \citenamefont {Johnson},
  \citenamefont {Proite}, \citenamefont {Henage}, \citenamefont {Walker},\ and\
  \citenamefont {Saffman}}]{yavuz2006fast}%
  \BibitemOpen
  \bibfield  {author} {\bibinfo {author} {\bibfnamefont {D.}~\bibnamefont
  {Yavuz}}, \bibinfo {author} {\bibfnamefont {P.}~\bibnamefont {Kulatunga}},
  \bibinfo {author} {\bibfnamefont {E.}~\bibnamefont {Urban}}, \bibinfo
  {author} {\bibfnamefont {T.~A.}\ \bibnamefont {Johnson}}, \bibinfo {author}
  {\bibfnamefont {N.}~\bibnamefont {Proite}}, \bibinfo {author} {\bibfnamefont
  {T.}~\bibnamefont {Henage}}, \bibinfo {author} {\bibfnamefont
  {T.}~\bibnamefont {Walker}},\ and\ \bibinfo {author} {\bibfnamefont
  {M.}~\bibnamefont {Saffman}},\ }\bibfield  {title} {\bibinfo {title} {Fast
  ground state manipulation of neutral atoms in microscopic optical traps},\
  }\href@noop {} {\bibfield  {journal} {\bibinfo  {journal} {Physical Review
  Letters}\ }\textbf {\bibinfo {volume} {96}},\ \bibinfo {pages} {063001}
  (\bibinfo {year} {2006})}\BibitemShut {NoStop}%
\bibitem [{\citenamefont {Morgado}\ and\ \citenamefont
  {Whitlock}(2021)}]{Morgado2021}%
  \BibitemOpen
  \bibfield  {author} {\bibinfo {author} {\bibfnamefont {M.}~\bibnamefont
  {Morgado}}\ and\ \bibinfo {author} {\bibfnamefont {S.}~\bibnamefont
  {Whitlock}},\ }\bibfield  {title} {\bibinfo {title} {Quantum simulation and
  computing with rydberg-interacting qubits},\ }\href
  {https://doi.org/10.1116/5.0036562} {\bibfield  {journal} {\bibinfo
  {journal} {AVS Quantum Science}\ }\textbf {\bibinfo {volume} {3}},\ \bibinfo
  {pages} {023501} (\bibinfo {year} {2021})},\ \Eprint
  {https://arxiv.org/abs/https://doi.org/10.1116/5.0036562}
  {https://doi.org/10.1116/5.0036562} \BibitemShut {NoStop}%
\bibitem [{\citenamefont {Unanyan}\ and\ \citenamefont
  {Fleischhauer}(2002)}]{Unanyan2002}%
  \BibitemOpen
  \bibfield  {author} {\bibinfo {author} {\bibfnamefont {R.~G.}\ \bibnamefont
  {Unanyan}}\ and\ \bibinfo {author} {\bibfnamefont {M.}~\bibnamefont
  {Fleischhauer}},\ }\bibfield  {title} {\bibinfo {title} {Efficient and robust
  entanglement generation in a many-particle system with resonant dipole-dipole
  interactions},\ }\href {https://doi.org/10.1103/PhysRevA.66.032109}
  {\bibfield  {journal} {\bibinfo  {journal} {Phys. Rev. A}\ }\textbf {\bibinfo
  {volume} {66}},\ \bibinfo {pages} {032109} (\bibinfo {year}
  {2002})}\BibitemShut {NoStop}%
\bibitem [{\citenamefont {Maxwell}\ \emph {et~al.}(2013)\citenamefont
  {Maxwell}, \citenamefont {Szwer}, \citenamefont {Paredes-Barato},
  \citenamefont {Busche}, \citenamefont {Pritchard}, \citenamefont {Gauguet},
  \citenamefont {Weatherill}, \citenamefont {Jones},\ and\ \citenamefont
  {Adams}}]{maxwell2013storage}%
  \BibitemOpen
  \bibfield  {author} {\bibinfo {author} {\bibfnamefont {D.}~\bibnamefont
  {Maxwell}}, \bibinfo {author} {\bibfnamefont {D.}~\bibnamefont {Szwer}},
  \bibinfo {author} {\bibfnamefont {D.}~\bibnamefont {Paredes-Barato}},
  \bibinfo {author} {\bibfnamefont {H.}~\bibnamefont {Busche}}, \bibinfo
  {author} {\bibfnamefont {J.~D.}\ \bibnamefont {Pritchard}}, \bibinfo {author}
  {\bibfnamefont {A.}~\bibnamefont {Gauguet}}, \bibinfo {author} {\bibfnamefont
  {K.~J.}\ \bibnamefont {Weatherill}}, \bibinfo {author} {\bibfnamefont
  {M.}~\bibnamefont {Jones}},\ and\ \bibinfo {author} {\bibfnamefont {C.~S.}\
  \bibnamefont {Adams}},\ }\bibfield  {title} {\bibinfo {title} {Storage and
  control of optical photons using rydberg polaritons},\ }\href@noop {}
  {\bibfield  {journal} {\bibinfo  {journal} {Physical review letters}\
  }\textbf {\bibinfo {volume} {110}},\ \bibinfo {pages} {103001} (\bibinfo
  {year} {2013})}\BibitemShut {NoStop}%
\bibitem [{\citenamefont {Paredes-Barato}\ and\ \citenamefont
  {Adams}(2014)}]{paredes2014all}%
  \BibitemOpen
  \bibfield  {author} {\bibinfo {author} {\bibfnamefont {D.}~\bibnamefont
  {Paredes-Barato}}\ and\ \bibinfo {author} {\bibfnamefont {C.}~\bibnamefont
  {Adams}},\ }\bibfield  {title} {\bibinfo {title} {All-optical quantum
  information processing using rydberg gates},\ }\href@noop {} {\bibfield
  {journal} {\bibinfo  {journal} {Physical review letters}\ }\textbf {\bibinfo
  {volume} {112}},\ \bibinfo {pages} {040501} (\bibinfo {year}
  {2014})}\BibitemShut {NoStop}%
\bibitem [{\citenamefont {Vatan}\ and\ \citenamefont
  {Williams}(2004)}]{Vatan2004}%
  \BibitemOpen
  \bibfield  {author} {\bibinfo {author} {\bibfnamefont {F.}~\bibnamefont
  {Vatan}}\ and\ \bibinfo {author} {\bibfnamefont {C.}~\bibnamefont
  {Williams}},\ }\bibfield  {title} {\bibinfo {title} {Optimal quantum circuits
  for general two-qubit gates},\ }\href
  {https://doi.org/10.1103/PhysRevA.69.032315} {\bibfield  {journal} {\bibinfo
  {journal} {Phys. Rev. A}\ }\textbf {\bibinfo {volume} {69}},\ \bibinfo
  {pages} {032315} (\bibinfo {year} {2004})}\BibitemShut {NoStop}%
\bibitem [{\citenamefont {Petersen}(2010)}]{Petersen2010}%
  \BibitemOpen
  \bibfield  {author} {\bibinfo {author} {\bibfnamefont {I.}~\bibnamefont
  {Petersen}},\ }\bibfield  {title} {\bibinfo {title} {Quantum control theory
  and applications: a survey},\ }\href
  {https://digital-library.theiet.org/content/journals/10.1049/iet-cta.2009.0508}
  {\bibfield  {journal} {\bibinfo  {journal} {IET Control Theory $\&$
  Applications}\ }\textbf {\bibinfo {volume} {4}},\ \bibinfo {pages} {2651}
  (\bibinfo {year} {2010})}\BibitemShut {NoStop}%
\bibitem [{\citenamefont {Choi}\ \emph {et~al.}(2014)\citenamefont {Choi},
  \citenamefont {Debnath}, \citenamefont {Manning}, \citenamefont {Figgatt},
  \citenamefont {Gong}, \citenamefont {Duan},\ and\ \citenamefont
  {Monroe}}]{choi2014optimal}%
  \BibitemOpen
  \bibfield  {author} {\bibinfo {author} {\bibfnamefont {T.}~\bibnamefont
  {Choi}}, \bibinfo {author} {\bibfnamefont {S.}~\bibnamefont {Debnath}},
  \bibinfo {author} {\bibfnamefont {T.}~\bibnamefont {Manning}}, \bibinfo
  {author} {\bibfnamefont {C.}~\bibnamefont {Figgatt}}, \bibinfo {author}
  {\bibfnamefont {Z.-X.}\ \bibnamefont {Gong}}, \bibinfo {author}
  {\bibfnamefont {L.-M.}\ \bibnamefont {Duan}},\ and\ \bibinfo {author}
  {\bibfnamefont {C.}~\bibnamefont {Monroe}},\ }\bibfield  {title} {\bibinfo
  {title} {Optimal quantum control of multimode couplings between trapped ion
  qubits for scalable entanglement},\ }\href@noop {} {\bibfield  {journal}
  {\bibinfo  {journal} {Physical review letters}\ }\textbf {\bibinfo {volume}
  {112}},\ \bibinfo {pages} {190502} (\bibinfo {year} {2014})}\BibitemShut
  {NoStop}%
\bibitem [{\citenamefont {Brown}\ \emph {et~al.}(2016)\citenamefont {Brown},
  \citenamefont {Kim},\ and\ \citenamefont {Monroe}}]{brown2016co}%
  \BibitemOpen
  \bibfield  {author} {\bibinfo {author} {\bibfnamefont {K.~R.}\ \bibnamefont
  {Brown}}, \bibinfo {author} {\bibfnamefont {J.}~\bibnamefont {Kim}},\ and\
  \bibinfo {author} {\bibfnamefont {C.}~\bibnamefont {Monroe}},\ }\bibfield
  {title} {\bibinfo {title} {Co-designing a scalable quantum computer with
  trapped atomic ions},\ }\href@noop {} {\bibfield  {journal} {\bibinfo
  {journal} {npj Quantum Information}\ }\textbf {\bibinfo {volume} {2}},\
  \bibinfo {pages} {1} (\bibinfo {year} {2016})}\BibitemShut {NoStop}%
\bibitem [{\citenamefont {Cruz}\ \emph {et~al.}(2019)\citenamefont {Cruz},
  \citenamefont {Fournier}, \citenamefont {Gremion}, \citenamefont {Jeannerot},
  \citenamefont {Komagata}, \citenamefont {Tosic}, \citenamefont
  {Thiesbrummel}, \citenamefont {Chan}, \citenamefont {Macris}, \citenamefont
  {Dupertuis} \emph {et~al.}}]{cruz2019efficient}%
  \BibitemOpen
  \bibfield  {author} {\bibinfo {author} {\bibfnamefont {D.}~\bibnamefont
  {Cruz}}, \bibinfo {author} {\bibfnamefont {R.}~\bibnamefont {Fournier}},
  \bibinfo {author} {\bibfnamefont {F.}~\bibnamefont {Gremion}}, \bibinfo
  {author} {\bibfnamefont {A.}~\bibnamefont {Jeannerot}}, \bibinfo {author}
  {\bibfnamefont {K.}~\bibnamefont {Komagata}}, \bibinfo {author}
  {\bibfnamefont {T.}~\bibnamefont {Tosic}}, \bibinfo {author} {\bibfnamefont
  {J.}~\bibnamefont {Thiesbrummel}}, \bibinfo {author} {\bibfnamefont {C.~L.}\
  \bibnamefont {Chan}}, \bibinfo {author} {\bibfnamefont {N.}~\bibnamefont
  {Macris}}, \bibinfo {author} {\bibfnamefont {M.-A.}\ \bibnamefont
  {Dupertuis}}, \emph {et~al.},\ }\bibfield  {title} {\bibinfo {title}
  {Efficient quantum algorithms for ghz and w states, and implementation on the
  ibm quantum computer},\ }\href@noop {} {\bibfield  {journal} {\bibinfo
  {journal} {Advanced Quantum Technologies}\ }\textbf {\bibinfo {volume} {2}},\
  \bibinfo {pages} {1900015} (\bibinfo {year} {2019})}\BibitemShut {NoStop}%
\bibitem [{\citenamefont {Saeedi}\ and\ \citenamefont
  {Pedram}(2013)}]{Saeedi2013}%
  \BibitemOpen
  \bibfield  {author} {\bibinfo {author} {\bibfnamefont {M.}~\bibnamefont
  {Saeedi}}\ and\ \bibinfo {author} {\bibfnamefont {M.}~\bibnamefont
  {Pedram}},\ }\bibfield  {title} {\bibinfo {title} {Linear-depth quantum
  circuits for $n$-qubit toffoli gates with no ancilla},\ }\href
  {https://doi.org/10.1103/PhysRevA.87.062318} {\bibfield  {journal} {\bibinfo
  {journal} {Phys. Rev. A}\ }\textbf {\bibinfo {volume} {87}},\ \bibinfo
  {pages} {062318} (\bibinfo {year} {2013})}\BibitemShut {NoStop}%
\bibitem [{Note3()}]{Note3}%
  \BibitemOpen
  \bibinfo {note} {We note that theoretical lower bound for the number of CNOT
  gates required for the $m$-bit Toffoli gate is $\Omega (m)$ \cite
  {shende2008cnot}}\BibitemShut {NoStop}%
\bibitem [{\citenamefont {James}\ \emph {et~al.}(2013)\citenamefont {James},
  \citenamefont {Witten}, \citenamefont {Hastie},\ and\ \citenamefont
  {Tibshirani}}]{james2013introduction}%
  \BibitemOpen
  \bibfield  {author} {\bibinfo {author} {\bibfnamefont {G.}~\bibnamefont
  {James}}, \bibinfo {author} {\bibfnamefont {D.}~\bibnamefont {Witten}},
  \bibinfo {author} {\bibfnamefont {T.}~\bibnamefont {Hastie}},\ and\ \bibinfo
  {author} {\bibfnamefont {R.}~\bibnamefont {Tibshirani}},\ }\href@noop {}
  {\emph {\bibinfo {title} {An introduction to statistical learning}}},\ Vol.\
  \bibinfo {volume} {112}\ (\bibinfo  {publisher} {Springer},\ \bibinfo {year}
  {2013})\BibitemShut {NoStop}%
\bibitem [{\citenamefont {Barredo}\ \emph {et~al.}(2018)\citenamefont
  {Barredo}, \citenamefont {Lienhard}, \citenamefont {De~Leseleuc},
  \citenamefont {Lahaye},\ and\ \citenamefont
  {Browaeys}}]{barredo2018synthetic}%
  \BibitemOpen
  \bibfield  {author} {\bibinfo {author} {\bibfnamefont {D.}~\bibnamefont
  {Barredo}}, \bibinfo {author} {\bibfnamefont {V.}~\bibnamefont {Lienhard}},
  \bibinfo {author} {\bibfnamefont {S.}~\bibnamefont {De~Leseleuc}}, \bibinfo
  {author} {\bibfnamefont {T.}~\bibnamefont {Lahaye}},\ and\ \bibinfo {author}
  {\bibfnamefont {A.}~\bibnamefont {Browaeys}},\ }\bibfield  {title} {\bibinfo
  {title} {Synthetic three-dimensional atomic structures assembled atom by
  atom},\ }\href@noop {} {\bibfield  {journal} {\bibinfo  {journal} {Nature}\
  }\textbf {\bibinfo {volume} {561}},\ \bibinfo {pages} {79} (\bibinfo {year}
  {2018})}\BibitemShut {NoStop}%
\bibitem [{\citenamefont {Deutsch}(1989)}]{deutsch1989quantum}%
  \BibitemOpen
  \bibfield  {author} {\bibinfo {author} {\bibfnamefont {D.~E.}\ \bibnamefont
  {Deutsch}},\ }\bibfield  {title} {\bibinfo {title} {Quantum computational
  networks},\ }\href@noop {} {\bibfield  {journal} {\bibinfo  {journal}
  {Proceedings of the Royal Society of London. A. Mathematical and Physical
  Sciences}\ }\textbf {\bibinfo {volume} {425}},\ \bibinfo {pages} {73}
  (\bibinfo {year} {1989})}\BibitemShut {NoStop}%
\bibitem [{\citenamefont {Suzuki}(1991)}]{suzuki1991general}%
  \BibitemOpen
  \bibfield  {author} {\bibinfo {author} {\bibfnamefont {M.}~\bibnamefont
  {Suzuki}},\ }\bibfield  {title} {\bibinfo {title} {General theory of fractal
  path integrals with applications to many-body theories and statistical
  physics},\ }\href@noop {} {\bibfield  {journal} {\bibinfo  {journal} {Journal
  of Mathematical Physics}\ }\textbf {\bibinfo {volume} {32}},\ \bibinfo
  {pages} {400} (\bibinfo {year} {1991})}\BibitemShut {NoStop}%
\bibitem [{\citenamefont {Childs}\ and\ \citenamefont
  {Su}(2019)}]{childs2019nearly}%
  \BibitemOpen
  \bibfield  {author} {\bibinfo {author} {\bibfnamefont {A.~M.}\ \bibnamefont
  {Childs}}\ and\ \bibinfo {author} {\bibfnamefont {Y.}~\bibnamefont {Su}},\
  }\bibfield  {title} {\bibinfo {title} {Nearly optimal lattice simulation by
  product formulas},\ }\href@noop {} {\bibfield  {journal} {\bibinfo  {journal}
  {Physical review letters}\ }\textbf {\bibinfo {volume} {123}},\ \bibinfo
  {pages} {050503} (\bibinfo {year} {2019})}\BibitemShut {NoStop}%
\bibitem [{\citenamefont {Jordan}\ \emph {et~al.}(2012)\citenamefont {Jordan},
  \citenamefont {Lee},\ and\ \citenamefont {Preskill}}]{Jordan2012}%
  \BibitemOpen
  \bibfield  {author} {\bibinfo {author} {\bibfnamefont {S.~P.}\ \bibnamefont
  {Jordan}}, \bibinfo {author} {\bibfnamefont {K.~S.~M.}\ \bibnamefont {Lee}},\
  and\ \bibinfo {author} {\bibfnamefont {J.}~\bibnamefont {Preskill}},\
  }\bibfield  {title} {\bibinfo {title} {Quantum algorithms for quantum field
  theories},\ }\href {https://doi.org/10.1126/science.1217069} {\bibfield
  {journal} {\bibinfo  {journal} {Science}\ }\textbf {\bibinfo {volume}
  {336}},\ \bibinfo {pages} {1130} (\bibinfo {year} {2012})},\ \Eprint
  {https://arxiv.org/abs/https://science.sciencemag.org/content/336/6085/1130.full.pdf}
  {https://science.sciencemag.org/content/336/6085/1130.full.pdf} \BibitemShut
  {NoStop}%
\bibitem [{\citenamefont {Lieb}\ and\ \citenamefont
  {Robinson}(1972)}]{lieb1972finite}%
  \BibitemOpen
  \bibfield  {author} {\bibinfo {author} {\bibfnamefont {E.~H.}\ \bibnamefont
  {Lieb}}\ and\ \bibinfo {author} {\bibfnamefont {D.~W.}\ \bibnamefont
  {Robinson}},\ }\bibfield  {title} {\bibinfo {title} {The finite group
  velocity of quantum spin systems},\ }in\ \href@noop {} {\emph {\bibinfo
  {booktitle} {Statistical mechanics}}}\ (\bibinfo  {publisher} {Springer},\
  \bibinfo {year} {1972})\ pp.\ \bibinfo {pages} {425--431}\BibitemShut
  {NoStop}%
\bibitem [{\citenamefont {Hastings}(2010)}]{hastings2010locality}%
  \BibitemOpen
  \bibfield  {author} {\bibinfo {author} {\bibfnamefont {M.~B.}\ \bibnamefont
  {Hastings}},\ }\bibfield  {title} {\bibinfo {title} {Locality in quantum
  systems},\ }\href@noop {} {\bibfield  {journal} {\bibinfo  {journal} {Quantum
  Theory from Small to Large Scales}\ }\textbf {\bibinfo {volume} {95}},\
  \bibinfo {pages} {171} (\bibinfo {year} {2010})}\BibitemShut {NoStop}%
\bibitem [{\citenamefont {Low}\ \emph {et~al.}(2016)\citenamefont {Low},
  \citenamefont {Yoder},\ and\ \citenamefont {Chuang}}]{low2016methodology}%
  \BibitemOpen
  \bibfield  {author} {\bibinfo {author} {\bibfnamefont {G.~H.}\ \bibnamefont
  {Low}}, \bibinfo {author} {\bibfnamefont {T.~J.}\ \bibnamefont {Yoder}},\
  and\ \bibinfo {author} {\bibfnamefont {I.~L.}\ \bibnamefont {Chuang}},\
  }\bibfield  {title} {\bibinfo {title} {Methodology of resonant equiangular
  composite quantum gates},\ }\href@noop {} {\bibfield  {journal} {\bibinfo
  {journal} {Physical Review X}\ }\textbf {\bibinfo {volume} {6}},\ \bibinfo
  {pages} {041067} (\bibinfo {year} {2016})}\BibitemShut {NoStop}%
\bibitem [{\citenamefont {Shende}\ and\ \citenamefont
  {Markov}(2008)}]{shende2008cnot}%
  \BibitemOpen
  \bibfield  {author} {\bibinfo {author} {\bibfnamefont {V.~V.}\ \bibnamefont
  {Shende}}\ and\ \bibinfo {author} {\bibfnamefont {I.~L.}\ \bibnamefont
  {Markov}},\ }\bibfield  {title} {\bibinfo {title} {On the cnot-cost of
  toffoli gates},\ }\href@noop {} {\bibfield  {journal} {\bibinfo  {journal}
  {arXiv preprint arXiv:0803.2316}\ } (\bibinfo {year} {2008})}\BibitemShut
  {NoStop}%
\end{thebibliography}

\end{document}